\documentclass[journal,onecolumn,font=12]{IEEEtran}

\usepackage[export]{adjustbox}
\usepackage[cmex10]{amsmath}
\usepackage{amssymb}
\usepackage{color}
\usepackage{bbm}
\usepackage{array}
\usepackage{cite}

\allowdisplaybreaks

\newtheorem{theorem}{Theorem}
\newtheorem{definition}{Definition}
\newtheorem{lemma}{Lemma}
\newtheorem{proposition}{Proposition}
\newtheorem{corollary}{Corollary}
\newtheorem{remark}{Remark}
\newtheorem{example}{Example}
\newcommand{\tp}{\textnormal{tp}}

\begin{document}

\title{Distributed Hypothesis Testing based on Unequal-Error Protection Codes}

\author{Sadaf Salehkalaibar, \emph{IEEE Member} and Mich\`ele Wigger, \emph{IEEE Senior Member}
	\thanks{S.~Salehkalaibar is  with the Department of Electrical and Computer Engineering, College of Engineering, University of Tehran, Tehran, Iran, s.saleh@ut.ac.ir,}
	\thanks{M.~Wigger is with   LTCI,  Telecom ParisTech, Universit\'e Paris-Saclay, 75013 Paris, michele.wigger@telecom-paristech.fr.}
	\thanks{Parts of the material in this paper was presented at  \emph{International Zurich Seminar, Zurich, Switzerland, February 2018.}}
}

\maketitle

\begin{abstract} Coding and testing schemes for binary hypothesis testing over noisy networks are proposed and their corresponding type-II error exponents are derived. When communication is over a discrete memoryless channel (DMC), our scheme combines Shimokawa-Han-Amari's hypothesis testing scheme with Borade's unequal error protection (UEP) for channel coding. A separate source channel coding architecture is employed. The resulting exponent is optimal for the newly introduced class of \emph{generalized testing against conditional independence}. When communication is over a MAC or a BC, our scheme combines  hybrid coding with UEP. The resulting error exponent over the MAC is optimal in the case of generalized testing against conditional independence with independent observations at the two sensors, when the MAC decomposes into two individual DMCs. In this case, separate source-channel coding is sufficient; this same conclusion holds also under arbitrarily  correlated  sensor observations when testing is against independence. For the BC, the error exponents region of hybrid coding with UEP exhibits a tradeoff between the exponents attained at the two decision centers. When both receivers aim at maximizing the error exponents under different hypotheses and the marginal distributions of the sensors' observations are different under these hypotheses, then this tradeoff can be mitigated with the following strategy. The sensor makes a tentative guess on the hypothesis, submits this  guess, and  applies our coding and testing scheme for the DMC only for the decision center  that is not interested in maximizing the exponent under the guessed hypothesis. 
\end{abstract}

\IEEEpeerreviewmaketitle

\section{Introduction}

Sensor networks are important parts of the future Internet of Things (IoT). In these networks, data collected at sensors is transmitted over a wireless medium to remote decision centers, which use this information to decide on one of multiple hypotheses. We follow previous  works in the information theory community \cite{Ahlswede, Han} and assume that the terminals observe   memoryless sequences that follow one of two possible joint distributions, depending on the underlying hypothesis $\mathcal{H}\in\{0,1\}$. The performance of the decision system is characterized by two error probabilities: the probability of type-I error of deciding on $\mathcal{H}=1$ when the true hypothesis is $\mathcal{H}=0$, and the probability of  type-II error of deciding on  $\mathcal{H}=0$ when the true hypothesis is $\mathcal{H}=1$. We consider asymmetric scenarios where one of the two errors (typically the type-II error) is more harmful than the other, and therefore a more stringent constraint on the asymptotic decay of this error probability is imposed. Specifically, the type-I error probability can decay to 0 arbitrarily slowly in the blocklength, whereas the type-II error probability is required to decay exponentially fast. The goal in our research is to find the largest possible type-II error exponent for a given distributed decision system. 

This problem statement has first been considered for the setup with a single sensor and a single decision center when communication is over a noiseless link of given capacity \cite{Ahlswede, Han}.  For this canonical problem, the optimal error exponent has been identified in the special cases of \emph{testing against independence} \cite{Ahlswede} and  \emph{testing against conditional independence}.  
In the former case, the joint distribution of the two sources under $\mathcal{H}=1$ equals the product of the two marginal distributions under $\mathcal{H}=0$. In the latter case, this product structure holds only conditional on a second observation at the decision center, which has same marginal distribution under both hypotheses. The optimal exponent for testing against conditional independence is achieved by the \emph{Shimokawa-Han-Amari (SHA)} scheme \cite{Amari}, which applies Wyner-Ziv source coding combined with  two local joint typicality tests at the sensor (between the quantized sequence and the sensor's observation) and at the decision center (between the quantized sequence and the decision center's observation). The decision center declares the alternative hypothesis $\mathcal{H}=1$ whenever one of the two joint typicality tests fails. To this end, the sensor sends a special $0$-message over the noiseless link to the decision center whenever its local typicality test fails.  The reason for sending this special $0$-message is that given the more stringent constraint on the type-II error probability, the decision center should decide on $\mathcal{H}=1$ in case of slightest doubt. 

The SHA scheme yields an achievable error exponent for  all distributed hypothesis testing problems (not only testing against conditional independence) \cite{Amari}, but it might not be optimal in general \cite{Weinberg}. 
The SHA scheme has been extended to various more involved setups such as  noiseless networks with multiple  sensors  and a single decision center \cite{Han, Wagner,Lai1}; networks where the sensor and the decision center can communicate interactively \cite{Kim, Piantanida}; multi-hop networks \cite{Michele3}, and  networks with multiple decision centers \cite{Michele, Michele3}.

The main focus of this paper is to extend above works to \emph{noisy channels}. In \cite{Gunduz}, it was shown that the optimal exponent for testing against conditional independence over a discrete memoryless channel (DMC) coincides with the optimal exponent for the same test over a noiseless link of rate equal to the capacity of the DMC.  
This performance is achieved by means of hybrid coding, \cite{Minero}, a joint source-channel coding scheme. A similar result is obtained also  for MACs with two individual DMCs connecting the two transmitters to the single receiver  \cite{Gunduz}. In this case, for testing against conditional independence,  separate source-channel coding achieves the same error exponent as when communication is over  noiseless links of same capacities as the DMCs. In these previous works, the optimal error exponent is thus not degraded because channels are noisy. Only   capacity matters. 

 In this paper, we propose  coding and testing schemes for general hypothesis testing over  three basic noisy networks: DMCs, MACs, and broadcast channels (BC).  They allow to treat issues related to multi-acces and to concurrent detections at multiple decision centers. Of course, there are many other interesting communication scenarios one could envision. In particular, multi-hop scenarios \cite{Michele2} are very relevant in practice.  Our schemes strictly improve over the previously proposed schemes, and they suggest that for general hypothesis tests, the transition law of the channel matters;  not only its capacity. 
 
  For DMCs, we propose a scheme that combines the SHA hypothesis testing scheme  in a separate source-channel coding architecture with Borade's Unequal Error Protection (UEP) \cite{Borade, Wornell}  coding that specially protects the   source-coding message $0$. At hand of an example, we show that without the UEP mechanism  the error exponent of our scheme degrades. We further show that the achieved exponent is optimal for a generalization of conditional testing against independence where the observations at the decision center can follow a different marginal distribution depending on the hypothesis.  We thus recover the result in \cite{Gunduz}, but with a separate source-channel coding architecture.

  The   error exponent achieved by our DMC scheme consists of three competing exponents. Two of them coincide with that of the noiseless setup \cite{Amari} when the rate of the noiseless link is replaced by the mutual information between the input and output of the channel. The third error exponent coincides with Borade's \emph{missed-detection} exponent  \cite{Borade}. Depending on the DMC, this third error exponent can be active or not. It is in particular not active for above described  generalized testing against conditional independence, illustrating why the optimal type-II error exponent in this setup only depends on the capacity of the DMC but not on its other properties.

Using hybrid coding  \cite{Minero} instead of separate source-channel coding,  above coding and testing scheme is extended to MACs. In this case, the error exponent achieved by our scheme is expressed in terms of   nine competing exponents. One of them corresponds to that of \cite{Amari}; three of them coincide with an incorrect decoding of the hybrid scheme; three of them correspond to the missed-detection exponents of the UEP scheme; and the other two correspond both to the UEP mechanism and incorrect decoding. The proposed coding scheme establishes the optimal error exponent of the generalized testing against conditional independence when the sources at the transmitters are independent under both hypotheses and the MAC decomposes into two individual DMCs. In this case, hybrid coding can be replaced by separate source-channel coding. Separate source-channel coding can in fact be shown to be sufficient to attain the optimal error exponent  for testing against independence over two individual DMCs. 

For the Gaussian version of this problem, i.e.,  jointly Gaussian sources and Gaussian MAC, we numerically evaluate the error exponents achieve by our coding and testing scheme. We show that this error exponent is close to a new upper bound on the optimal   exponent that we derive based on Witsenhausen's max-correlation argument \cite{Witsenhausen}.

 The last part of this manuscript studies  distributed hypothesis testing  over a BC. Two scenarios can be envisioned here: the two receivers wish to  maximize the error exponent under the same hypothesis, or they wish to maximize the exponents related to two different hypotheses. The first scenarios has previously been studied in \cite{Michele3} for the special case of testing against conditional independence. The second scenario was considered in \cite{Michele2} for the special case of a common noiseless  link from the transmitter to all receivers. We propose coding and testing schemes for both scenarios. Our scheme for the first scenario combines hybrid coding with UEP. The resulting exponents have a similar form as for the MAC, but they exhibit tradeoff between the exponents that can be attained at the two receivers. This tradeoff mostly  stems from the tradeoff that is inherent to any scheme for lossy transmission of a source over a BC with receiver side-information. The same scheme can also be applied to the second scenario when the marginal distributions at the sensor are the same under both hypotheses. 
 
We propose a different scheme for the second scenario when the marginal distributions of the observations at the sensor are different under the two hypotheses. In this case, we suggest that the sensor first performs a tentative decision on the hypothesis. Then, if the sensor thinks that $\mathcal{H}=0$, it sends this guess to both receivers using an UEP mechanism and  continues to apply the previously proposed coding and testing scheme over a DMC to the receiver that is  interested in maximizing the exponent under $\mathcal{H}=1$. If the sensor thinks  $\mathcal{H}=1$,  it will code for the receiver interested in maximizing the exponent under $\mathcal{H}=0$. The  error exponent region corresponding to this scheme, is built on  four competing error exponents at each receiver; two of them coincide with the exponents in  the noiseless setup \cite{Michele2}; one of them with Borade's missed-detection exponent; the fourth corresponds to the event that a decision center wrongly decodes the sensor's tentative decision in favour of the other hypothesis. In this case, the error exponents region achieved by our scheme exhibit only a wek tradeoff between the two exponents. That means, the exponents region is approximately rectangular, and each decision center gets almost the same performance as if the other center was not present.
 
 We conclude this	 introduction with a summary of the main contributions of the paper and remarks on notation. 

\subsection{Contributions}

The main contributions of the paper are as follows.
\begin{itemize}
	\item A coding and testing scheme for DMCs is proposed (Theorem~\ref{thm2noisy} in Section~\ref{sec:p2p}). The scheme is based on separate source-channel coding and unequal error protection (UEP). A matching converse  is derived for generalized testing against conditional independence (Theorem \ref{cor:extended_conditional} in Section~\ref{sec:p2p}), thus establishing the optimal exponent for this case. The employed UEP mechanism allows to significantly improve the error exponent in some cases (Fig.~\ref{noUEP} in Section~\ref{ex-num:p2p}).
	\item A coding and testing scheme  for  MACs is proposed  (Theorem~\ref{macthm} in Section~\ref{MACsection}). The scheme is based on hybrid coding and unequal error protection. A matching converse is derived for generalized testing against conditional independence  over an orthogonal MAC when the sources are independent under both hypotheses (Theorem~\ref{thm2opt} in Section~\ref{MACsection}). In this special case, separate source-channel coding is sufficient. Separate source-channel coding is shown to be optimal also for testing against independence under  arbitrarily correlated sensor observations when the MAC decomposes into two orthogonal DMCs from each of the sensors to the decision center (Proposition~\ref{separation-ortho} in Section~\ref{MACsection}). The results on the MAC are concluded with the study of a Gaussian example, where the error exponent achieved by our scheme numerically matches a newly derived upper bound on the optimal error exponent (Corollary	\ref{cor-gaussian-separate} and Theorem~\ref{gout-thm} in Section~\ref{ex-Gaussian:MAC}, see also Fig.~\ref{fig:MACplot}).
	\item Two different coding and testing schemes for BCs are proposed (Theorem~\ref{thm-BC} in Section~\ref{BCsection}), depending on whether both receivers are interested in the exponent under the same hypothesis or on whether the  marginal pmf of the source observations is the same under both hypotheses. 
	In some cases,   the sensor can make a reasonable guess of the  hypothesis, allowing it to focus on a single decision center. In this case, there is almost no tradeoff in exponents between the two exponents, and the  performance at each decision center is close to the performance of a setup where  the other decision center is  not present.
	\end{itemize}

\subsection{Notation}

We mostly follow the notation in \cite{ElGamal}.  Random variables are denoted by capital letters, e.g., $X,$ $Y,$ and their realizations by lower-case letters, e.g., $x,$ $y$.  Script symbols  such as $\mathcal{X}$ and $\mathcal{Y}$ stand for alphabets of  random variables, and $\mathcal{X}^n$ and $\mathcal{Y}^n$ for the corresponding $n$-fold Cartesian products.  Sequences of random variables $(X_i,...,X_j)$ and realizations $(x_i,\ldots, x_j)$ are  abbreviated by $X_i^j$ and $x_{i}^j$.  When $i=1$, then we also use $X^j$ and $x^j$ instead of $X_1^j$ and $x_{1}^j$. 

We write the probability mass function (pmf) of a discrete random variable $X$  as $P_X$; to indicate the pmf under hypothesis $\mathcal{H}=1$, we also use $Q_X$.  The conditional pmf of $X$ given $Y$  is written as  $P_{X|Y}$, or as $Q_{X|Y}$ when $\mathcal{H}=1$. 
The term   $D(P\| Q)$ stands for  the Kullback-Leibler (KL) divergence between two pmfs $P$ and $Q$ over the same alphabet.  
We use $\tp(\mathbf{a},\mathbf{b})$ to denote the \emph{joint type} of the pair $(\mathbf{a},\mathbf{b})$, and cond\_tp$(\mathbf{a}|\mathbf{b})$ for the conditional type of $\mathbf{a}$ given $\mathbf{b}$. For a joint type $\pi_{ABC}$ over alphabet $\mathcal{A}\times \mathcal{B}\times \mathcal{C}$, we denote by $I_{\pi_{ABC}}(A;B|C)$ the mutual information assuming that the random triple $(A,B,C)$ has  pmf  $\pi_{ABC}$; similarly for the entropy $H_{\pi_{ABC}}(A)$ and the conditional entropy $H_{\pi_{ABC}} (A|B)$. Sometimes we abbreviate $\pi_{ABC}$ by $\pi$. Also, when $\pi_{ABC}$ has been defined and is clear from the context, we write $\pi_{A}$ or $\pi_{AB}$ for the corresponding subtypes. When the type $\pi_{ABC}$ coincides with the actual pmf of a triple $(A,B,C)$, we  omit the subscript and simply write $H(A)$, $H(A|B)$, and $I(A;B|C)$.

For a given $P_X$ and a constant $\mu>0$, let   $\mathcal{T}_{\mu}^n(P_X)$ be the set of \emph{$\mu$-typical sequences} in $\mathcal{X}^n$.
Similarly,   $\mathcal{T}_{\mu}^n(P_{XY})$ stands for the set of \emph{jointly $\mu$-typical sequences}.
The expectation operator is written as $\mathbb{E}[\cdot]$.  We  abbreviate \emph{independent and identically distributed} by \emph{i.i.d.}.
The $\log$ function is  taken with base 2. Finally, in our justifications, we use (DP) and (CR) for ``data processing inequality" and ``chain rule".

%

\begin{figure}[b!]
	\centering
	\includegraphics[scale=0.3]{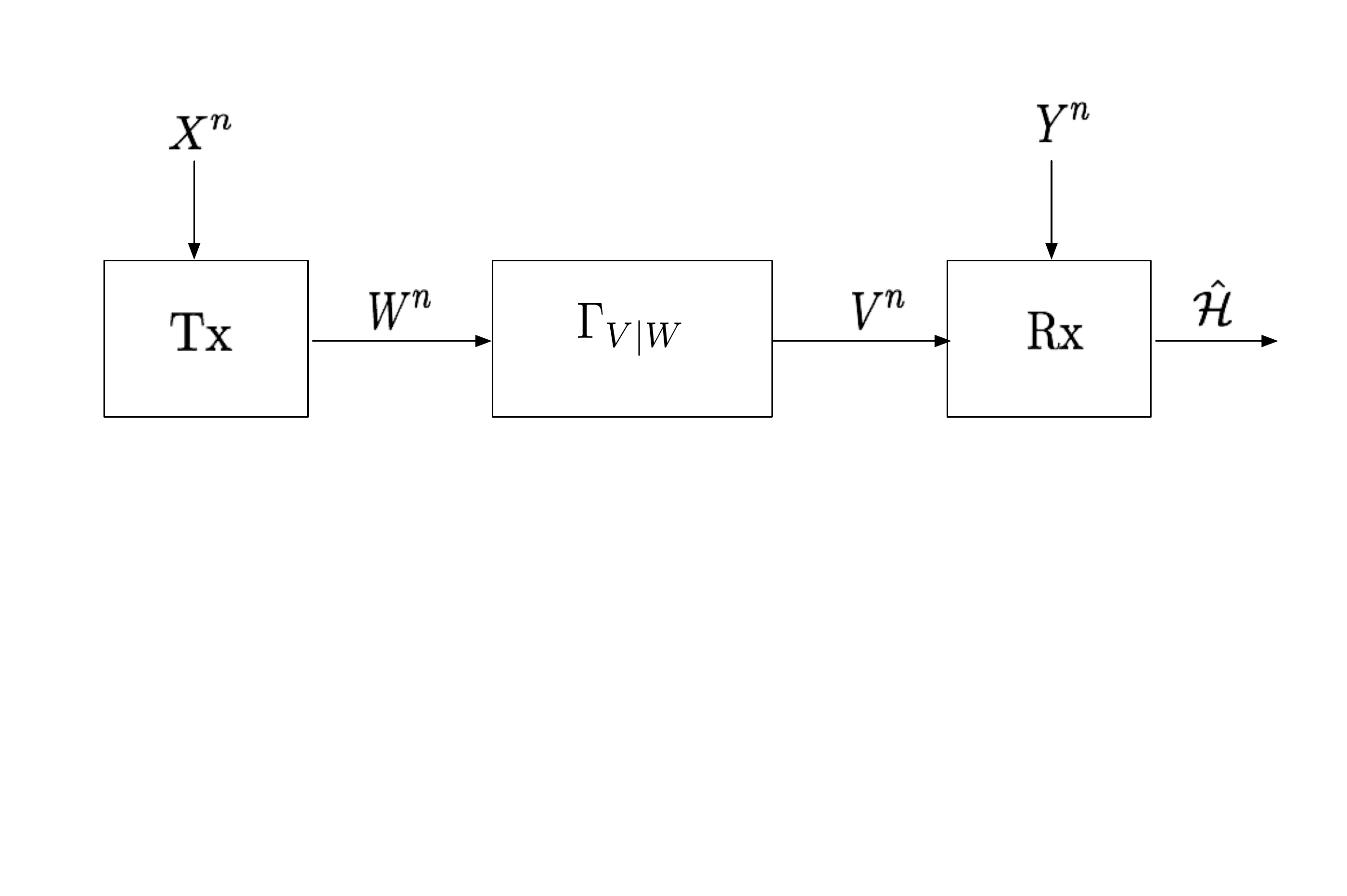}
	\vspace{-2.5cm}
	\caption{Hypothesis testing over a DMC $\Gamma_{V|W}$.}
	\label{fig1}
\end{figure}

\section{Hypothesis Testing over Discrete Memoryless Channels}\label{sec:p2p}

\subsection{System Model}

Consider the distributed hypothesis testing problem in Fig.~\ref{fig1}, where a transmitter observes  source sequence $X^n$ and a receiver   source sequence $Y^n$.  Under the null hypothesis:
\begin{align}
\mathcal{H}=0\colon (X^n,Y^n)\quad \text{i.i.d.}\; \sim P_{XY},
\end{align}
and under the alternative hypothesis:
\begin{align}
\mathcal{H}=1\colon (X^n,Y^n)\quad \text{i.i.d.}\; \sim Q_{XY}.
\end{align}
for two given pmfs $P_{XY}$ and $Q_{XY}$. The transmitter can communicate with the receiver over $n$ uses of a discrete memory channel $(\mathcal{W},\mathcal{V},\Gamma_{V|W})$ where $\mathcal{W}$ denotes the finite channel input alphabet and  $\mathcal{V}$ the finite channel output alphabet. 
Specifically, the transmitter feeds inputs 
\begin{equation}
W^n=f^{(n)}(X^n)
\end{equation}
to the channel, where $f^{(n)}$ denotes the chosen 
(possibly stochastic) encoding function 
\begin{equation}
f^{(n)}:\mathcal{X}^n\to \mathcal{W}^n. 
\end{equation} 

The receiver  observes the ouputs $V^n$, where for a given input $W_t=w_t$,
\begin{align}
V_t \sim \Gamma_{V|W}(\cdot |w_t),\qquad t\in\{1,\ldots,n\}.
\end{align}
Based on the sequence of channel outputs $V^n$ and the source sequence $Y^n$, the  receiver  decides on the hypothesis $\mathcal{H}$. That means, it  produces the guess 
\begin{equation}
\hat{\mathcal{H}}=g^{(n)}(V^n,Y^n),
\end{equation}
by means of a decoding function 
\begin{equation}
g^{(n)} \colon \mathcal{V}^n\times \mathcal{Y}^n\to \{0,1\}.
\end{equation}
\begin{definition}\label{deftype1}
	For each $\epsilon \in (0,1)$, an exponent $\theta$ is said $\epsilon$-achievable, if for each sufficiently large blocklength $n$, there exist encoding and decoding functions $(f^{(n)}, g^{(n)})$ such that the corresponding type-I  and type-II error probabilities at the receiver
	\begin{align}
	\alpha_{n}&\stackrel{\Delta}{=} \Pr[\hat{\mathcal{H}}=1|\mathcal{H}=0],
	\\
	\beta_{n}&\stackrel{\Delta}{=} \Pr[\hat{\mathcal{H}}=0|\mathcal{H}=1],
	\end{align}
	satisfy 
	\begin{align}
	\alpha_{n}&\leq \epsilon,\label{def1bb}
	\end{align}
	and 
	\begin{align}
	-\varlimsup_{n\to \infty}\frac{1}{n}\log\beta_{n}&\geq \theta. \label{def2bb}
	\end{align} 
\end{definition}
The goal  is to maximize the type-II error exponent $\theta$. 
\subsection{Coding and Testing Scheme}\label{sec:scheme}

\begin{figure}[b!]
	\centering
	\includegraphics[scale=0.26]{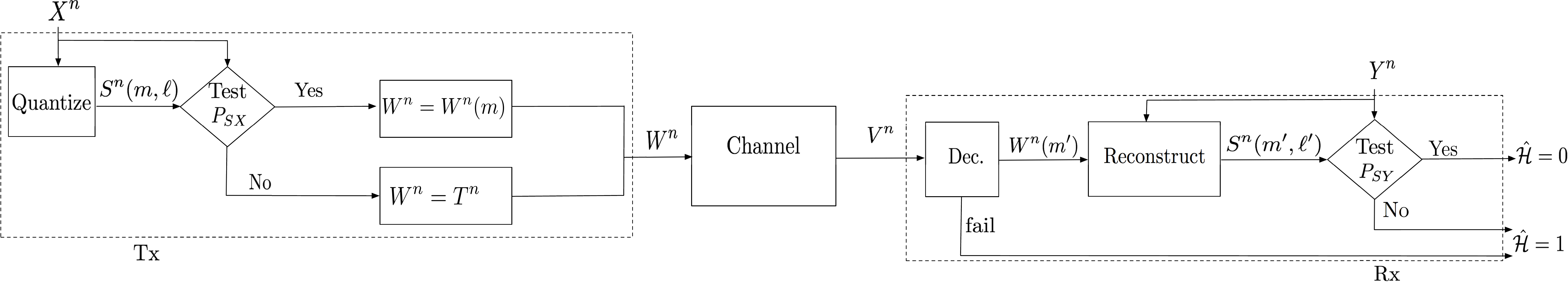}
	\caption{Coding and testing scheme for hypothesis testing over a DMC.}
	\label{coding-p2p}
\end{figure}

We describe a coding and testing scheme for this setup, see Fig.~\ref{coding-p2p}. The analysis of the scheme is postponed to Appendix~\ref{sec:proof}.

\noindent\underline{\textit{Preparations:}}
Choose a large positive integer $n$, an auxiliary distribution $P_T$ over $\mathcal{W}$, a conditional channel input distribution $P_{W|T}$, and a conditional source distribution $P_{S|X}$ over a finite auxiliary alphabet $\mathcal{S}$ so that 
\begin{align}
I(S;X)&< I(S;Y)+I(V;W|T),\label{nois3}
\end{align}
where  mutual informations in this section are calculated according to the following joint distribution
 \begin{align}P_{SXYWVT}=P_{S|X}\cdot P_{XY}\cdot P_T\cdot P_{W|T} \cdot \Gamma_{V|W}.\label{dist}\end{align}
Then, choose a sufficiently small $\mu>0$.  If $I(S;X) < I(W;V|T)$, let 
	\begin{align}
	R &=I(S;X)+\mu,\label{nois0a}\\
	R'&=0.\label{nois1a}
	\end{align}
 If $I(S;X)\geq I(W;V|T)$,  let 
 \begin{align}
 R&=I(W;V|T)-\mu,\label{nois2}\\
 R'&=I(S;X)-I(W;V|T)+2\mu.\label{nois0}
 \end{align}

\noindent\underline{\textit{Code Construction:}}
Construct a random codebook  \begin{equation}\mathcal{C}_S=\big\{S^n(m,\ell)\colon m\in \{1,...,\lfloor 2^{nR}\rfloor\}, \ell\in \{1,...,\lfloor 2^{nR'}\rfloor\}\big\},
\end{equation} by  independently drawing all codewords  i.i.d. according to $P_S(s)=\sum_{x\in\mathcal{X}} P_{X}(x) P_{S|X}(s|x)$. 

Generate a  sequence $T^n$ i.i.d. according to $P_T$.
Construct a random codebook  $$\mathcal{C}_W=\big\{W^n(m):m\in \{1,...,\lfloor 2^{nR}\rfloor\}\big\}$$ superpositioned on $T^n$ where each codeword is drawn  independently  according to $P_{W|T}$ conditioned on $T^n$. Reveal the realizations of the codebooks and the realization of the time-sharing sequence $T^n=t^n$ to all terminals.

Our scheme is based on separate source and channel coding.

\noindent\underline{\textit{Transmitter:}} Given that it observes the source sequence $X^n=x^n$, the transmitter looks for a pair $(m,\ell)$ that satisfies 
\begin{align}
(s^n(m,\ell),x^n)\in \mathcal{T}_{\mu/2}^n(P_{SX}).
\end{align} 
If successful, it picks one of these pairs uniformly at random and sends the codeword $w^n(m)$ over the channel. Otherwise it sends the sequence of inputs $t^n$ over the channel.

\noindent\underline{\textit{Receiver:}} Assume that  $V^n=v^n$ and $Y^n=y^n$. The receiver first looks for an index $m'\in \{1,\ldots,\lfloor 2^{nR} \rfloor\}$ so that 
\begin{align}
(t^n,w^n(m'),v^n) \in \mathcal{T}_{\mu}^n(P_{TWV}).
\end{align}
If it is not successful, it declares $\hat{\mathcal{H}}=1$. Otherwise, it randomly picks one of the indices  $\ell'\in\{1,\ldots,\lfloor 2^{nR} \rfloor\}$ that  satisfy:
\begin{align}
&H_{\text{tp}(s^n(m',\ell'),y^n)}(S|Y) = \min_{\tilde{\ell}\in \{1,...,\lfloor 2^{nR'}\rfloor\}} H_{\text{tp}(s^n(m',\tilde{\ell}),y^n)}(S|Y),\label{MI}
\end{align}
and checks whether
\begin{align}
(s^n(m',\ell'),y^n)&\in \mathcal{T}_{\mu}^n(P_{SY}).
\end{align} 
If successful,  it declares $\hat{\mathcal{H}}=0$. Otherwise, it declares $\hat{\mathcal{H}}=1$.

\subsection{Results on the Error Exponent}
The coding and testing scheme described in the previous section allows to establish the following theorem.

\medskip

\begin{theorem}\label{thm2noisy} Every error exponent $\theta\geq 0$ that satisfies the following condition \eqref{eq:condss} is achievable:
	\begin{align}\label{eq:condss}
\theta &\leq  \max_{\substack{P_{S|X},   P_{TW} \colon \\[.4ex]
		I(S;X|Y) \leq I(W;V|T)		}}\min\big\{\theta^{\text{standard}},\ \theta^{\text{dec}},\ \theta^{\text{miss}}\big\},
\end{align}
where for given (conditional) pmfs $P_{S|X}$ and $P_{TW}$ we define the joint pmf
 \begin{align}P_{SXYWVT}=P_{S|X}\cdot P_{XY}\cdot P_T\cdot P_{W|T} \cdot \Gamma_{V|W}.\label{dist2}
 \end{align}
and the exponents
\begin{align}
\theta^{\text{standard}} &:= \min_{\substack{\tilde{P}_{SXY}:\\\tilde{P}_{SX}=P_{SX}\\\tilde{P}_{SY}=P_{SY}}} D(\tilde{P}_{SXY}\|P_{S|X}Q_{XY}),\label{tht1}\\[2ex]
\theta^{\text{dec}} &:=  \min_{\substack{\tilde{P}_{SXY}:\\\tilde{P}_{SX}=P_{SX}\\\tilde{P}_{Y}=P_{Y}\\H(S|Y)\leq H_{\tilde{P}}(S|Y)}}   D(\tilde{P}_{SXY} \|P_{S|X}Q_{XY}) +I(V ;W|T)-I(S;"X|Y),\label{tht2}\\[2ex]
\theta^{\text{miss}} &:= D(P_{ Y}\|Q_{Y})+I(V;W|T)-I(S;X|Y) +  \sum_{t\in\mathcal{W}} P_T(t) \cdot D(P_{V|T=t}\| \Gamma_{V|W=t}).\label{tht3}
\end{align}
Here, mutual informations and the conditional marginal pmf $P_{V|T}$  are calculated with respect to the joint distribution in \eqref{dist2}.

\end{theorem}
\begin{IEEEproof}
See Appendix~\ref{sec:proof}.
\end{IEEEproof}
\medskip
\begin{lemma}
	It suffices to consider the auxiliary random variable $S$ over an alphabet  $\mathcal{S}$ that is of size $|\mathcal{S}|= |\mathcal{X}|+2$. For the special case of $P_Y=Q_Y$, it suffices to consider $|\mathcal{S}|= |\mathcal{X}|+1$.
	\end{lemma}
	\begin{IEEEproof}
	Based on Carath\'eodory's theorem. Omitted.
	\end{IEEEproof}
	\bigskip 
	
Our coding and testing scheme  combines the SHA hypothesis testing scheme for a noiseless link  \cite{Amari} with Borade's UEP channel coding that protects the $0$-message (which indicates that the transmitter decides on $\mathcal{H}=1$) better than the other messages \cite{Wornell, Borade}.  
 In fact, since here we are only interested in the type-II error exponent, the receiver should decide on $\mathcal{H}=0$ only if the transmitter also shares this opinion. 

The expressions in Theorem~\ref{thm2noisy} show three competing error exponents.   In \eqref{tht1} and \eqref{tht2}, we recognize the two competing error exponents of the SHA scheme for the noiseless setup: $\theta^{\text{standard}}$ is the exponent associated to the event that the receiver reconstructs the correct binned codeword and decides on $\hat{\mathcal{H}}=0$ instead of $\mathcal{H}=1$, and $\theta^{\text{dec}}$ is associated to the event that either the binning or the noisy channel introduces a decoding error followed by a wrong decision on the hypothesis. The exponent $\theta^{\text{miss}}$ in \eqref{tht3} is new and can be associated to the event that the specially protected $0$-message is wrongly decoded followed by a wrong decision on the hypothesis. We remark in particular that $\theta^{\text{miss}}$ contains the term
\begin{equation}\label{eq:redalert}
E_{\textnormal{miss}} :=\sum_{t\in\mathcal{W}} P_T(t) \cdot D(P_{V|T=t}\|P_{V|W=t}),
\end{equation}  which represents the largest possible \emph{miss-detection exponent} for a single specially protected message at a rate $I(W;V|T)$  \cite[Th. 34]{Borade}.


Which of the three exponents  $\theta^{\text{standard}},\theta^{\text{dec}},\theta^{\text{miss}}$ is smallest depends on the source and channel parameters  and of the choice of $P_{S|X}$ and $P_{TW}$. Notice that the third error exponent $\theta^{\text{miss}}$ is inactive for DMCs with large miss-detection exponent \eqref{eq:redalert}, such as binary symmetric channels (BSC) with small cross-over probability. It is also inactive for certain types of sources, irrespective of the employed DMC. This is the content of the next remark.
\begin{remark}\label{rem:no_red}
	For source distributions $P_{XY}$ and $Q_{XY}$  where  irrespective of  the choice of the auxiliary distribution $P_{S|X}$:
	\begin{align}\label{eq:zero_condition}
	\min_{\substack{\phantom{[}\tilde{P}_{SXY}:\\ \tilde{P}_{SX}=P_{SX}\\\tilde{P}_Y=P_Y\\H(S|Y)\leq H_{\tilde{P}}(S|Y)}} \mathbb{E}_{P_Y}[  D(\tilde{P}_{SX|Y}\|P_{S|X}Q_{X|Y}) ]= 0,
	\end{align}  
	error exponent $\theta^{\textnormal{miss}}$  is never smaller than  $\theta^{\textnormal{dec}}$, and  therefore non-active. In this case, it is best to choose $W$ the capacity-achieving input distribution and $T$ a constant.  So, under condition~\eqref{eq:zero_condition}, Theorem~\ref{thm2noisy}  results in:
	\begin{align}\label{eq:d2}
	\theta &\leq  \max_{\substack{P_{S|X}\colon \\[.4ex]
			I(S;X|Y) \leq C		}}\min\big\{\theta^{\textnormal{standard}} ,\theta^{\textnormal{dec}}\big\},
	\end{align}
	where 
	\begin{IEEEeqnarray}{rCl}
		\theta^{\textnormal{standard}} &:=& \min_{\substack{\tilde{P}_{SXY}:\\ \tilde{P}_{SX}=P_{SX}\\ \tilde{P}_{SY}=P_{SY}}} D(\tilde{P}_{SXY}\|Q_{XY}P_{S|X}) \label{eq:standard},\\[1ex]
		\theta^{\textnormal{dec}} &:= & D(P_Y\|Q_Y) +C-I(S;X|Y).\label{eq:binning}
	\end{IEEEeqnarray}
	
	This exponent coincides  with the Shimokawa-Han-Amari exponent \cite{Amari} for these source distributions when communication is rate-limited to the capacity $C$ of the DMC.
\end{remark}

We consider a special case where the expression in \eqref{eq:d2} can be further simplified and the resulting exponent can be proved to be optimal. 
\medskip

\begin{theorem} \label{cor:extended_conditional}
	If there exists a function $f$ from $\mathcal{Y}$ to an auxiliary domain $\mathcal{Z}$ so that  
	\begin{equation}\label{eq:two_conditions}
	\textnormal{under \ } \mathcal{H}=1 \colon \quad X \to f(Y) \to Y ,
	\end{equation}
 the pair $(X,f(Y))$ has the same distribution under both hypotheses, then the optimal error exponent is:
	\begin{equation}
	\theta^*=   D(P_Y\|Q_Y) + \max_{\substack{P_{S|X} \colon \\ I(S;X|f(Y)) \leq C}} I(S;Y|f(Y)),\label{optimal-p2p-Ex}
	\end{equation}
	where $C$ denotes the capacity of the DMC.
\end{theorem}
\begin{IEEEproof} See Appendix~\ref{app:proof_of_corollary}.
\end{IEEEproof}
\medskip

This theorem recovers the optimal error exponents for testing against conditional independence over a noisy channel \cite[Lemma 5]{Gunduz}  or over a noiseless link \cite[Theorem 1]{Wagner}. 
\medskip

Now, we specialize Theorem~\ref{cor:extended_conditional} to Gaussian sources. 	
\begin{example}[Theorem~\ref{cor:extended_conditional} for Gaussian sources]\label{p2p-Gaussian} For given $\rho_0 \in [0,1]$, define the two covariance matrices
	\begin{align}
	\mathbf{K}_{XY}^0=\left[ \begin{array}{cc} 1 & \rho_0 \\ \rho_0  & 1 \end{array} \right] \qquad \textnormal{and}\qquad 	\mathbf{K}_{XY}^1=\left[ \begin{array}{cc} 1 & 0 \\ 0  & 1 \end{array} \right].
	\end{align}
	
	Under the null hypothesis, 
	\begin{align}
	\mathcal{H}=0\colon\qquad (X,Y)\sim \mathcal{N}(0,\mathbf{K}_{XY}^0),
	\end{align}
	and under the alternative hypothesis, 
	\begin{align}
	\mathcal{H}=1\colon\qquad(X,Y)\sim \mathcal{N}(0,\mathbf{K}_{XY}^1).
	\end{align}
	Moreover, assume that the transmitter communicates to the receiver over a DMC of capacity $C$.
	This setup is a special case of Theorem~\ref{cor:extended_conditional}. Appendix~\ref{Gp2p-proof} shows that in this case, the optimal error exponent in \eqref{optimal-p2p-Ex} evaluates to:
	\begin{align}\label{eq:exponent_Ex1}
	\theta^*=\frac{1}{2}\log  \left(\frac{1}{1-\rho_0^2+\rho_0^2\cdot 2^{-2C}}\right).
	\end{align}
	This result recovers as a special case  the optimal exponent  for testing against independence of Gaussian sources over a noiseless link in \cite[Corollary 7]{Wagner}.
\end{example}

\medskip

 \begin{proposition} The result of Theorem \ref{cor:extended_conditional} remains valid when there is instantaneous noise-free  feedback from the receiver to the transmitter. 
	\end{proposition}
\begin{IEEEproof} A close inspection reveals  that the converse proof of the theorem  remains valid even with feedback.
	\end{IEEEproof}

\medskip

\subsection{Numerical Example to Theorem~\ref{thm2noisy}}\label{ex-num:p2p}
We now present an example and evaluate the largest type-II error exponents attained by Theorem~\ref{thm2noisy} for this example. We also show that depending on the parameters of the sources or the channel, a different error exponent $\theta^{\text{standard}},\theta^{\text{dec}},$ or $\theta^{\text{miss}}$ is active.
 Let under the null hypothesis
	\begin{align}
	\mathcal{H}=0 \colon \qquad  X&\sim \text{Bern}(p_0) ,\;\;\;\;\;Y= X\oplus N_0,\;\;\;\;\;\nonumber\\ N_0&\sim \text{Bern}(q_0),
	\end{align}
	for $N_0$ independent of $X$.
	Under the alternative hypothesis:
	\begin{align}
	\mathcal{H}=1 \colon \qquad X\sim \text{Bern}(p_1),\;\;\;\;\; Y\sim \text{Bern}(p_0\star q_0),\label{exh1}
	\end{align}
	with $X$ and $Y$ independent. Assume that  $\Gamma_{V|W}$ is a binary symmetric channel (BSC) with cross-over probability $r\in[0,1/2]$. 
	
	For this example, $P_Y=Q_Y$ and Theorem~\ref{thm2noisy} simplifies to: 	
		\begin{align}\label{eq:cond}
\theta &\leq  \max_{\substack{P_{S|X},   P_{TW} \colon \\[.4ex]
		I(S;X|Y) \leq I(W;V|T)		}}\min\big\{\theta^{\text{standard}},\theta^{\text{dec}},\theta^{\text{miss}}\big\},
\end{align} where
\begin{align}
	\theta^{\text{standard}} &\leq  D(P_{X}\|Q_X)+I(S;Y),\\
	\theta^{\text{dec}} &\leq  D(P_X\|Q_X)+I(V ;W|T) +I(S;Y)-I(S;X),\\
	\theta^{\text{miss}} &\leq  \sum_{t\in\mathcal{W}} P_T(t) D(P_{V|T=t}\|P_{V|W=t}) + I(V ;W|T) +I(S;Y)-I(S;X).\label{eq:miss}
	\end{align}
Depending on the parameters of the setup and the choice of the auxiliary distributions, either of the exponents $\theta^{\text{standard}},\theta^{\text{dec}}$, or $\theta^{\text{miss}}$ is active. For example,   when the cross-over probability of the BSC is large, $r\geq 0.4325$, then
	\begin{align}
	D(P_X\|Q_X) &\geq \sum_{t\in\mathcal{W}} P_T(t) D(P_{V|T=t}\|\Gamma_{V|W=t})  +I(V ;W|T),
	\end{align}
	and  irrespective of the choice of the random variables $S, T, W$ the exponent $\theta^{\text{miss}}$ is smaller than $\theta^{\text{standard}}$ and $\theta^{\text{dec}}$. Since by the Markov chain $S-X-Y$, we have $I(S;Y)-I(S;X) <0$, it is then optimal  to choose $S$ constant and $(T,W)$ so as to maximize the sum 
	\begin{equation}
\sum_{t\in\mathcal{W}} P_T(t) D(P_{V|T=t}\|\Gamma_{V|W=t}) +I(V ;W|T) = \sum_{t,w\in\mathcal{W}} P_{TW}(t,w) D( \Gamma_{V|W=w} \| \Gamma_{V|W=t} ).
	\end{equation}That means, choose  $W$ and $T$ deterministically equal to two maximally distinguishable inputs. Since on a BSC there are only two inputs ($0$ and $1$) and the channel law is completely symmetric with respect to these inputs, for $r \in (0.4325,0.5)$ the largest error exponent achieved by our scheme is:
	\begin{equation}
\hat{\theta}:=  \max_{\substack{P_{S|X},P_{TW}:\\I(S;X|Y)\leq I(W;V|T)}} \min \{\theta^{\text{standard}},\theta^{\text{dec}},\theta^{\text{miss}}\}= D(P_{V|W=0}\|P_{V|W=1}) = (1-2r)\log\frac{1-r}{r}.\label{eq:hat_theta}
	\end{equation} 
For example, when $r=\frac{4}{9}$,  one  obtains  $\hat\theta=0.0358= \frac{1}{9}\log \frac{5}{4}$. 
	
In contrast, when the cross-over probability of the BSC is small, the miss-detection  exponent \eqref{eq:redalert} is large and the exponent $\theta^{\text{miss}}$ is never active irrespective of the choice of the auxiliary random variable $S$. The overall exponent is then determined by the smaller of $\theta^{\text{standard}}$ and $\theta^{\text{dec}}$, and in particular by a choice  $S,X,W$ that makes the two equal. In this case, for a scenario with parameters $p_0=0.2, q_0=0.3, p_1=0.4$, and $r=0.1$, the largest exponent achieved by our scheme is $\theta=0.19$. 

\begin{figure}[t]
	\centering
	\vspace{-2cm}
	\includegraphics[scale=0.5]{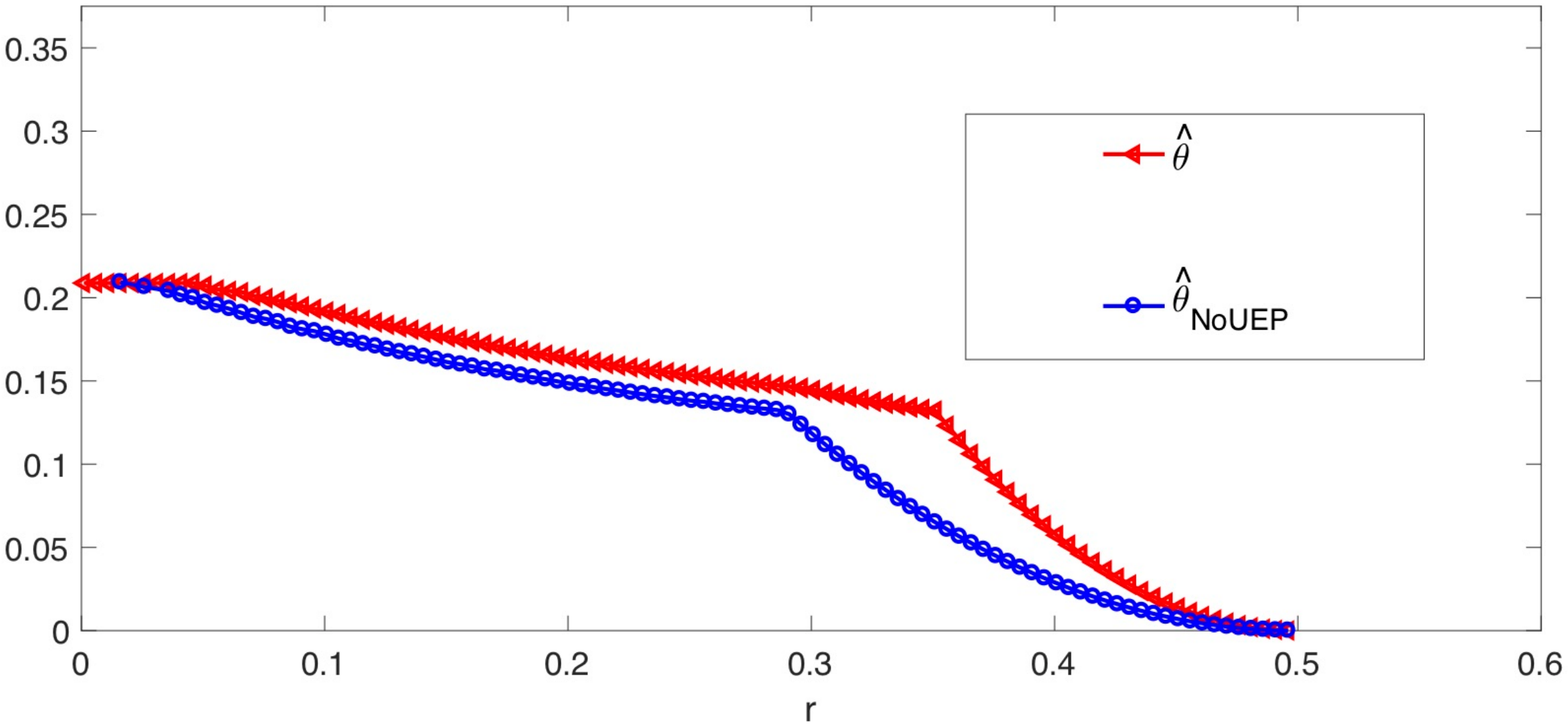}
	\vspace{-2cm}
	\caption{The  achievable error exponents with and without unequal error protection, $\hat{\theta}$ in \eqref{eq:hat_theta} and $ \hat{\theta}_{\textnormal{NoUEP}}$ in \eqref{eq:cond_noUEP}, for  the proposed example with $p_0=0.2$, $p_1=0.4$ and $q_0=0.3$}
	\label{noUEP}
\end{figure}

In the following, we study the maximum error exponent achieved by our scheme $\hat{\theta}$ in function of the  channel cross-over probability $r$.
This dependency  is shown in Figure \ref{noUEP}, and Table \ref{table1} indicates which of the three exponents $\theta^{\text{standard}},\theta^{\text{dec}},$ $\theta^{\text{miss}}$ is smallest. 
 Notice that for $r\geq 0.296$,  error exponent $\theta^{\text{miss}}$ is  smallest, and for $r\leq 0.046$,  error exponent $\theta^{\text{standard}}$ is smallest.
 
An important feature of our scheme is the UEP mechanism used to send the $0$-message. In fact, if the $0$-message had been sent using an ordinary codeword  from codebook $\mathcal{C}_W$, then exponent $	\theta^{\text{miss}}$ in \eqref{eq:miss} had to  be replaced by the smaller exponent  
\begin{align}
\theta_{\text{no-UEP}}^{\text{miss}} = D(P_Y\| Q_Y)+I(V;W)-I(S;X|Y).
\end{align}
Notice that $\theta_{\text{no-UEP}}^{\text{miss}}  \leq \theta^{\text{dec}}$ and thus without UEP  our coding and testing scheme would achieve only  exponents that satisfy
		\begin{align}\label{eq:cond_noUEP}
		\theta &\leq \hat{\theta}_{\textnormal{NoUEP}}:= \max_{\substack{P_{S|X},   P_{TW} \colon \\[.4ex]
				I(S;X|Y) \leq I(W;V|T)		}}\min\big\{\theta^{\text{standard}},\ \theta_{\text{no-UEP}}^{\text{miss}} \big\},
		\end{align} Figure \ref{noUEP} also shows the exponent in \eqref{eq:cond_noUEP}. 

\begin{table}[t!]
	\centering
	
		\begin{tabular}{|c|c|c|}
		\hline
		& & \\
		&$0\leq r\leq 0.286$& $0.286\leq r\leq 0.5$\\ & & \\
		\hline
		& & \\
		$\hat{\theta}_\text{NoUEP}$  & $\theta^{\text{miss}}_{\text{no-UEP}}=\theta^{\text{standard}}$ & $\theta^{\text{miss}}_{\text{no-UEP}}\leq\theta^{\text{standard}}$  \\
		& & \\
		\hline\hline
	\end{tabular}
		
		\begin{tabular}{ |c| c |  c | c  | c |}				
			\hline
			& & & & \\
			&$0\leq r\leq 0.046$ & $0.046\leq r\leq 0.296$ & $0.296\leq r\leq 0.351$ & $0.351\leq r\leq 0.5$ \\
			& &  & & \\
			\hline 
			& &  & & \\
			$\hat{\theta}$& $\theta^{\text{standard}}\leq \min\{\theta^{\text{dec}},\theta^{\text{miss}}\}$ & $\theta^{\text{dec}}=\theta^{\text{standard}}\leq \theta^{\text{miss}}$ & $\theta^{\text{standard}}=\theta^{\text{miss}}\leq \theta^{\text{dec}}$ & $\theta^{\text{miss}}\leq \min\{\theta^{\text{standard}},\theta^{\text{dec}}\}$\\ 
			&  &  & & \\
			\hline 
			\end{tabular}
	\vspace{3mm}
\caption{The smallest error exponent as a function of $r$}\label{table1}
\end{table}

\section{Hypothesis Testing over Multi-Access Channels}\label{MACsection}

\subsection{System Model}

\begin{figure}[b]
	\centering	\includegraphics[scale=0.3]{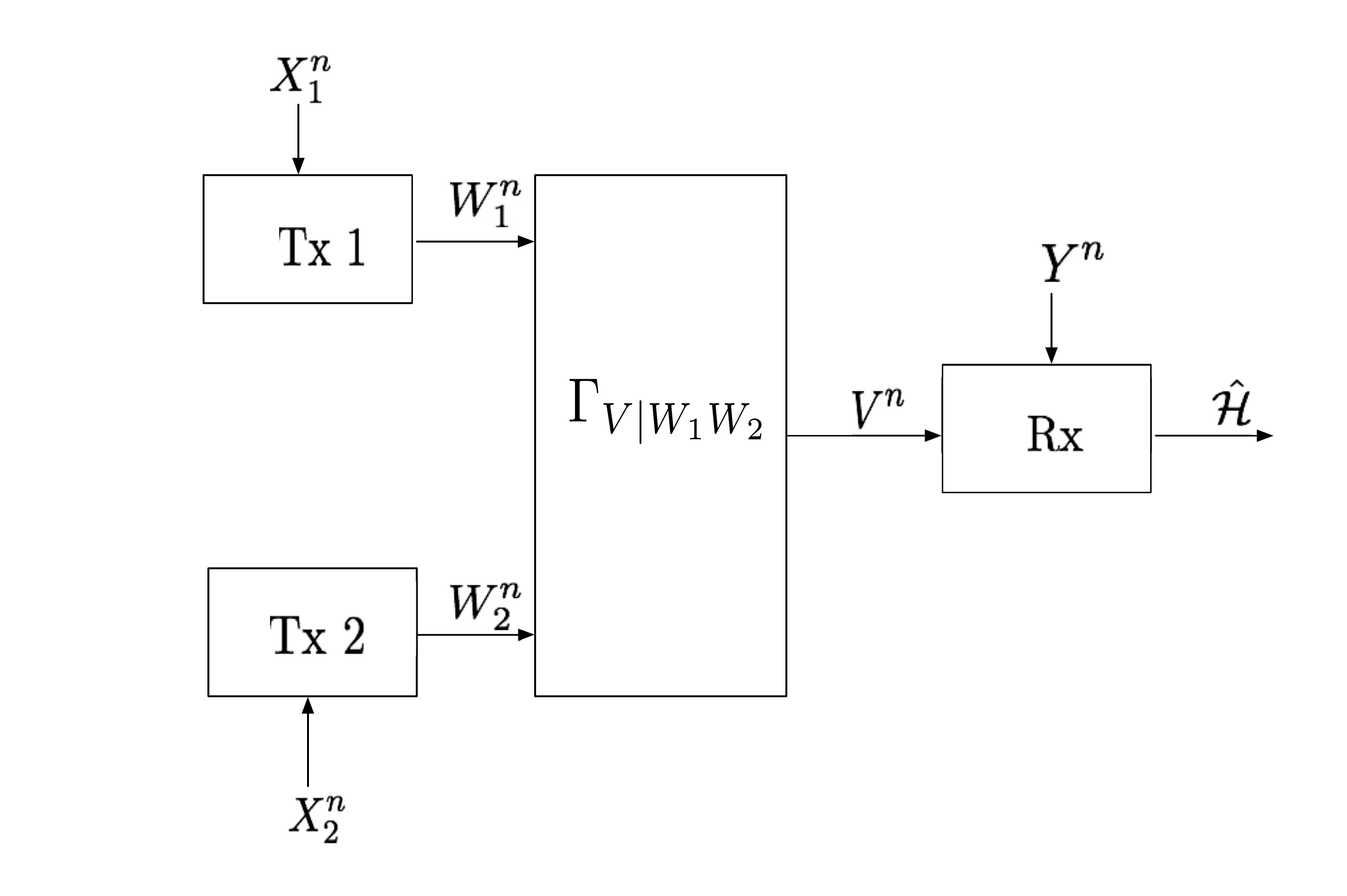}
	\caption{Hypothesis testing over a noisy MAC}
	\label{fig22}
\end{figure}

Consider a setup with two sensors  that communicate to a single decision center over a discrete memoryless  multiple-access channel (MAC), see Fig.~\ref{fig22}. The channel is described by the quadruple $(\mathcal{W}_1\times \mathcal{W}_2,\mathcal{V},\Gamma_{V|W_1,W_2})$, where $\mathcal{W}_1$ and $\mathcal{W}_2$ denote the finite channel input alphabets and  $\mathcal{V}$ denotes the finite channel output alphabet.   Each transmitter $i$  ($i=1,2$) observes the sequence $X_i^n$ and produces channel inputs $W_i^n$ as 
\begin{equation}
W_i^n=f_i^{(n)}(X_i^n)
\end{equation} by means of a possibly stochastic encoding function 
\begin{equation}f_i^{(n)}\colon \mathcal{X}_i^n\to \mathcal{W}^n.
\end{equation} The receiver observes the corresponding channel outputs $V^n$ as well as the source sequence $Y^n$. Under the null hypothesis
\begin{align}
\mathcal{H}=0 \colon \;\;\; (X_1^n,X_2^n,Y^n)\sim \text{i.i.d.}\qquad P_{X_1X_2Y},
\end{align}
and under the alternative hypothesis
\begin{align}
\mathcal{H}=1 \colon \;\;\; (X_1^n,X_2^n,Y^n)\sim \text{i.i.d.}\qquad Q_{X_1X_2Y},
\end{align}
for two given pmfs $P_{X_1X_2Y}$ and $Q_{X_1X_2Y}$. The receiver should decide on the hypothesis $\mathcal{H}$. Besides $Y^n$, it also observes the MAC ouputs $V^n$, where for given inputs $W_{1,t}=w_{1,t}$ and $W_{2,t}=w_{2,t}$, 
\begin{align}
V_t \sim \Gamma_{V|W_1,W_2}( \cdot | w_{1,t}, w_{2,t}),\qquad t\in\{1,\ldots,n\},
\end{align}It thus produces the guess 
\begin{equation}
\hat{\mathcal{H}}=g^{(n)}(V^n,Y^n)
\end{equation} using a decoding function 
\begin{equation}\mathcal{V}^n\times \mathcal{Y}^n\to \{0,1\}.
\end{equation}

\medskip 
\begin{definition}\label{deftype1}
	For each $\epsilon \in (0,1)$, an exponent $\theta$ is said $\epsilon$-achievable, if for each sufficiently large blocklength $n$, there exist encoding and decoding functions $(f^{(n)}, g^{(n)})$ such that the corresponding type-I  and type-II error probabilities at the receiver
	\begin{align}
	\alpha_{n}&\stackrel{\Delta}{=} \Pr[\hat{\mathcal{H}}=1|\mathcal{H}=0],
	\\
	\beta_{n}&\stackrel{\Delta}{=} \Pr[\hat{\mathcal{H}}=0|\mathcal{H}=1],
	\end{align}
	satisfy 
	\begin{align}
	\alpha_{n}&\leq \epsilon,\label{def1bb}
	\end{align}
	and 
	\begin{align}
	-\varlimsup_{n\to \infty}\frac{1}{n}\log\beta_{n}&\geq \theta. \label{def2bb}
	\end{align} 
\end{definition}
The goal  is to maximize the type-II error exponent $\theta$. 

\subsection{Coding and Testing Scheme}\label{sec:MACcoding}

\begin{figure}[b!]
	\centering
	\includegraphics[scale=0.26]{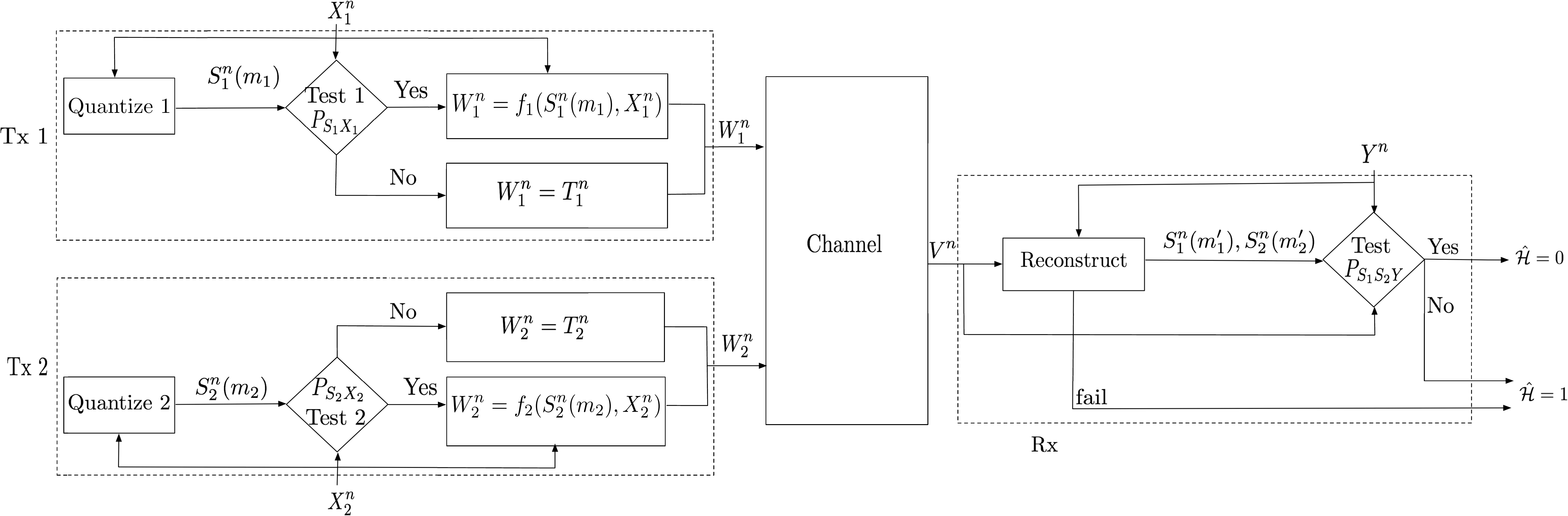}
	\caption{Coding and testing scheme for hypothesis testing over a MAC.}
	\label{coding-MAC}
\end{figure}

We describe a coding and testing scheme for distributed hypothesis testing over a noisy MAC, see Fig.~\ref{coding-MAC}.

\noindent\underline{\textit{Preparations:}}
Choose a sufficiently large blocklength $n$, auxiliary alphabets $\mathcal{S}_1$ and $\mathcal{S}_2$, and functions 
\begin{equation}\label{eq:function}
f_i\colon \mathcal{S}_i\times \mathcal{X}_i\to \mathcal{W}_i, \qquad i\in\{1,2\},
\end{equation}
and define the shorthand notation
\begin{IEEEeqnarray}{rCl}\label{eq:channel2}
\Gamma_{V|S_1S_2X_1X_2}(v|s_1,s_2,x_1,x_2): = \Gamma_{V|W_1,W_2}(v| f_1(s_1,x_1), f_2(s_2,x_2)), \qquad \forall s_1 \in \mathcal{S}_1, s_2\in\mathcal{S}_2, x_1\in \mathcal{X}_1,x_2\in\mathcal{X}_2.
\end{IEEEeqnarray}
Choose then a distribution $P_{T_1T_2}$ over $\mathcal{W}_1\times \mathcal{W}_2$, and for $i\in\{1,2\}$, a conditional  distribution $P_{S_i|X_iT_1T_2}$ over  $\mathcal{S}_i$ in a way that:
\begin{subequations}\label{eq:Iconditions}
	\begin{align}
	I(S_1;X_1|T_1,T_2)&< I(S_1;S_2,Y,V|T_1,T_2),\label{nois3mac}\\
	I(S_2;X_2|T_1,T_2)&< I(S_2;S_1,Y,V|T_1,T_2),\label{nois5mac}\\
	I(S_1,S_2;X_1,X_2|T_1,T_2)&<  I(S_1,S_2;Y,V|T_1,T_2)\label{nois6mac}
	\end{align}
\end{subequations}
when these mutual informations and all subsequent mutual informations in this section are evaluated according to the joint pmf
 \begin{IEEEeqnarray}{rCl}
 P_{S_1S_2X_1X_2YT_1T_2V}  & =  & P_{S_1|X_1T_1T_2} \cdot P_{S_2|X_2T_1T_2}\cdot  P_{X_1X_2Y} \cdot P_{T_1T_2} \cdot \Gamma_{V|S_1S_2X_1X_2}.\label{distmac}
\end{IEEEeqnarray}
Further, choose $\mu>0$ and positive rates:
\begin{align}
R_i&= I(S_i;X_i|T_1,T_2)+\mu,\;\;\;\;\;\;\;\;\; i\in\{1,2\},\label{nois1mac}
\end{align}
so that the following three conditions hold:\begin{subequations}
	\begin{align}
	R_1 &< I(S_1;S_2,Y,V|T_1,T_2),\label{nois2mac}\\
	R_2&< I(S_2;S_1,Y,V|T_2,T_2),\label{nois32mac}\\
	R_1+R_2 &< I(S_1,S_2;Y,V|T_1,T_2)+I(S_1;S_2|T_1,T_2).\label{nois4mac}
	\end{align}
\end{subequations}

\noindent\underline{\textit{Code Construction:}} Generate a pair of sequences $T_1^n=(T_{1,1},\ldots,T_{1,n})$ and $T_2^n=(T_{2,1},\ldots,T_{2,n})$ by independently drawing each pair $(T_{1,k},T_{2,k})$ according to $P_{T_1T_2}(.,.)$.
For $i\in\{1,2\}$, construct a random codebook  
\begin{equation}\mathcal{C}_{S_i}=\big\{S_i^n(m_i)\colon m_i\in \{1,...,\lfloor 2^{nR_i}\rfloor\}\big\},
\end{equation} superpositioned on $(T_1^n,T_2^n)$ by  independently drawing the $k$-th component of each codeword according to the conditional law $\ P_{S_i|T_1T_2}(\cdot |x_i,t_1,t_2)$ when $X_{i,k}=x_i, T_{1,k}=t_1$, and $T_{2,k}=t_2$.
Reveal the realizations of the codebooks and the realizations $(t_1^n,t_2^n)$ of $(T_1^n,T_2^n)$ to all terminals.

Our scheme is based on hybrid coding. 

\noindent\underline{\textit{Transmitter $i\in\{1,2\}$:}} Given  source sequence $X_i^n=x_i^n$,  Transmitter~$i$ looks for an index $m_i$ that satisfies 
\begin{align}
(s_i^n(m_i),x_i^n,t_1^n,t_2^n)\in \mathcal{T}_{\mu/2}^n(P_{S_iX_iT_1T_2}).
\end{align} 
If successful, it picks one of these indices uniformly at random and sends the sequence $w_i^n$ over the channel, where
$$w_{i,k}=f_{i}(s_{i,k}(m_i),x_{i,k}),\;\;\;\;\; k \in \{1,\ldots, n\},$$
and where $s_{i,k}(m_{i})$ denotes the $k$-th component of codeword $s_i^{n}(m_i)$.
Otherwise, Transmitter $i$ sends $t_i^n$ over the channel.

\noindent\underline{\textit{Receiver:}} Assume that the receiver observes the sequences $V^n=v^n$ and $Y^n=y^n$. It first searches for a pair of indices  $(m'_1,m'_2)$ that satisfies the condition:
\begin{align}
&H_{\text{tp}(s_1^n(m'_1),s_2^n(m'_2),y^n,t_1^n,t_2^n,v^n)}(S_1,S_2|Y,T_1,T_2,V) =\min_{\substack{\tilde{m}_1,\tilde{m}_2}}  H_{\text{tp}(s_1^n(\tilde{m}_1),s_2^n(\tilde{m}_2),y^n,t_1^n,t_2^n,v^n)}(S_1,S_2|Y,T_1,T_2,V).
\end{align}
It picks one such pair at random and checks whether the chosen pair $(m_1',m_2')$ satisfies
\begin{align}
(s_1^n(m'_1),s_2^n(m'_2),y^n,t_1^n,t_2^n,v^n)&\in \mathcal{T}_{\mu}^n(P_{S_1S_2YT_1T_2V}).
\end{align} 
If successful,  it declares $\hat{\mathcal{H}}=0$. Otherwise, it declares $\hat{\mathcal{H}}=1$.

\subsection{Results on the Error Exponent}
The coding and testing scheme described in the previous section yields Theorem \ref{macthm} ahead. For given (conditional) pmfs $P_{T_1T_2}$, $P_{S_1|X_1T_1T_2}$, and $P_{S_2|X_2T_1T_2}$, and  functions  $f_1$ and $f_2$  as in \eqref{eq:function}, let the conditional and joint pmfs  $\Gamma_{V|S_1S_2X_1X_2}$ and $P_{S_1S_2X_1X_2YW_1W_2VT_1T_2}$ be as in  \eqref{eq:channel2} and \eqref{distmac}. Define also for all  $s_1 \in \mathcal{S}_1$, $s_2 \in \mathcal{S}_2$, $t_1 \in \mathcal{T}_1$, $t_2 \in \mathcal{T}_2$, $x_1 \in \mathcal{X}_1$, $x_2 \in \mathcal{X}_2$, and $v\in\mathcal{V}$:
\begin{IEEEeqnarray}{rCl}
\Gamma^{(1)}_{V|T_1S_2X_2}(v|t_1,s_2,x_2) &:= &\Gamma_{V|W_1W_2}(v| t_1, f_2(s_2,x_2)) \\
\Gamma^{(2)}_{V|S_1X_1T_2}(v|s_1,x_1,t_2) &:= &\Gamma_{V|W_1W_2}(v|  f_1(s_1,x_1), t_2) \\
\Gamma^{(12)}_{V|T_1T_2}(v|t_1,t_2) &:= &\Gamma_{V|W_1W_2}(v| t_1, t_2),
\end{IEEEeqnarray}
and the following nine exponents:
\begin{align}
\theta^{\text{standard}}&:=\!\!\!\!\!\!\!\!\min_{\substack{\tilde{P}_{S_1S_2X_1X_2YT_1T_2V}:\\\tilde{P}_{S_iX_iT_1T_2}=P_{S_iX_iT_1T_2}, \; i\in\{1,2\},\\\tilde{P}_{S_1S_2YT_1T_2V}=P_{S_1S_2YT_1T_2V}}}\!\!\!\!\!\!\!\!\!\!\!\!\! D\left(\tilde{P}_{S_1S_2X_1X_2YT_1T_2V}\|P_{S_1|X_1T_1T_2} P_{S_2|X_2T_1T_2} Q_{X_1X_2Y}P_{T_1T_2}\Gamma_{V|S_1S_2X_1X_2}\right),\nonumber
\\&\label{thnoiseless}\\[-1ex]
\theta^{\text{dec,1}} &:= \!\!\!\!\!\! \min_{\substack{\tilde{P}_{S_1S_2X_1X_2YT_1T_2V}:\\\tilde{P}_{S_iX_iT_1T_2}=P_{S_iX_iT_1T_2}, \; i\in\{1,2\},\\\tilde{P}_{S_2YT_1T_2V}=P_{S_2YT_1T_2V}\\H(S_1|S_2,Y,T_1,T_2,V)\leq H_{\tilde{P}}(S_1|S_2,Y,T_1,T_2,V)}} \hspace{-0.5cm}D\left(\tilde{P}_{S_1S_2X_1X_2 YT_1T_2V}\|P_{S_1|X_1T_1T_2} P_{S_2|X_2T_1T_2} Q_{X_1X_2Y} P_{T_1T_2}  \Gamma_{V|S_1S_2X_1X_2}  \right)\nonumber\\[-6ex]
& \hspace{8cm} +I(S_1;Y,V|S_2,T_1,T_2)-I(S_1;X_1|S_2,T_1,T_2),\label{th1noisy}\\[3.5ex]
\theta^{\text{dec,2}} &:= \!\!\!\!\!\!\!\!\!\!\!\!\min_{\substack{\tilde{P}_{S_1S_2X_1X_2YT_1T_2V}:\\\tilde{P}_{S_iX_iT_1T_2}=P_{S_iX_iT_1T_2}, \; i\in\{1,2\},\\\tilde{P}_{S_1YT_1T_2V}=P_{S_1YT_1T_2V}\\H(S_2|S_1,Y,T_1,T_2,V)\leq H_{\tilde{P}}(S_2|S_1,Y,T_1,T_2,V)}} D\left(\tilde{P}_{S_1S_2X_1X_2 YT_1T_2V}\|P_{S_1|X_1T_1T_2} P_{S_2|X_2T_1T_2} Q_{X_1X_2Y} P_{T_1T_2} \Gamma_{V|S_1S_2X_1X_2}  \right)\nonumber\\[-6ex]
& \hspace{8cm}+I(S_2;Y,V|S_1,T_1,T_2)-I(S_2;X_2|S_1,T_1,T_2),\label{th2noisy}\\[3.5ex]
\theta^{\text{dec},12} &:=\!\!\!\!\!\! \min_{\substack{\tilde{P}_{S_1S_2X_1X_2YT_1T_2V}:\\\tilde{P}_{S_iX_iT_1T_2}=P_{S_iX_iT_1T_2}, \; i\in\{1,2\},\\\tilde{P}_{YT_1T_2V}=P_{YT_1T_2V}\\H(S_1,S_2|Y,T_1,T_2,V)\leq H_{\tilde{P}}(S_1,S_2|Y,T_1,T_2,V)}} D\left(\tilde{P}_{S_1S_2X_1X_2 YT_1T_2V}\|P_{S_1|X_1T_1T_2}  P_{S_2|X_2T_1T_2} Q_{X_1X_2Y} P_{T_1T_2}  \Gamma_{V|S_1S_2X_1X_2} \right)\nonumber\\[-6ex]
&\hspace{8cm} +I(S_1,S_2;Y,V|T_1,T_2)-I(S_1,S_2;X_1,X_2|T_1,T_2),\label{th3noisy}\\[3.5ex]
\theta^{\text{miss},1\textnormal{a}} &:=\min_{\substack{\tilde{P}_{S_2X_2YT_1T_2V}:\\\tilde{P}_{S_2X_2T_1T_2}=P_{S_2X_2T_1T_2}\\\tilde{P}_{YT_1T_2V}=P_{YT_1T_2V}\\H(S_2|Y,T_1,T_2,V)\leq H_{\tilde{P}}(S_2|Y,T_1,T_2,V)}} D\left(\tilde{P}_{S_2X_2YT_1T_2V}\|P_{S_2|X_2T_1T_2} Q_{X_2Y}  P_{T_1T_2}    \Gamma^{(1)}_{V|T_1S_2X_2}  \right)\nonumber\\[-6ex]
&\hspace{8cm}+
I(S_1,S_2;V,Y|T_1,T_2)-I(S_1,S_2;X_1,X_2|T_1,T_2), \label{eq:missed1a}\\[6ex] 
\theta^{\text{miss},1\textnormal{b}} &:=\min_{\substack{\tilde{P}_{S_1S_2X_2YT_1T_2V}:\\\tilde{P}_{S_2X_2T_1T_2}=P_{S_2X_2T_1T_2}\\\tilde{P}_{S_2YT_1T_2V}=P_{S_2YT_1T_2V}}} D\left(\tilde{P}_{S_2X_2YT_1T_2V}\|P_{S_2|X_2T_1T_2}  Q_{X_2Y}  P_{T_1T_2} \Gamma^{(1)}_{V|T_1S_2X_2} \right)\nonumber\\[-4ex]
&\hspace{8cm} +
I(S_1;V,Y|S_2,T_1,T_2)-I(S_1;X_1|S_2,T_1,T_2),\label{eq:missed1b}\\[4ex]
\theta^{\text{miss},2\textnormal{a}} &:=\min_{\substack{\tilde{P}_{S_1X_1YT_1T_2V}:\\\tilde{P}_{S_1X_1T_1T_2}=P_{S_1X_1T_1T_2}\\\tilde{P}_{YT_1T_2V}=P_{YT_1T_2V}\\H(S_1|Y,T_1,T_2,V)\leq H_{\tilde{P}}(S_1|V,Y,T_1,T_2)}} D\left(\tilde{P}_{S_1X_1YVT_1T_2}\|P_{S_1|X_1T_1T_2}   Q_{X_1Y} P_{T_1T_2} \Gamma^{(2)}_{V|S_1X_1T_2}  \right)\nonumber\\[-6ex]
&\hspace{8cm}+
I(S_1,S_2;V,Y|T_1,T_2)-I(S_1,S_2;X_1,X_2|T_1,T_2),\label{eq:missed2a}\\[6ex] 
\theta^{\text{miss},2\textnormal{b}} &:=\min_{\substack{\tilde{P}_{S_1X_1YT_1T_2V}:\\\tilde{P}_{S_1X_1T_1T_2}=P_{S_1X_1T_1T_2}\\\tilde{P}_{S_1YT_1T_2V}=P_{S_1YT_1T_2V}}} D\left(\tilde{P}_{S_1X_1YT_1T_2V}\|P_{S_1|X_1T_1T_2}  Q_{X_1Y} P_{T_1T_2}  \Gamma^{(2)}_{V|S_1X_1T_2} \right)\nonumber\\[-4ex]
&\hspace{8cm} +I(S_2;V,Y|S_1,T_1,T_2)-I(S_2;X_2|S_1,T_1,T_2),\label{eq:missed2b}\\[4ex]
\theta^{\text{miss},12} &:= \mathbb{E}_{P_{T_1T_2}}\left[D\left(P_{  YV|T_1T_2}\| Q_Y\Gamma^{(12)}_{V|T_1T_2} \right)\right]+I(S_1,S_2;Y,V|T_1,T_2)-I(S_1,S_2;X_1,X_2|T_1,T_2),\label{eq:missed12}
\end{align}
where mutual informations and  the conditional pmf $P_{VY|T_1T_2}$ are  calculated according to the joint pmf  $P_{S_1S_2X_1X_2YVT_1T_2}$ in \eqref{distmac}. 
\medskip
\begin{theorem}\label{macthm} Error exponent $\theta\geq 0$ is achievable, if it satisfies
	\begin{align}\label{eq:condmac}
	\theta &\leq   \max 
	\; \min\left\{\theta^{\textnormal{standard}}, \theta^{\textnormal{dec},1}, \theta^{\textnormal{dec},2},\theta^{\textnormal{dec},12}, \theta^{\textnormal{miss},1\textnormal{a}},  \theta^{\textnormal{miss},1\textnormal{b}}, \theta^{\textnormal{miss},2\textnormal{a}},\theta^{\textnormal{miss},2\textnormal{b}}, \theta^{\textnormal{miss},12}\right\},
	\end{align} where the maximization is over all (conditional) pmfs $P_{T_1T_2}$, $P_{S_1|X_1T_1T_2}$, and $P_{S_2|X_2T_1T_2}$, and  functions  $f_1$ and $f_2$  as in \eqref{eq:function} so that the conditions in \eqref{eq:Iconditions} are satisfied with strict inequalities ``$<$" replaced by non-strict inequalities ``$\leq$". 
\end{theorem}
\begin{IEEEproof} See Appendix \ref{macsecproof}.
\end{IEEEproof}
\medskip
\begin{remark}\label{rem:separate}
The error exponents in the preceding theorem are obtained by means of the hybrid coding scheme described in the previous subsection~\ref{sec:MACcoding}. As usual, choosing the auxiliary random variables $S_1=(W_1, \bar{S}_1)$ and $S_2=(W_2, \bar{S}_2)$ and the tuple $(T_1,T_2,W_1,W_2)$ independent of the tuple $(\bar{S}_1, \bar{S}_2, X_1, X_2)$, is equivalent to replacing the hybrid coding scheme by a separate source-channel coding scheme. Specifically, $(\bar{S}_1, \bar{S}_2)$ then correspond to the source random variables and $(T_1,T_2,W_1,W_2)$ to the channel coding random variables.  Similarly to the transmission of correlated sources over a MAC, restricting to separate source-channel coding is strictly suboptimal. As Theorem~\ref{thm2opt} and Proposition~\ref{separation-ortho} ahead show, it can achieve the optimal exponent in some cases.

Choosing the auxiliary random variables $S_1$ and $S_2$ constant and $W_1=f_1(X_1)$ and $W_2=f_2(X_2)$, corresponds to uncoded transmission. 
\end{remark}
\medskip
\begin{remark}
	Notice that the solution to the minimization problem in \eqref{eq:missed1a} is smaller than the solution to the minimization problem in \eqref{eq:missed1b}. (In fact, the constraints are less stringent since $\tilde{P}_{S_2YT_1T_2V}=P_{S_2YT_1T_2V}$ implies $\tilde{P}_{YT_1T_2V}=P_{YT_1T_2V}$ and $H(S_2|Y,T_1,T_2,V) \leq H_{\tilde{P}}(S_2|T_1,T_2,V)$.) In the same way, the solution to the minimization problem in \eqref{eq:missed1b} is smaller than the solution to the minimization in \eqref{eq:missed2b}. 
	However, since the difference of mutual informations in \eqref{eq:missed1a} is larger than the one in \eqref{eq:missed1b}, and the one in \eqref{eq:missed2a} is larger than the one in \eqref{eq:missed2b}, it is \`a priori not clear which of these exponents is smallest.  
	
	A similar reasoning shows that the solution to the minimization problem in \eqref{th3noisy} is smaller than the solutions to the minimization problems in \eqref{thnoiseless}, \eqref{th1noisy}, and \eqref{th2noisy}, but the difference of mutual informations is larger. It is thus again unclear which of these exponents is smallest.
\end{remark}
\medskip

In analogy to Remark~\ref{rem:no_red}, it can be shown that also in this MAC setup the missed-detection exponents are sometimes not active. This is in particular the case for the following case of generalized testing against conditional independence.

\medskip

\begin{corollary}\label{corr-mac-dif-marginal}
	Consider the special case where $Y=(\bar{Y},Z)$ and under the alternative hypothesis $\mathcal{H}=1$:
	\begin{align}
	Q_{X_1X_2\bar{Y}Z} = P_{X_1X_2Z}\cdot Q_{\bar{Y}|Z},\label{h1}
	\end{align}
	In this case, any error exponent $\theta\geq 0$ is achievable that satisfies
	\begin{equation}\label{eq:condmacspecial}
\theta \leq \max\; \left(  \mathbb{E}_{{P}_{ZT_1T_2V}} \big[ D({P}_{\bar{Y}|ZT_1T_2V} \|  Q_{\bar{Y}|Z} ) \big] +I(S_1,S_2;\bar{Y}|Z,T_1,T_2,V)\right),
\end{equation}
where   the maximization is over all (conditional) pmfs $P_{S_1|X_1T_1T_2}$, and $P_{S_2|X_2T_1T_2}$, and  functions  $f_1$ and $f_2$ as in  \eqref{eq:function}  that satisfy the following conditions:
\begin{subequations}\label{eq:cond_ind}
	\begin{IEEEeqnarray}{rCl}
	I(S_1;X_1|S_2,Z,T_1,T_2) &\leq &I(S_1;V|S_2,Z,T_1,T_2),\label{con-ind1}\\
	I(S_2;X_2|S_1,Z,T_1,T_2) &\leq &I(S_2;V|S_1,Z,T_1,T_2),\label{con-ind2}\\
	I(S_1,S_2;X_1,X_2|Z,T_1,T_2) &\leq & I(S_1,S_2;V|Z,T_1,T_2),\label{con-ind3}
	\end{IEEEeqnarray}
	\end{subequations}
and all  mutual informations and the conditional pmf $P_{\bar{Y}|ZT_1T_2V}$ are  calculated with respect to the joint pmf
	\begin{align}
	P_{S_1S_2X_1X_2\bar{Y}ZT_1T_2V} = P_{S_1|X_1T_1T_2} \cdot P_{S_2|X_2T_1T_2}  \cdot  P_{X_1X_2\bar{Y}Z}  \cdot  P_{T_1T_2}\cdot  
	\Gamma_{V|S_1S_2X_1X_2}.
	\end{align}
\end{corollary}
\begin{IEEEproof} See Appendix \ref{remark-dif-marginal}. 
\end{IEEEproof}
\medskip

For  testing against conditional independence,  i.e.,  \begin{equation}Q_{\bar{Y}|Z}=P_{\bar{Y}|Z}, \label{eq:condind}
\end{equation}
 and when communication  is over noiseless links of given rates, Corollary~\ref{corr-mac-dif-marginal} recovers as a special case the result in \cite[Theorem 1]{Wagner}.  Similarly, for testing against independence, i.e.,  when 
 \begin{equation}
 Q_{X_1X_2Y} = P_{X_1X_2} P_Y,
 \end{equation}
  and when the MAC $\Gamma_{V|W_1W_2}$ decomposes  into two orthogonal DMCs $\Gamma_{V_1|W_1}$ and $\Gamma_{V_2|W_2}$, i.e.,    
 \begin{subequations}\label{eq:orthogonal_channel}
\begin{IEEEeqnarray}{rCl}
	V & = & (V_1,V_2)\\
\Gamma_{V_1V_2|W_1W_2}(v_1,v_2|w_1,w_2) &=& \Gamma_{V_1|W_1}(v_1|w_1)\cdot \Gamma_{V_2|W_2}(v_2|w_2), \label{eq:ortho_channel}
\end{IEEEeqnarray} 
\end{subequations} then specializing Corollary~\ref{corr-mac-dif-marginal} to  separate source-channel coding  recovers the achievable error exponent  in \cite[Theorem 6]{Gunduz}. 

\medskip

Im fact, specializing Corollary~\ref{corr-mac-dif-marginal} to separate source-channel coding, by Remark~\ref{rem:separate}, results in the following achievability result.

\begin{corollary}\label{corr-mac-dif-marginal_separate}
Reconsider the setup in Corollary~\ref{corr-mac-dif-marginal}. 
Using separate source-channel coding, any error exponent $\theta\geq 0$ is achievable that satisfies
	\begin{equation}\label{eq:condmacspecial}
	\theta \leq   \mathbb{E}_{{P}_{Z}} \big[ D({P}_{\bar{Y}|Z} \|  Q_{\bar{Y}|Z} ) \big] + \max\; I(\bar S_1,\bar S_2;\bar{Y}|Z),
	\end{equation}
	where   the maximization is over all (conditional) pmfs $P_{\bar S_1|X_1}$, $P_{\bar S_2|X_2}$, $P_{T_1T_2}$,  $P_{W_1|T_1T_2}$, and $P_{W_2|T_1T_2}$ that satisfy the following conditions:
	\begin{subequations}\label{eq:cond_ind}
		\begin{IEEEeqnarray}{rCl}
			I(\bar S_1;X_1|\bar S_2,Z) &\leq &I(W_1;V|W_2,T_1,T_2),\label{con-ind1}\\
			I(\bar S_2;X_2|\bar S_1,Z) &\leq &I(W_2;V|W_1,T_1,T_2),\label{con-ind2}\\
			I(\bar S_1,\bar S_2;X_1,X_2|Z) &\leq & I(W_1,W_2;V|T_1,T_2 ),\label{con-ind3}
		\end{IEEEeqnarray}
	\end{subequations}
and	where all  mutual informations are calculated with respect to the joint pmf
	\begin{align}
	P_{\bar S_1\bar S_2X_1X_2\bar{Y}ZT_1T_2W_1W_2V} = P_{\bar S_1|X_1} \cdot P_{\bar S_2|X_2}  \cdot  P_{X_1X_2\bar{Y}Z}  \cdot   
	P_{T_1T_2}\cdot P_{W_1|T_1T_2} \cdot P_{W_2|T_1T_2}\cdot
	\Gamma_{V|W_1W_2}.
	\end{align}
\end{corollary}
\medskip

This corollary recovers, for example, the optimal error exponent in \cite[Corollary 4]{Wagner} for the Gaussian one-helper hypothesis testing against independence problem where communication takes place over two individual noiseless links.  As shown in \cite[Corollary 4]{Wagner}, in this case the exponent of Corollary~\ref{corr-mac-dif-marginal_separate} is optimal. 
The following theorem proves that the  exponent in Corollary~\ref{corr-mac-dif-marginal_separate} is also optimal for generalized testing against conditional independence when the sources are independent under both hypotheses.

\medskip

\begin{theorem}\label{thm2opt} Consider  generalized testing against conditional independence with independent sources, i.e.,  
\begin{IEEEeqnarray}{rCl}
P_{X_1X_2Y} &= & P_{X_1} \cdot P_{X_2} \cdot P_{Y|X_1X_2} \\
Q_{X_1X_2Y} &= & P_{X_1}\cdot  P_{X_2}\cdot P_{Y},
\end{IEEEeqnarray}
and assume that communication from the sensors to the decision center takes place over two orthogonal DMCs $\Gamma_{V_1|W_1}$ and  $\Gamma_{V_2|W_2}$ as defined in \eqref{eq:orthogonal_channel}.  Let $C_1$ and $C_2$ denote the capacities of the two DMCs $\Gamma_{V_1|W_1}$ and  $\Gamma_{V_2|W_2}$. The optimal error exponent is:
	\begin{align}
	\theta^*= \hspace{0.1cm} D(P_Y\|Q_Y)+\hspace{-0.1cm}\max_{\substack{P_{\bar{S}_i|X_i},P_{W_i}, i\in\{1,2\}\\I(\bar{S}_1;X_1|\bar{S}_2)\leq C_1\\I(\bar{S}_2;X_2|\bar{S}_1)\leq C_2\\I(\bar{S}_1,\bar{S}_2;X_1,X_2)\leq C_1+C_2 }} \hspace{-0.1cm} I(\bar{S}_1,\bar{S}_2;Y).\label{eq:exp_opt}
	\end{align}
\end{theorem}
\begin{IEEEproof} 
	Achievability follows directly by specializing Corollary~\ref{corr-mac-dif-marginal_separate} to $Z$ a constant and thus $\bar{Y}=Y$. The converse is proved in Appendix~\ref{app:conv}.
\end{IEEEproof}
\medskip

We specialize above theorem to an example with independent Gaussian sources. 
\begin{example}[Theorem~\ref{thm2opt} for Gaussians]\label{ex:Gauss_MAC}
Let $X_1$ and $X_2$ be independent standard Gaussians under both hypotheses. Under the null hypothesis, 
\begin{align}
\mathcal{H}=0\colon \qquad\qquad Y= X_1+X_2+N_0,\qquad N_0\sim \mathcal{N}(0,\sigma_0^2),\label{gdef2orth}
\end{align}
for an $N_0$ independent of $(X_1,X_2)$ and for a given nonnegative variance $\sigma_{0}^2 >0$. Under the alternative hypothesis,
\begin{align}
\mathcal{H}=1\colon \qquad\qquad Y\sim \mathcal{N}(0,\sigma_y^2),\qquad  \textnormal{independent of }(X_1,X_2),\label{gsource2orth}
\end{align}
for a given nonnegative variance $\sigma_{y}^2>0$. 
Further assume an orthogonal MAC as in \eqref{eq:ortho_channel} with the two individual DMCs of  capacities $C_1$ and $C_2$. 

The described setup is a special case of the setup considered in Theorem~\ref{thm2opt}. Appendix \ref{gcap-proof} shows that  in this case, the optimal exponent in  \eqref{eq:exp_opt} evaluates to:
\begin{align}
	\theta^*= \frac{1}{2}\log \left( \frac{\sigma_y^2}{ 2^{-2C_1}+ 2^{-2C_2}+\sigma_0^2} \right)+\left(\frac{2+\sigma_0^2}{2\sigma_y^2}-\frac{1}{2}\right)\cdot \log e.
	\end{align} 
\end{example}

\medskip

Theorem~\ref{thm2opt} shows that separate source-channel coding is optimal for generalized  testing against conditional independence over two orthogonal channels. The following proposition extends this result to all joint source distributions  $P_{X_1X_2}$. The proposition also provides a multi-letter characterization of the optimal error exponent in this case.

\medskip
\begin{proposition}\label{separation-ortho} Consider testing against independence over an orthogonal MAC, i.e., assume that  \eqref{eq:condind}--\eqref{eq:ortho_channel} hold.
	Then, the optimal error exponent is given by 
	\begin{IEEEeqnarray}{rCl}
		\theta^*&=& D(P_Y\|Q_Y)+\lim_{n\to\infty} \frac{1}{n} \max I(S_1^n,S_2^n;Y^n),
		\end{IEEEeqnarray}
	where the maximization is over all $P_{S_1^n|X_1^n}$ and $P_{S_2^n|X_2^n}$ satisfying:
	\begin{IEEEeqnarray}{rCl}
 \lim_{n\to\infty} \frac{1}{n} I(X_1^n;S_1^n|S_2^n)	& \leq &	C_1,\\
 \lim_{n\to\infty} \frac{1}{n} I(X_2^n;S_2^n|S_1^n)& \leq &	C_2,\\
	 \lim_{n\to\infty} \frac{1}{n} I(X_1^n,X_2^n;S_1^n,S_2^n)& \leq &	C_1+C_2.
		\end{IEEEeqnarray}
	\end{proposition}
\begin{IEEEproof} Achievability can be shown  in a similar way as Theorem~\ref{thm2opt}. The converse  proof follows similar arguments as in \cite[Theorem 2.4]{Luo}. It is  detailed out in Appendix~\ref{separation-proof} for completeness. 
	\end{IEEEproof}

\subsection{Correlated Gaussian Sources over a Gaussian MAC}\label{ex-Gaussian:MAC}
In this last subsection of Section~\ref{MACsection}, we focus on testing against independence over a Gaussian MAC when the sources are jointly Gaussian (but not necessarily independent as in Example~\ref{ex:Gauss_MAC}.
Consider   a symmetric  Gaussian setup where under  both hypotheses: 
	\begin{equation}
	(X_1,X_2)\sim \mathcal{N}(0,\mathbf{K}_{X_1X_2})\label{gsource1}
	\end{equation}
	for a positive semidefinite covariance matrix
	\begin{align}
	\mathbf{K}_{X_1X_2} = \left[ \begin{array}{cc} 1 & \rho \\ \rho  & 1 \end{array} \right], \qquad 0\leq \rho \leq 1.\label{gdef1}
	\end{align}
	Assume as in Example~\ref{ex:Gauss_MAC} that under the null hypothesis,
	\begin{align}
	\mathcal{H}=0\colon \qquad\qquad Y= X_1+X_2+N_0,\qquad N_0\sim \mathcal{N}(0,\sigma_0^2),\label{gdef2}
	\end{align}
	for $N_0$ independent of $(X_1,X_2)$ and for  $\sigma_{0}^2 >0$, and under the alternative hypothesis,
	\begin{align}
	\mathcal{H}=1\colon \qquad\qquad Y\sim \mathcal{N}(0,\sigma_y^2),\qquad  \textnormal{independent of }(X_1,X_2),\label{gsource2}
	\end{align}
	for  $\sigma_{y}^2>0$.

	Communication  takes place over the Gaussian MAC 
	\begin{align}
	V= W_{1}+W_{2}+N,
	\end{align}
	where the noise $N$ is zero-mean Gaussian of variance $\sigma^2>0$,  independent of the inputs $(W_{1},W_{2})$. Each transmitter's input sequence is subject to an average block-power  constraint  $P$. 
	
	The described setup corresponds to generalized testing against conditional independence. We can thus use  Corollary \ref{corr-mac-dif-marginal} to obtain an achievable error exponent for this problem. 
The above choice of random variables yields the following result on the achievable error exponent.
	
	\begin{corollary}\label{cor-gaussian-separate} For the described Gaussian setup
		any error exponent $\theta\geq 0$ is achievable that satisfies the following condition:
		\begin{align}
		\theta &\leq   \max 		\; \frac{1}{2}\log  \frac{\sigma_y^2}{\frac{2\xi^2(1+\rho)\sigma^2}{2\xi^2(\alpha-\beta)^2\cdot (1+\rho)+\sigma^2(1+\rho+\xi^2)}+\sigma_0^2 }+\left(\frac{\sigma_0^2+2+2\rho}{2\sigma_y^2}-\frac{1}{2}\right)\cdot \log e,\label{ghyb-ach-theta}
		\end{align}
		where  the maximization is over all $\xi^2, \alpha^2,\beta^2,\gamma^2\geq 0$ satisfying 
		\begin{align}
		\gamma^2+\alpha^2+\beta^2\xi^2\leq P,
		\end{align}
		and 
		\begin{subequations}\label{eq:condhyb}
		\begin{align}
		\frac{(1+\xi^2)^2-\rho^2}{(1+\xi^2)\cdot \xi^2}&\leq \frac{\sigma^2+ 2P-\gamma^2+2\alpha^2\rho-\frac{(\alpha\cdot (1+\rho)+\beta \cdot \xi^2)^2}{1+\xi^2}}{\sigma^2+\frac{2(\alpha-\beta)^2\cdot (1+\rho)\xi^2}{1+\rho+\xi^2}} ,\label{ghyb-cons1}\\[1.2ex]
		\frac{(1+\xi^2)^2-\rho^2}{\xi^4} &\leq  \frac{\sigma^2+2P+2\alpha^2\rho}{\sigma^2+\frac{2(\alpha-\beta)^2\cdot (1+\rho)\xi^2}{1+\rho+\xi^2}}.\label{ghyb-cons2}
		\end{align}
				\end{subequations}
	\end{corollary}
	\begin{IEEEproof}  See Appendix \ref{app:hybrid}.
	\end{IEEEproof}

	\medskip
The following theorem provides an upper bound on the optimal error exponent.
\begin{theorem}\label{gout-thm} For the proposed Gaussian setup, the optimal error exponent $\theta^*$  satisfies 
	\begin{IEEEeqnarray}{rCl}
		\theta^* \leq\frac{1}{2}\cdot \left[\log \left(\frac{\sigma_y^2}{\frac{ 2(1+\rho) \sigma^2}{2P(1+\rho)+\sigma^2}+\sigma_0^2 } \right)+\left(\frac{2+2\rho+\sigma_0^2}{\sigma_y^2}-1\right)\cdot \log e\right]
		\end{IEEEeqnarray}
	\end{theorem}
\begin{IEEEproof} See Appendix \ref{gout-proof}.
	\end{IEEEproof}
\begin{figure}[t]
	\centering
	\includegraphics[width = 9cm, height=5cm]{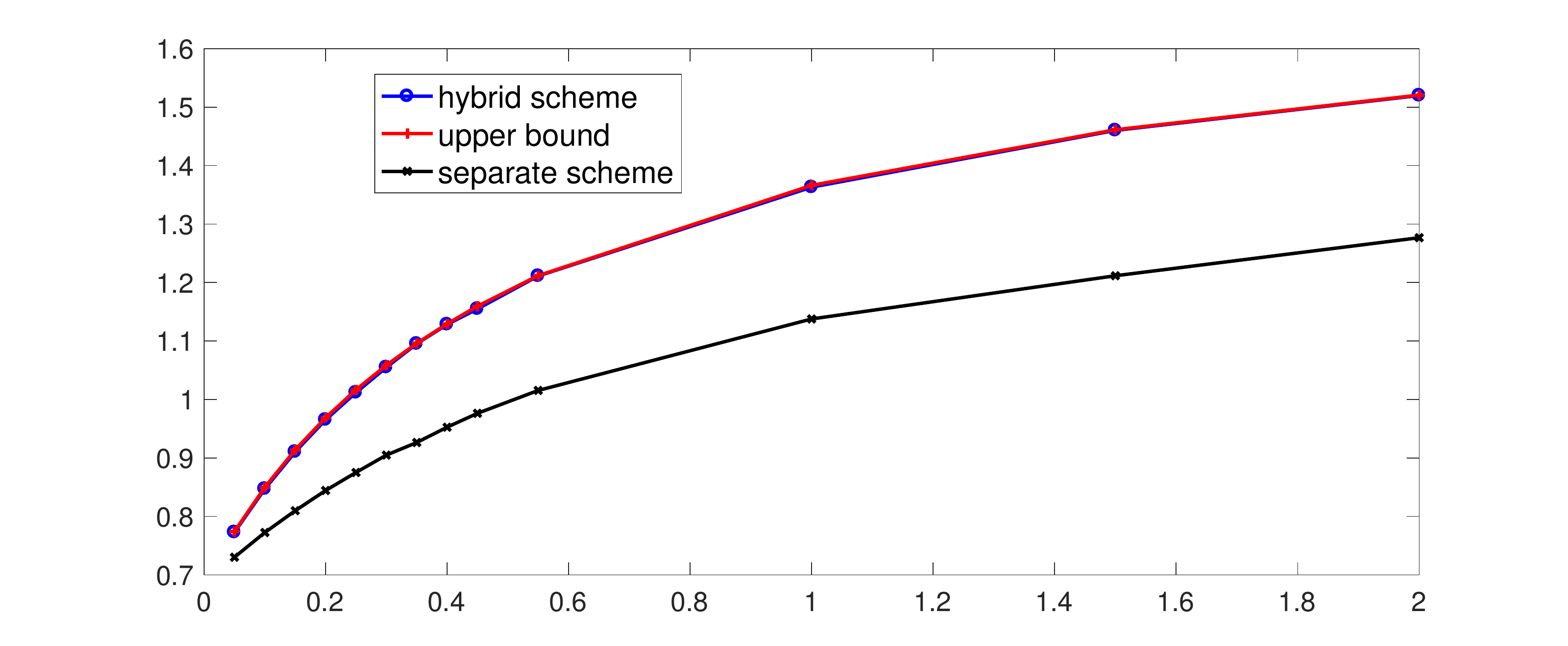}
	\caption{Upper and lower bounds on the optimal exponent $\theta^*$  of the proposed Gaussian example for $\rho=0.8$, $\sigma_0^2=1$, $\sigma_y^2=1.5$ and $\sigma^2=1$.}
	\label{fig:MACplot}
\end{figure}
	Figure~\ref{fig:MACplot} compares the presented upper and lower bounds on the optimal error exponent $\theta^*$. They are very close for the considered setup. For comparison, the figure also shows the exponent that is achieved with the same choice of source variables but with separate source-channel coding. That means, by specializing the exponent in \eqref{ghyb-ach-theta} to $\alpha=\beta=0$.

\section{Hypothesis Testing over Broadcast Channels}\label{BCsection}
\subsection{System Model}
\begin{figure}[b]
	\centering
	\includegraphics[scale=0.4]{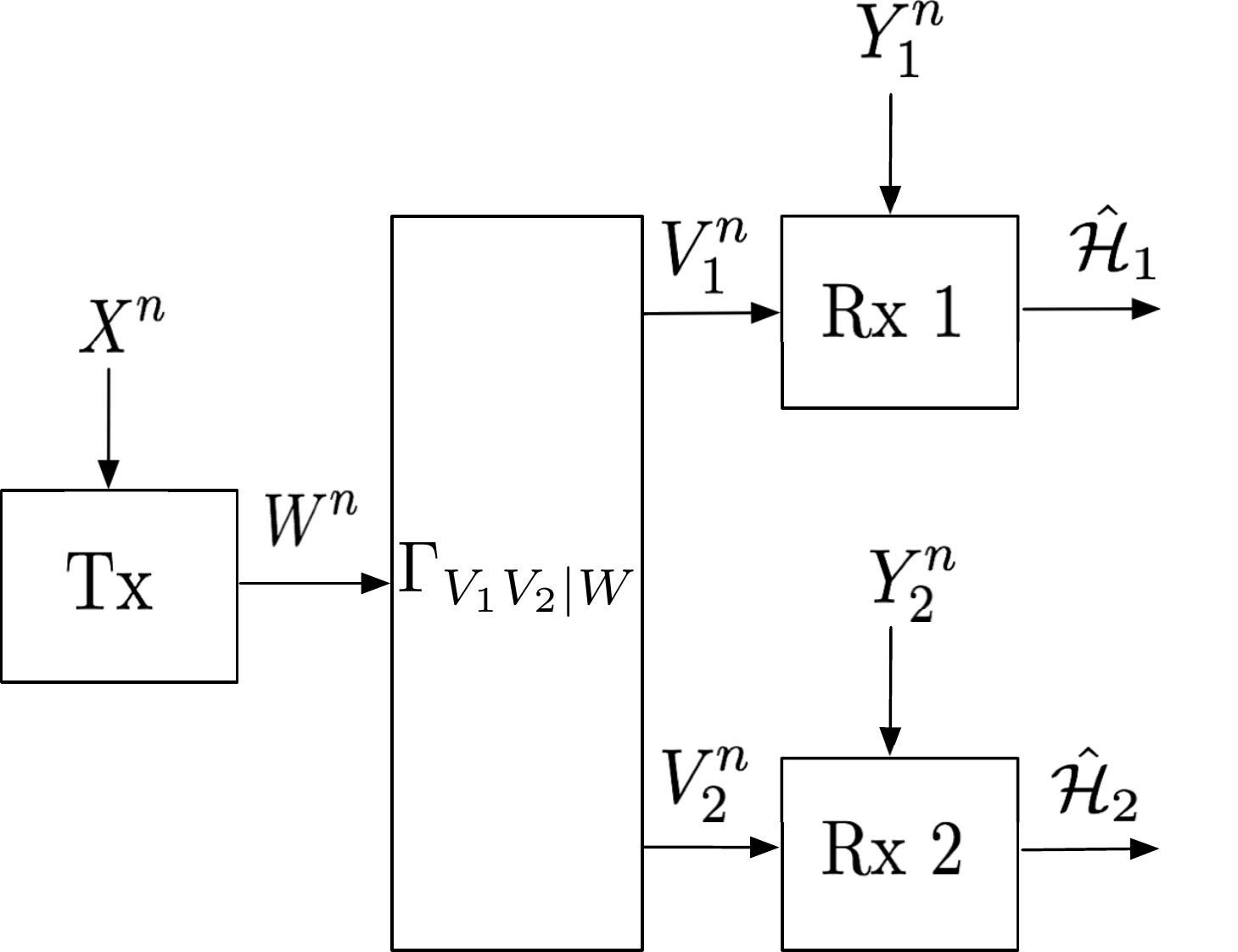}
	\caption{Hypothesis testing over a noisy BC.}
	\label{figureBC}
	\end{figure}

Consider the distributed hypothesis testing problem in Fig.~\ref{figureBC}, where a transmitter observes sequence $X^n$, Receiver $1$  sequence $Y_1^n$, and Receiver~2 sequence $Y_2^n$. Under the null hypothesis:
\begin{align}
\mathcal{H}=0\colon (X^n,Y_1^n,Y_2^n)\quad \text{i.i.d.}\; \sim P_{XY_1Y_2},
\end{align}
and under the alternative hypothesis:
\begin{align}
\mathcal{H}=1\colon (X^n,Y_1^n,Y_2^n)\quad \text{i.i.d.}\; \sim Q_{XY_1Y_2},
\end{align}
for two given pmfs $P_{XY_1Y_2}$ and $Q_{XY_1Y_2}$. The transmitter can communicate with the receivers over $n$ uses of a discrete memoryless broadcast channel $(\mathcal{W},\mathcal{V}_1\times \mathcal{V}_2,P_{V_1V_2|W})$ where $\mathcal{W}$ denotes the finite channel input alphabet and  $\mathcal{V}_1$ and  $\mathcal{V}_2$, the finite channel output alphabets. 
Specifically, the transmitter feeds inputs 
\begin{equation}
W^n=f^{(n)}(X^n),
\end{equation}
to the channel, where $f^{(n)}$ denotes the chosen 
(possibly stochastic) encoding function 
\begin{equation}
f^{(n)}:\mathcal{X}^n\to \mathcal{W}^n. 
\end{equation} 
Each Receiver $i\in\{1,2\}$ observes the BC ouputs $V_i^n$, where for a given input $W_t=w_t$,
\begin{align}
(V_{1,t},V_{2,t})\sim \Gamma_{V_1V_2|W}(\cdot , \cdot |w_t),\qquad t\in\{1,\ldots,n\}.
\end{align}
Based on the sequence of channel outputs $V_i^n$ and the source sequence $Y_i^n$,  Receiver $i$  decides on the hypothesis $\mathcal{H}$. That means, it  produces the guess 
\begin{equation}
\hat{\mathcal{H}}_i=g^{(n)}(V_i^n,Y_i^n),
\end{equation}
for a chosen decoding function 
\begin{equation}
g_i^{(n)} \colon \mathcal{V}_i^n\times \mathcal{Y}_i^n\to \{0,1\}.
\end{equation}

\newcommand{\h}{\mathsf{h}}
\newcommand{\hnot}{\bar{\h}}
There are different possible scenarios regarding the requirements on error probabilities. As in previous sections, we assume that each receiver is interested in only one of the two exponents. For each $i\in\{1,2\}$, let $\h_i\in\{0,1\}$ be the hypothesis whose error exponent Receiver~$i$ wishes to maximize, and $\hnot_i$ the other hypothesis, i.e., $\hnot_i\in\{0,1\}$ and $\h_i \neq \hnot_i$. (The values of $\h_1$ and $\h_2$ are fixed and part of the problem statement.) We then have:
\begin{definition}\label{deftype1BC}
	For each $\epsilon \in (0,1)$, an exponent pair $(\theta_1,\theta_2)$ is said $\epsilon$-achievable, if for each sufficiently large blocklength $n$, there exist encoding and decoding functions $(f^{(n)}, g_1^{(n)}, g_2^{(n)})$ such that:
		\begin{align}
	\alpha_{1,n}&\stackrel{\Delta}{=} \Pr[\hat{\mathcal{H}}_1=\h_1|\mathcal{H}=\hnot_1],\qquad\qquad \alpha_{2,n}\stackrel{\Delta}{=} \Pr[\hat{\mathcal{H}}_2=\h_2|\mathcal{H}=\hnot_2],
	\\
	\beta_{1,n}&\stackrel{\Delta}{=} \Pr[\hat{\mathcal{H}}_1=\hnot_1|\mathcal{H}=\h_1],\qquad\qquad \beta_{2,n}\stackrel{\Delta}{=} \Pr[\hat{\mathcal{H}}_2=\hnot_2|\mathcal{H}=\h_2],
	\end{align}
	satisfy 
	\begin{align}
	\alpha_{i,n}&\leq \epsilon,\qquad\qquad i\in\{1,2\},\label{def1bc}
	\end{align}
	and 
	\begin{align}
	-\varlimsup_{n\to \infty}\frac{1}{n}\log\beta_{i,n}&\geq \theta_i,\qquad\qquad i\in\{1,2\}. \label{def2bc}
	\end{align} 
\end{definition}

\medskip

\begin{remark} \label{rem:BCequiv}
Notice that both $\alpha_{1,n}$ and $\beta_{1,n}$ depend of the BC law $\Gamma_{V_1V_2|W}$ only through the conditional marginal distribution $\Gamma_{V_1|W}$. Similarly, $\alpha_{2,n}$ and $\beta_{2,n}$ only depend on $\Gamma_{V_2|W}$. Furthermore, the error exponents region depends on the joint laws $P_{XY_1Y_2}$ and $Q_{XY_1Y_2}$ only through their marginal laws $P_{XY_1}$, $P_{XY_2}$, $Q_{XY_1}$, and $Q_{XY_2}$.  Therefore, when $P_X=Q_X$,  it is possible to relabel some of the marginals $P_{XY_1}$, $P_{XY_2}$, $Q_{XY_1}$, and $Q_{XY_2}$  without changing the exponents region and so that both receivers aim at maximizing the error exponent under hypothesis $\mathcal{H}=1$, i.e., $\h_1=\h_2=1$. Assume for example that $\h_1=0$ and $\h_2=1$. Then  by relabelling  $P_{XY_1}$ as $Q_{XY_1}$ and vice versa, the new setup for $\h_1=\h_2=1$ has same exponents region as the original setup.
\end{remark}

\medskip
To simplify notation in the sequel, we use the following shorthand notations for the pmfs $P_{XY_1Y_2}$ and $Q_{XY_1Y_2}$.
\begin{subequations}\label{eq:pqi}
For each $i\in\{1,2\}$:
\begin{equation}
\textnormal{if} \;\; \h_i=0  \quad \Longrightarrow\quad  \left(p_{XY_1Y_2}^{i}: =P_{XY_1Y_2}\quad \textnormal{and} \quad q_{XY_1Y_2}^{i}: =Q_{XY_1Y_2} \right)
\end{equation} 
and 
\begin{equation}
\textnormal{if} \;\; \h_i=1  \quad \Longrightarrow \quad \left(p_{XY_1Y_2}^{i}: =Q_{XY_1Y_2}\quad \textnormal{and} \quad q_{XY_1Y_2}^{i} : =P_{XY_1Y_2}\right).
\end{equation} 
\end{subequations}

We propose two coding schemes. One for the case when 
\begin{equation}\label{eq:px_equal}
\forall x\in \mathcal{X} \colon \; p_{X}^{1} (x) = p_{X}^{2}(x), 
\end{equation}
and one for the case when 
\begin{equation}\label{eq:px_different}
\exists  x\in \mathcal{X} \colon \; p_{X}^{1} \neq  p_{X}^{2}.
\end{equation}
Notice that \eqref{eq:px_equal} always holds when $\h_1=\h_2$. In fact, by Remark~\ref{rem:BCequiv}, given \eqref{eq:px_equal} we can focus on the case $\h_1=\h_2$. In contrast, given \eqref{eq:px_different}, then obviously $\h_1\neq \h_2$. 

\subsection{Coding and Testing  Scheme when $p_{X}^{1} = p_X^{2}$}
In this case, the scheme is based on hybrid source-channel coding. 
Choose a large positive integer $n$, auxiliary alphabets $\mathcal{S}$, $\mathcal{U}_1,$ and $\mathcal{U}_2$, and a function 
 \begin{align}f\colon \mathcal{S}\times\mathcal{U}_1\times\mathcal{U}_2\times \mathcal{X}\to \mathcal{W}.\label{function-BC}\end{align}
Then, define the shorthand notation:
\begin{align}
\Gamma_{V_1V_2|SU_1U_2X}:=\Gamma_{V_1V_2|W}(v_1,v_2|f(s,u_1,u_2,x)), \qquad \forall s\in\mathcal{S}, u_1\in\mathcal{U}_1, u_2\in\mathcal{U}_2, x\in\mathcal{X},
\end{align}
and choose 
an auxiliary distribution $P_T$ over $\mathcal{W}$, a conditional distribution $P_{SU_1U_2|XT}$ over  $\mathcal{S} \times \mathcal{U}_1\times \mathcal{U}_2$ so that for $i\in \{1,2\}$:
\begin{subequations}\label{constraints_same_marginal}
\begin{align}
I_{p^{i}}(S,U_i;X|T) &< I_{p^{i}}(S,U_i;Y_i,V_i|T),\\
I_{p^{i}}(U_i;X|S,T) &< I_{p^{i}}(U_i;Y_i,V_i|S,T),\\
I_{p^1}(S,U_1;X|T)+I_{p^1}(S,U_2;X|T)+I_{p^1}(U_1;U_2|S,T) &< I_{p^1}(S,U_1;Y_1,V_1|T)+I_{p^2}(S,U_2;Y_2,V_2|T),\\
I_{p^1}(U_1;X|S,T)+I_{p^1}(U_2;X|S,T)+I_{p^1}(U_1;U_2|S,T) &< I_{p^1}(U_1;Y_1,V_1|S,T)+I_{p^2}(U_2;Y_2,V_2|S,T),\\
I_{p^1}(U_1;X|S,T)+I_{p^1}(S,U_2;X|T)+I_{p^1}(U_1;U_2|S,T) &< I_{p^1}(U_1;Y_1,V_1|S,T)+I_{p^2}(S,U_2;Y_2,V_2|T),\\
I_{p^1}(S,U_1;X|T)+I_{p^1}(U_2;X|S,T)+I_{p^1}(U_1;U_2|S,T) &< I_{p^1}(S,U_1;Y_1,V_1|T)+I_{p^2}(U_2;Y_2,V_2|S,T),
\end{align}
\end{subequations}
where the mutual informations in this section are calculated according to the following joint distribution
\begin{align}\label{eq:joint_pi}
p^i_{SU_1U_2XY_1Y_1TV_1V_2}&=P_{SU_1U_2|XT}\cdot p^{i}_{XY_1Y_2}\cdot P_T \cdot \Gamma_{V_1V_2|SU_1U_2X}.
\end{align}
Then, choose a positive $\mu$ and rates $R_0,R_1,R_2$ so that 
\begin{subequations}\label{eq:rate_constraints_marton}
\begin{align}
R_0 &=I_{p^1}(S;X|T)+\mu,\\
R_i&> I_{p^1}(U_i;X|S,T),\qquad i\in\{1,2\},\\
R_1+R_2 &> I_{p^1}(U_1;X|S,T)+I_{p^1}(U_2;X|S,T)+I_{p^1}(U_1;U_2|S,T),
\end{align}
and
\begin{align}
R_0+R_i &\leq I_{p^{i}}(S,U_i;Y_i,V_i|T),\\
R_i &\leq I_{p^{i}}(U_i;Y_i,V_i|S,T).
\end{align}
\end{subequations}

Generate a  sequence $T^n$ i.i.d. according to $P_T$ and construct a random codebook  $$\mathcal{C}_S=\big\{S^n(m_0):m_0\in \{1,...,\lfloor 2^{nR_0}\rfloor\}\big\}$$ superpositioned on $T^n$ where each codeword is drawn  independently  according to $p^1_{S|T}$ conditioned on $T^n$. Then, for each index $m_0$ and $i\in\{1,2\}$, randomly generate a codebook
$$\mathcal{C}_{U_i}(m_0)=\big\{U_i^n(m_0,m_i):m_i\in \{1,...,\lfloor 2^{nR_i}\rfloor\}\big\}$$ superpositioned on $(T^n,S^n(m_0))$ by drawing each entry of the 
$n$-length codeword $U_i^n(m_0,m_i)$ i.i.d. according to the conditional pmf $p^1_{U_i|ST}(.|S_k(m_0),T)$ where $S_k(m_0)$ denotes the $k$-th symbol of $S^n(m_0)$. Reveal the realizations of the codebooks and the  sequence $T^n$ to all terminals.

\noindent\underline{\textit{Transmitter:}} Given that it observes the source sequence $X^n=x^n$, the transmitter looks for indices $(m_0,m_1,m_2)$ that satisfy 
\begin{align}
\left(s^n(m_0),u_1^n(m_0,m_1),u_2^n(m_0,m_2),x^n,t^n\right)\in \mathcal{T}_{\mu/2}^n\left(p^1_{SU_1U_2XT}\right).
\end{align} 
If successful, it picks one of these indices uniformly at random and sends the codeword $w^n$ over the channel, where
\begin{align}
w_{k}=f\left(s_k(m_0),u_{1,k}(m_0,m_1),u_{2,k}(m_0,m_2),x_k\right),\qquad k\in\{1,\ldots,n\},
\end{align}
and where $(s_k(m_0),u_{1,k}(m_0,m_1),u_{2,k}(m_0,m_2))$ denote the $k$-the components of codewords $(s^n(m_0),u_1^n(m_0,m_1),u_2^n(m_0,$ $m_2))$. Otherwise, it sends the sequence of inputs $t^n$ over the channel.

\noindent\underline{\textit{Receiver~$i\in\{1,2\}$:}}    After observing  $V_i^n=v_i^n$ and $Y_i^n=y_i^n$, Receiver~$i\in\{1,2\}$ looks for indices $m'_0 \in \{1,\ldots,\lfloor 2^{nR_0} \rfloor\}$ and $m'_i \in \{1,\ldots,\lfloor 2^{nR_i} \rfloor\}$ that satisfy the following conditions: 
\begin{enumerate}
	\item
	\begin{align}
	(s^n(m'_0),u_i^n(m'_0,m'_i),y_i^n,t^n,v_i^n) \in \mathcal{T}_{\mu}^n(p^i_{SU_iY_iTV_i}).
	\end{align}
	\item
	\begin{align}
	&H_{\text{tp}\left(s^n(m'_0),u_i^n(m'_0,m'_i),y_i^n,t^n,v_i^n\right)}(S,U_i|Y_i,T,V_i)=  \min_{\tilde{m}_0,\tilde{m}_i}H_{\text{tp}\left(s^n(\tilde{m}_0),u_i^n(\tilde{m}_0,\tilde{m}_i),y_i^n,t^n,v_i^n\right)}(S,U_i|Y_i,T,V_i),\label{MI-bc}
	\end{align}
\end{enumerate}
If successful,  Receiver~$i$ declares $\hat{\mathcal{H}}_i = \hnot_i$. Otherwise, it declares $\hat{\mathcal{H}}_i=\h_i$.

\subsection{Coding and Testing Scheme when $p_{X}^{1} \neq p_{X}^{2}$}\label{BC-feedback}

\begin{figure}[b!]
	\centering
	\includegraphics[scale=0.26]{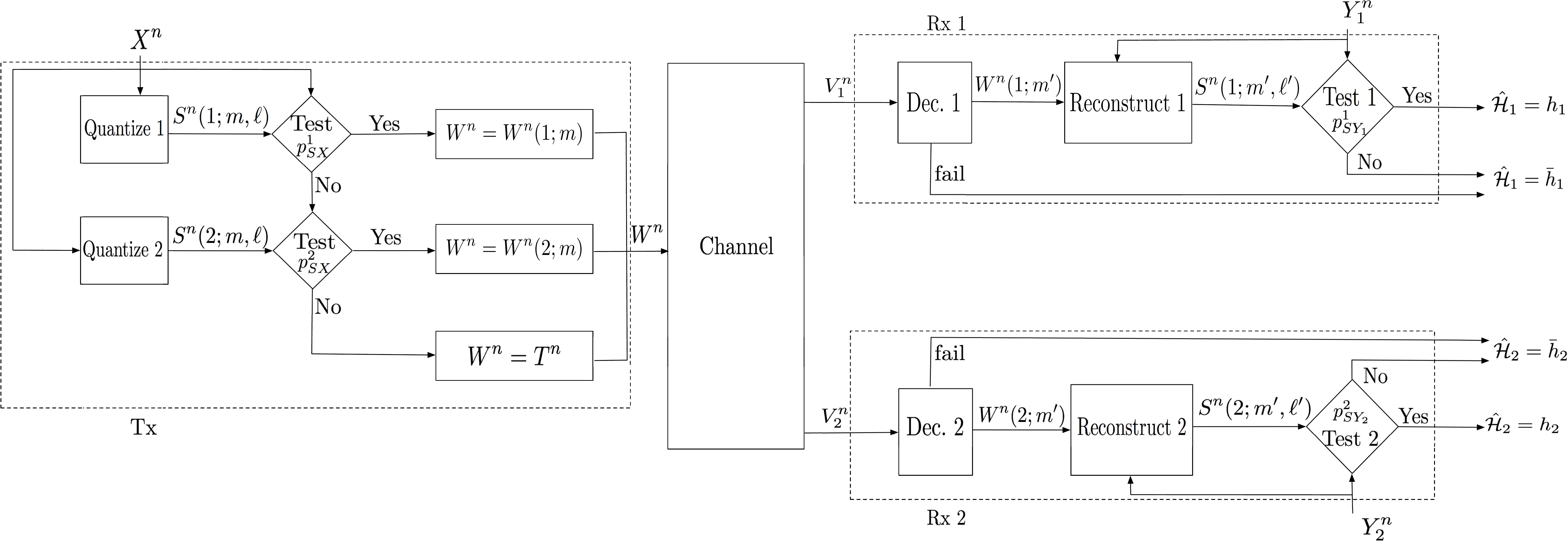}
	\caption{Coding and testing scheme for hypothesis testing over a BC.}
	\label{coding-BC}
\end{figure}

In this case, separate source-channel coding is applied. The main feature here is that the transmitter can make a tentative decision on $\mathcal{H}$ and accordingly use a different source and channel codes, see Fig.~\ref{coding-BC}. Details are as follows.

Fix $\mu > 0$, a sufficiently large blocklength $n$, auxiliary distributions $p_{T}$, $p^{1}_{T_1|T}$ and $p^{2}_{T_2|T}$ over $\mathcal{W}$, conditional channel input distributions $p^{1}_{W|TT_1}$ and $p^{2}_{W|TT_2}$, and conditional pmfs $p^{1}_{S|X}$ and $p^{2}_{S|X}$  over a finite auxiliary alphabet $\mathcal{S}$  
such that for each $i\in\{1,2\}$:
\begin{equation}
I_{p^{i}}(S;X|Y_i) <  I_{p^{i}}(W;V_i|T,T_i).\label{constraint-BC}
\end{equation}
The mutual information in \eqref{constraint-BC} is calculated according to the joint distribution:
\begin{align}
p^i_{SXY_1Y_2TT_iWV_1V_2} &=  p^i_{S|X}\cdot p^i_{XY_1Y_2}\cdot p_{T}\cdot p^i_{T_i|T}\cdot p^i_{W|TT_i}\cdot \Gamma_{V_1V_2|W}.\label{dist-P-def}
\end{align}
For each $i\in\{1,2\}$, if  $I_{p^{i}}(S;X) < I_{p^{i}}(W;V_i|T,T_i)$, choose rates
\begin{align}
R_i &:=I_{p^{i}}(S;X)+\mu,\label{eq:RiBC}\\
R'_i &:=0.
\end{align}
If $I_{p^{i}}(S;X)\geq I_{p^{i}}(W;V_i|T,T_i)$, then choose rates
\begin{align}
R_i&:=I_{p^{i}}(W;V_i|T,T_i)-\mu,\label{rate-BC1}\\
R'_i&:=I_{p^{i}}(S;X)-I_{p^{i}}(W;V_i|T,T_i)+2\mu.\label{rate-BC2}
\end{align}
Again, all mutual informations in \eqref{eq:RiBC}--\eqref{rate-BC2} are calculated with respect to the pmf in \eqref{dist-P-def}.

\underline{\textit{Code Construction}:}  Generate a sequence $T^{n}=(T_1,\ldots,T_{n})$ by independently drawing each component $T_k$ according to $p_{T}$. For each $i\in\{1,2\}$, generate a sequence $T_i^n=(T_{i,1},\ldots,T_{i,n})$ by independently drawing each $T_{i,k}$ according to $p^i_{T_i|T}(.|t)$ when  $T_k=t$.  Also, construct a random codebook  
\begin{equation}
\mathcal{C}_{W}^i=\big\{W^{n}(i;m)\colon m\in \{1,...,\lfloor 2^{nR_i}\rfloor\}\big\}
\end{equation}
superpositioned on $(T^{n},T_i^n)$ where  the $k$-th symbol  $W_{k}(i;m)$ of codeword $W^{n}(i;m)$ is drawn  independently  of all codeword symbols according to $p^i_{W|TT_i}(\cdot|t,t_i)$ when $T_k=t$ and $T_{i,k}=t_{i}$. Finally, construct a 
random codebook
\begin{align}
\mathcal{C}_{S}^i &=\{S^n(i;m,\ell)\colon m\in \{1,\ldots,\lfloor 2^{nR_i} \rfloor\},\ell\in \{1,\ldots,\lfloor 2^{nR'_i} \rfloor\}\}, \quad i\in\{1,2\},
\end{align}
by independently drawing the $k$-th component $S_{k}(i;m,\ell)$ of codeword $S^n(i;m,\ell)$   according to the marginal pmf $p^i_{S}$.

Reveal all codebooks and the realizations $t^n,t_1^n,t_2^n$ of the sequences $T^n, T_1^n, T_2^n$ to all terminals.

\underline{\textit{Transmitter}:} Given source sequence $X^n=x^n$, the transmitter looks for indices $(i,m,\ell) \in \{1,2\}\times\{1,\ldots,\lfloor 2^{nR_1}\rfloor\} \times  \{1,\ldots,\lfloor 2^{nR'_i}\rfloor \}$ such that codeword $s^n(i;m,\ell)$ from codebook $\mathcal{C}_{S}^i$ satisfies
\begin{align}
(s^n(i;m,\ell),x^n)\in \mathcal{T}_{\mu/2}^n(p^i_{SX}),\label{typ}
\end{align}
and the corresponding codeword $w^n(i;m)$ from codebook $\mathcal{C}_W^i$ satisfies the following:
\begin{align}
(t^n,t_i^n,w^n(i;m))\in\mathcal{T}_{\mu/2}^n(p^i_{TT_iW}).\label{typ2}
\end{align}
(Notice that  when $\mu$ is sufficiently small, then Condition~\eqref{typ} can be satisfied for at most one value $i\in\{1,2\}$, because  $p_X^{1} \neq p^2_X$.)  
If successful, the transmitter  picks  uniformly at random one of the triples $(i,m,\ell)$ that satisfy \eqref{typ}, and it sends the sequence $w^n(i;m)$ over the channel. 
If no triple satisfies Conditions \eqref{typ} and \eqref{typ2}, then the transmitter sends the sequence $t^n$ over the channel.

\underline{\textit{Receiver $i\in\{1,2\}$}:} Receives $v_i^{n}$ and  checks whether there exist indices $(m',\ell')$ such that the following three conditions are satisfied:
\begin{enumerate}
	\item
	\begin{align}
	(t^n,t_i^n,w^n(i;m'),v_i^n)\in\mathcal{T}_{\mu}^n(p^i_{TT_iWV_i}), \label{check1-rec1}
	\end{align}
	\item
	\begin{align}
	H_{\text{tp}(s^n(i;m',\ell'),y_i^{n})}(S|Y_i) = \min_{\tilde{\ell}} H_{\text{tp}(s^n(i;m',\tilde{\ell}),y_i^{n})}(S|Y_i),\label{check2-rec1}
	\end{align}
	\item 
	\begin{align}
	(s^n(i;m',\ell'),y_i^{n})\in \mathcal{T}_{\mu}^n(p^i_{SY_i}). \label{check3-rec1}
	\end{align}
\end{enumerate}
If successful, it declares $\hat{\mathcal{H}}_i=\hnot_i$. Otherwise, it declares $\hat{\mathcal{H}}_i=\h_i$. 


\subsection{Result on the Error Exponent}

The coding and testing schemes described in the previous two subsections yield the following two theorems.
\begin{theorem}\label{thm:BCequal}
If $p_X^{1}=p_X^{2}$, i.e., \eqref{eq:px_equal} holds, then  the union of all nonnegative error exponent pairs $(\theta_1,\theta_2)$ satisfying the following condition are achievable:
\begin{subequations}\label{tmain-equal}
\begin{align}
\theta_i &\leq \min\left\{ \theta_{\text{standard},i}, \; \theta^a_{\text{dec},i},\;\theta^b_{\text{dec},i},\; \theta_{\text{miss},i} \right\},\qquad i\in\{1,2\},\\
\theta_1+\theta_2 &\leq \min\Big\{\theta_{\text{standard},1} +\theta_{\text{standard},2}, \;\; \theta_{\text{standard},1} +\theta^a_{\text{dec},2},\;\;\theta_{\text{standard},1}+\theta^b_{\text{dec},2},\nonumber\\*&\hspace{4.5cm} \theta_{\text{standard},2}+\theta^a_{\text{dec},1},\;\;\theta_{\text{standard},2}+\theta^b_{\text{dec},1},\;\;\theta_{\text{miss},1}+\theta_{\text{miss},2}\Big\}-I_{p^1}(U_1;U_2|S,T),\\
\theta_1+\theta_2 &\leq \min\left\{ \theta^a_{\text{dec},1},\;\theta^b_{\text{dec},1} \right\}+\min\left\{\theta^a_{\text{dec},2},\;\theta^b_{\text{dec},2} \right\}-2I_{p^1}(U_1;U_2|S,T),
\end{align}
\end{subequations}
where the union is over pmfs $P_{T}$, $P_{SU_1U_2|XT}$ and the function $f$ in \eqref{function-BC} so that the joint pmfs $p^{1}, p^2, q^1,q^2$ defined in \eqref{eq:pqi} and \eqref{eq:joint_pi} 
satisfy \eqref{constraints_same_marginal} for $i\in\{1,2\}$, and where  the eight exponents in \eqref{tmain-equal} are defined as 
\begin{align}
\theta_{\text{standard},i} &:= \min_{\substack{\tilde{P}_{SU_iXY_iTV_i}:\\\tilde{P}_{SU_iXT}=p^i_{SU_iXT}\\\tilde{P}_{SU_iY_iTV_i}={p}^{i}_{SU_iY_iTV_i}}} D\left(\tilde{P}_{SU_iXY_iTV_i}\Big\| p^i_{SU_i|X}q^i_{XY_i}P_T\Gamma_{V_i|SU_1U_2X}\right)  ,\\[1.2ex]
\theta^a_{\text{dec},i} &:=\!\!\!\!\!\!\!\!\hspace{-0.2cm} \min_{\substack{\tilde{P}_{SU_iXY_iTV_i}:\\\tilde{P}_{SU_iXT}=p^i_{SU_iXT}\\\tilde{P}_{Y_iTV_i}=p^i_{Y_iTV_i}\\ H_{p^i}(S,U_i|Y_i,T,V_i)\leq H_{\tilde{P}}(S,U_i|Y_i,T,V_i)}}\hspace{-1cm} D\left(\tilde{P}_{SU_iXY_iTV_i} \Big\| p^i_{SU_i|X}q^i_{XY_i}P_T\Gamma_{V_i|SU_1U_2X}\right)-I_{p^i}(S,U_i;X|T)+ I_{p^i}(S,U_i;Y_i,V_i|T),\\[0.5ex]
\theta^b_{\text{dec},i} &:=\!\!\!\!\!\!\!\!\hspace{-0.2cm} \min_{\substack{\tilde{P}_{SU_iXY_iTV_i}:\\\tilde{P}_{SU_iXT}=p^i_{SU_iXT}\\\tilde{P}_{SY_iTV_i}=p^i_{SY_iTV_i}\\ H_{p^i}(U_i|S,Y_i,T,V_i)\leq H_{\tilde{P}}(U_i|S,Y_i,T,V_i)}}\hspace{-1cm} D\left(\tilde{P}_{SU_iXY_iTV_i}\Big\| p^i_{SU_i|X}q^i_{XY_i}P_T\Gamma_{V_i|SU_1U_2X}\right)-I_{p^i}(U_i;X|S,T)+ I_{p^i}(U_i;Y_i,V_i|S,T),\\[0.5ex]
\theta_{\text{miss},i} &:= \mathbb{E}_{P_T}\left[ D\left(p^i_{Y_iV_i|T}\Big\| q_{Y_i}^i \Gamma_{V_i|W=T}\right)\right]-I_{p^i}(S,U_i;X|T)+ I_{p^i}(S,U_i;Y_i,V_i|T).
\end{align}
\end{theorem}
\begin{IEEEproof} The proof is similar to the proof of Theorem~\ref{macthm}. In particular, error exponent $\theta_{\text{standard},i}$ corresponds to the event that  Receiver $i$ decodes the correct cloud and satellite codewords but wrongly decides on $\hat{\mathcal{H}}_i=0$.  In contrast, error exponents $\theta_{\text{dec},i}^a$ and $\theta_{\text{dec},i}^b$ correspond to the events that Receiver~$i$ wrongly decides on $\hat{\mathcal{H}}_i=0$ after wrongly decoding both the cloud center and the satellite or only the satellite. Error exponent $\theta_{\text{miss},i}$ corresponds to the miss-detection event. Because of the implicit rate-constraints in \eqref{eq:rate_constraints_marton}, the final constraints in \eqref{tmain-equal} are obtained by eliminating the rates $R_0, R_1,R_2$ by means of Fourier-Motzkin elimination. 
	\end{IEEEproof}
\medskip

For each $i\in\{1,2\}$, exponents $\theta_{\text{standard},i}, \theta_{\text{dec},i}^a,  \theta_{\text{dec},i}^b$,  and $\theta_{\text{miss},i}$ have the same form as the three exponents in Theorem~\ref{thm2noisy} for the DMC. There is however a tradeoff between the two exponents $\theta_1$ and $\theta_2$ in above theorem because they share the same choice of the auxiliary pmfs $P_T$ and $P_{SU_1U_2|XT}$ and the function $f$. In \cite{Michele3}, the above setup is studied in the special case of testing against conditional independence, and the mentioned tradeoff is illustrated through a Gaussian example. It is further proved that in some special cases, above theorem yields the optimal exponent. 

\medskip
\begin{theorem}\label{thm-BC} If $p_X^{1}\neq p_X^{2}$, i.e., \eqref{eq:px_different} holds,  then  all error exponent pairs $(\theta_1,\theta_2)$ satisfying the following condition are achievable:
	\begin{align}
	\theta_i &\leq \min\{ \theta_{\text{standard},i},\;\theta_{\text{dec},i},\;\theta_{\text{cross},i},\;\theta_{\text{miss},i} \},\qquad i\in\{1,2\},\label{tmain}
	\end{align}
	where the union is over pmfs $p^i_{S|X}$, $p_{T}$, $p_{T_i|T}^i$, and $p^i_{W|T_i}$, for $i\in\{1,2\}$, so that the joint pmfs $p^1,p^2,q^1,q^2$ defined through \eqref{eq:pqi}  and \eqref{dist-P-def} satisfy constraints \eqref{constraint-BC}, and where the exponents in \eqref{tmain} are defined as:
	\begin{align}
\theta_{\text{standard},i} &:= \min_{\substack{\tilde{P}_{SXY_i}:\\\tilde{P}_{SX}=p^i_{SX}\\\tilde{P}_{SY_i}={p}^{i}_{SY_i}}} D(\tilde{P}_{SXY_i}\| p^i_{S|X}q^i_{XY_i}) ,\\[2ex]
\theta_{\text{dec},i} &:=\!\!\!\!\!\!\!\! \min_{\substack{\tilde{P}_{SXY_i}:\\\tilde{P}_{SX}=p^i_{SX}\\\tilde{P}_{Y_i}=p^i_{Y_i}\\ H_{p^i}(S|Y_i)\leq H_{\tilde{P}}(S|Y_i)}}\!\!\!\!\!\!\! D(\tilde{P}_{SXY_i}\| p^i_{S|X}q^i_{XY_i})-I_{p^i}(S;X|Y_i)+ I_p^i(W;V_i|T,T_i),\\[2ex]
\theta_{\text{miss},i} &:= D(p^i_{Y_i}\|q^i_{Y_i})+\mathbb{E}_{p_T}\left[ D\big(p^{i}_{V_i|T}\| \Gamma_{V_i|W=T}\big)\right]-I_{{p^i}}(S;X|Y_i)+I_p^i(W;V_i|T,T_i),\\[2ex]
	\theta_{\text{cross},i} &:=
	\min_{\substack{\tilde{P}_{SXY_i}:\\ \tilde{P}_{Y_i}=p^i_{Y_i}\\ H_{p^i}(S|Y_i)\leq H_{\tilde{P}}(S|Y_i)}}\!    \mathbb{E}_{q^i_{XS}}\left[ D\left(\tilde{P}_{Y_i|XS}\|q^{i}_{Y_i|X}\right)\right] - I_{p^i}(S;X|Y_i)  + I_{p^i}(W;V_i|T,T_i) \nonumber \\[2ex]
	&\hspace{2cm} +\min_{\substack{\tilde{P}_{TT_iW}:\\ \tilde{P}_{TW}=q^i_{TW}\\\tilde{P}_{TT_i}={p}^i_{TT_i}}} \mathbb{E}_{\tilde{P}_{TT_iW}}\left[D(p^i_{V_i|TT_i} \| \Gamma_{V_i|W}) \right].
	\end{align}
\end{theorem}
\begin{IEEEproof} See Appendix \ref{BC-proof}.\end{IEEEproof}

\medskip

In Theorem~\ref{thm-BC},  the exponent triple $\theta_{\text{standard},1}, \theta_{\text{dec},1}, \theta_{\text{miss},1}$ can be optimized over the pmfs $p^1_{T_1|T}$  and $p^1_{W|T,T_1}$ and independently thereof the exponent triple $\theta_{\text{standard},2}, \theta_{\text{dec},2}, \theta_{\text{miss},2}$ can be optimized over the pmfs $p^2_{T_2|T}$  and $p^2_{W|T,T_2}$. (The pmf $p^T$ is common to both optimizations.) Therefore, whenever the two additional exponents $\theta_{\text{cross},1}$ and $\theta_{\text{cross},2}$ are not active, in Theorem~\ref{thm-BC} there is (almost) no tradeoff between the two exponents  $\theta_1$ and $\theta_2$. In other words, the same exponents $\theta_1$ and $\theta_2$ can be attained as in a system where the transmitter communicates over individual DMCs $\Gamma_{V_1|W}$ and $\Gamma_{V_2|W}$   to the two receivers. 

Exponent  $\theta_{\text{cross},1}$ corresponds to the event when the transmitter  sends a codeword from code $\mathcal{C}_{W}^2$, but Receiver~$1$ decides that a codeword from $\mathcal{C}_{W}^1$ was sent and the corresponding source codeword (from source codebook $\mathcal{C}_S^2$) satisfies the minimum conditional entropy and the typicality check with the observed source sequence $y_1^n$.  Similarly for error exponent $\theta_{\text{cross},2}$.
Notice that setting $T_i$ constant, decreases error exponent $\theta_{\text{cross},i}$.

 For  the special case where the BC consists of a common noiseless link,  Theorem~\ref{thm-BC} has been proved in \cite{Michele2} (More precisely, \cite{Michele2} considers the more general case with $K\geq 2$ receivers and $M\geq K$ hypotheses.) In this case, the exponents $(\theta_{\text{miss},1},\theta_{\text{cross},1})$ and $(\theta_{\text{miss},2},\theta_{\text{cross},2})$  are not active.%

\section{Summary and Discussion}\label{sec:summary}
The paper proposes coding and testing schemes for distributed binary hypothesis testing    over DMCs, MACs, and BCs when each decision center aims at maximizing  a single error exponent. Our schemes recover previous optimality results for testing against conditional independence when terminals are connected by noisefree  links or  DMCs. They are in fact  optimal for a more  general testing setup that we term \emph{generalized testing against conditional independence}. To prove this, we derive new information-theoretic converse bounds. In all   these cases, separate source-channel coding suffice.  

Our schemes apply hybrid coding (in case of MAC and BC) and UEP mechanisms  to specially protect the transmission of single bits (typically the tentative guesses of the sensor nodes). These features can significantly improve the achieved error exponents. 

In this work, we have focused on the most basic communication channels: DMC, MAC, BC. Similar investigations can be performed for more involved networks. 
\appendices

\section{Proof of Theorem~\ref{thm2noisy}}\label{sec:proof}

The proof of the theorem is based on the scheme in Section~\ref{sec:scheme}.  Fix a choice of the blocklength $n$, the small positive $\mu$, and the  (conditional) pmfs $P_T, P_{W|T},$ and $P_{S|X}$ so that \eqref{nois3} holds. 
Assume that $I(S;X)\geq I(W;V|T)$, in which  case the rates $R$ and $R'$ are chosen as in \eqref{nois2} and \eqref{nois0}. Also, set for convenience of notation:
\begin{IEEEeqnarray}{rCl}
P_{S'}(s) &=&P_{S}(s), \qquad \forall s \in \mathcal{S}, \\
P_{W'|T}(w|t) & =& P_{W|T}(w|t), \qquad \forall t \in \mathcal{T}, \ w \in \mathcal{W}.
\end{IEEEeqnarray}
Let $\mathcal{P}_{\mu,\text{type-I}}^n$ be the subset of types $\pi_{SS'XY}\in\mathcal{P}^n$ that simultaneously satisfy the following  conditions for all $(s,s',x,y)\in \mathcal{S}\times\mathcal{S}\times \mathcal{X}\times \mathcal{Y}$:
\begin{align}
|\pi_{SX}(s,x)-P_{SX}(s,x)| &\leq \mu/2,\\
|\pi_{SY}(s,y)-P_{SY}(s,y)| & \leq \mu,\\
|\pi_{S'}(s)-P_{S'}(s)|  &\leq \mu,
\end{align}
and 
\begin{align}
H_{\pi_{S'Y}}(S'|Y)&\leq H_{\pi_{SY}}(S|Y). 
\end{align}
Notice that 
\begin{equation}\label{eq:Pstar_limit}\mathcal{P}_{\mu,\text{type-I}}^n \to \mathcal{P}_{\text{type-I}}^* \qquad \textnormal{as }\; \mu \to 0 \textnormal{ \; and\; } n \to \infty,
\end{equation}
where 
\begin{IEEEeqnarray}{rCl}\label{eq:Pstar}
\mathcal{P}_{\text{type-I}}^* := \big\{ \tilde{P}_{SS'XY} \colon \tilde{P}_{SX}=P_{SX} \textnormal{ and } \tilde{P}_{SY}=P_{SY} \textnormal{ and } \tilde{P}_{S'}=P_S \textnormal{ and }  H_{\tilde{P}_{S'Y}}(S'|Y) \leq H_{\tilde{P}_{SY}}(S|Y)  \big\}.
\end{IEEEeqnarray}

Consider now the type-I error probability averaged over the random code construction.  Let $(M,L)$ be the indices of the codeword chosen at the transmitter, if they exist, and define the following events:
\begin{align}
\mathcal{E}_{\text{Tx}}&\colon \{ \nexists (m,\ell) \colon (S^n(m,\ell),X^n)\in\mathcal{T}_{\mu/2}^n(P_{SX})\}\\
\mathcal{E}_{\text{Rx}}^{(1)}&\colon \{  (S^n(M,L),Y^n)\notin\mathcal{T}_{\mu}^n(P_{SY}) \}\\ 
\mathcal{E}_{\text{Rx}}^{(2)}&\colon \{ \exists m' \neq M \colon (T^n,W^n(m'),V^n)\in\mathcal{T}_{\mu}^n(P_{TWV}) \}\\ 
\mathcal{E}_{\text{Rx}}^{(3)}&\colon \{ \exists \ell'\neq L\colon  H_{\text{tp}(s^n(M,\ell'),y^n)}(S|Y)= \min_{\tilde{\ell} } H_{\text{tp}(s^n(M,\tilde{\ell}),y^n)}(S|Y)\} . 
\end{align}
We obtain for all sufficiently small values of $\mu$ and sufficiently large blocklengths $n$:
\begin{align}
\mathbb{E}_{\mathcal{C}}[\alpha_n ]
&\leq \Pr\left[\mathcal{E}_{\text{Tx}}\right]+\Pr\left[\mathcal{E}_{\text{Rx}}^{(1)} \Big |\mathcal{E}_{\text{Tx}}^{c}\right] +\Pr\left[\mathcal{E}_{\text{Rx}}^{(2)} \Big|\mathcal{E}_{\text{Rx}}^{(1)c},\mathcal{E}_{\text{Tx}}^{c}\right] + \Pr\left[\mathcal{E}_{\text{Rx}}^{(3)} \Big| \mathcal{E}_{\text{Rx}}^{(1)c},\mathcal{E}_{\text{Rx}}^{(2)c}, \mathcal{E}_{\text{Tx}}^{c}\right] \label{eq:d}\\
&\leq  \epsilon/8 + \epsilon/8 + \epsilon/8 + \epsilon/8\\&=\epsilon/2,
\end{align}
where the first summand of \eqref{eq:d} is upper bounded by means of the covering lemma \cite{ElGamal} and using rate constraints \eqref{nois2} and \eqref{nois0};  the second by means of the Markov lemma \cite{ElGamal}; the third by  following a similar set of inequalities as in  \cite[Appendix~H]{Michele}:  
\begin{IEEEeqnarray}{rCl}
\lefteqn{ \Pr\left[\mathcal{E}_{\text{Rx}}^{(2)}\Big|\mathcal{E}_{\text{Rx}}^{(1)c},\mathcal{E}_{\text{Tx}}^{(0)c}\right] }  \nonumber \\[1ex]
& =&\Pr\Big[ H_{\text{tp}(S^n(M,L),Y^n)}(S|Y) \geq \min_{\tilde{\ell}\neq L} H_{\text{tp}(S^n(M,\tilde{\ell}),Y^n)}(S|Y)\; \big| \nonumber\\&&\qquad\qquad\qquad\qquad\qquad \; (S^n(M,L),Y^n)\in \mathcal{T}_{\mu}^n(P_{SY}),\; \; (S^n(M,L),X^n)\in \mathcal{T}_{\mu/2}^n(P_{SX}),\; \;S^n(M,\tilde{\ell}) \in\mathcal{T}_{\mu/2}^n(P_S) \Big]\nonumber\\[1.2ex]
& \stackrel{(a)}{=}&\Pr\Big[ H_{\text{tp}(S^n(1,1),Y^n)}(S|Y) \geq \min_{\tilde{\ell}> 1} H_{\text{tp}(S^n(1,\tilde{\ell}),Y^n)}(S|Y)\; \big| \nonumber\\
&&\qquad\qquad\qquad\qquad\qquad \;  (S^n(1,1),Y^n)\in \mathcal{T}_{\mu}^n(P_{SY}),\;\;(S^n(1,1),X^n)\in \mathcal{T}_{\mu/2}^n(P_{SX}),\;\; S^n(1,\tilde{\ell}) \in\mathcal{T}_{\mu/2}^n(P_S), \;\;M=L=1 \Big]\nonumber\\[1.2ex]
&= &\Pr\Big[ \bigcup_{\tilde{\ell} > 1} \big\{H_{\text{tp}(S^n(1,1),Y^n)}(S|Y) \geq  H_{\text{tp}(S^n(1,\tilde{\ell}),Y^n)}(S|Y)\big\}\; \big| \nonumber\\[-1.5ex]&&\qquad\qquad\qquad\qquad\qquad \;  (S^n(1,1),Y^n)\in \mathcal{T}_{\mu}^n(P_{SY}),\;\;(S^n(1,1),X^n)\in \mathcal{T}_{\mu/2}^n(P_{SX}),\; \;S^n(1,\tilde{\ell}) \in\mathcal{T}_{\mu/2}^n(P_S),\;\; M=L=1 \Big]\nonumber\\[1.2ex]
&\stackrel{(b)}{\leq} & \sum_{\substack{\pi_{SS'Y}\\\in\mathcal{P}_{\mu,\text{type-I}}^{n}}}\; \; \sum_{\tilde{\ell}=2}^{\lfloor 2^{nR'}\rfloor} \; \; \sum_{\substack{s^n,s'^n,y^n:\\\text{tp}(s^n,s'^{n},y^n)\\=\pi_{SS'Y}}} \Pr\Big[S^n(1,1)=s^n, S^n(1,\tilde{\ell})=s'^n,Y^n=y^n\; \big| \;\nonumber \\		
&& \hspace{3.5cm} (S^n(1,1),Y^n)\in \mathcal{T}_{\mu}^n(P_{SY}),\;\;(S^n(1,1),X^n)\in \mathcal{T}_{\mu/2}^n(P_{SX}),\; \; S^n(1,\tilde{\ell}) \in\mathcal{T}_{\mu/2}^n(P_S), \;\; M=L=1 \Big]\nonumber\\[1.2ex]
&\stackrel{(c)}{\leq}& \sum_{\substack{\pi_{SS'Y}\\\in\mathcal{P}_{\mu,\text{type-I}}^{n}}} \;\; \sum_{\tilde{\ell}=2}^{\lfloor 2^{nR'}\rfloor}\;\sum_{\substack{s^n,s'^n,y^n:\\\text{tp}(s^n,s'^{n},y^n)\\=\pi_{SS'Y}}} \Pr\Big[S^n(1,1)=s^n,Y^n=y^n\; \big| \nonumber\\
&&\qquad\qquad\qquad\;\;\;\; \; \qquad\;\; (S^n(1,1),Y^n)\in \mathcal{T}_{\mu}^n(P_{SY}),\;\; (S^n(1,1),X^n)\in \mathcal{T}_{\mu/2}^n(P_{SX}),\; \; S^n(1,\tilde{\ell}) \in\mathcal{T}_{\mu/2}^n(P_S), \;\;M=L=1 \Big]\nonumber \\[1.2ex]
&& \hspace{1cm}\cdot \Pr \Big[S^n(1,\tilde{\ell})=s'^n \big| \nonumber\\&& \qquad\qquad\qquad\qquad\qquad (S^n(1,1),Y^n)\in \mathcal{T}_{\mu}^n(P_{SY}),\;\;(S^n(1,1),X^n)\in \mathcal{T}_{\mu/2}^n(P_{SX}),\; \; S^n(1,\tilde{\ell}) \in\mathcal{T}_{\mu/2}^n(P_S),\;\; M=L=1\Big]\nonumber\\[1.2ex]
&\stackrel{(d)}{\leq}& \sum_{\pi_{SS'Y}\in\mathcal{P}_{\mu,\text{type-I}}^{n}}\sum_{\tilde{\ell}=2}^{\lfloor 2^{nR'}\rfloor}\sum_{\substack{s^n,y^n,s'^n:\\\text{tp}(s^n,s'^{n},y^n)=\pi_{SS'Y}}} 2^{-nH_{\pi}(S,Y)}\cdot 2^{-nH_{\pi}(S')}\nonumber\\[0.8ex]
&\stackrel{(e)}{\leq}&\sum_{\pi_{SS'Y}\in\mathcal{P}_{\mu,\text{type-I}}^{n}}\sum_{\tilde{\ell}=2}^{\lfloor 2^{nR'}\rfloor}\;\;\;\;\;\;\;\;\;\;\; 2^{nH_{\pi}(S,S',Y)}\cdot 2^{-nH_{\pi}(S,Y)}\cdot 2^{-nH_{\pi}(S')}\nonumber\\[0.8ex]
&=&\sum_{\pi_{SS'Y}\in\mathcal{P}_{\mu,\text{type-I}}^{n}}  2^{n(R'-I_{\pi}(S';Y,S))}\nonumber\\[0.8ex]
&\leq&\sum_{\pi_{SS'Y}\in\mathcal{P}_{\mu,\text{type-I}}^{n}}  2^{n(R'-I_{\pi}(S';Y))}\nonumber\\[0.8ex]
&\stackrel{(f)}{\leq} & (n+1)^{|\mathcal{S}|^2\cdot |\mathcal{Y}|}\cdot \max_{\pi_{SS'Y}\in\mathcal{P}_{\mu,\text{type-I}}^{n}}  2^{n(R'-I(S;Y)+\delta_n(\mu))} \label{eq:lastss}  \\
& \stackrel{(g)}{\leq} & \epsilon/8,
\end{IEEEeqnarray}
where $\delta_n(\mu)$ is a function that tends to $0$ as $\mu \to 0$ and  $n \to \infty$. The inequalities are justified as follows:
\begin{itemize}
	\item 
	$(a)$: holds by the symmetry in the code construction;
	
	\item $(b)$: holds by the union bound;
	
	\item $(c)$:  holds because the codebook's codewords are drawn independently of each other;
	\item $(d)$: holds because all $2^{nH_{\pi}(S,S',Y)}$ tuples $(s^n,s'^n,y^n)$ of the same type $\pi$ have same conditional probability and similarly all $2^{nH_{\pi}(S'|Y)}$ sequences $s'^n$ of the same joint type have same conditional probability;
	\item  
	 $(e)$:   holds by standard  arguments on types;
	\item $(f)$:  holds because $|\mathcal{P}_{\mu,\text{type-I}}^{n}| \leq (n+1)^{|\mathcal{S}|^2 \cdot |\mathcal{Y}|}$, because $H_{\pi}(S'|Y) \leq H_{\pi}(S|Y)$, by\eqref{eq:Pstar_limit}, and  by the continuity of the entropy function; and
	\item $(g)$: holds for all sufficiently large $n$ and small $\mu$   because    $R'< I(S;Y)$ and $\delta_n(\mu) \to 0$ as  $n\to \infty$ and $\mu \to 0$.
\end{itemize}

Now, consider the type-II error probability averaged over the random code construction. For all $m,m'\in\{1,\ldots, \lfloor 2^{nR} \rfloor \}$ and $\ell, \ell' \in \{1,\ldots, \lfloor 2^{nR'} \rfloor \}$  define events:
\begin{align}
\mathcal{E}_{\text{Tx}}(m,\ell)&\colon \{(S^n(m,\ell),X^n)\in \mathcal{T}_{\mu/2}^n(P_{SX}) , \; \; W^n(m) \textnormal{ is sent}\},\\
\mathcal{E}_{\text{Rx}}(m',\ell')&\colon \{(S^n(m',\ell'),Y^n)\in \mathcal{T}_{\mu}^n(P_{SY}),\; (T^n,W^n(m'),V^n)\in \mathcal{T}_{\mu}^n(P_{TWV}),\nonumber\\&\qquad \qquad \qquad\qquad\qquad\qquad\qquad\qquad\qquad\qquad\qquad\qquad H_{\text{tp}(S^n(m',\ell'),Y^n)}(S|Y) = \min_{\tilde{\ell}} H_{\text{tp}(S^n(m',\tilde{\ell}),Y^n)}(S|Y)\},
\end{align}
and notice that  
\begin{align}
\mathbb{E}_{\mathcal{C}}[\beta_n] = \Pr[\hat{\mathcal{H}}=0|\mathcal{H}=1] = \Pr\left[\bigcup_{m',\ell'}\;\;\mathcal{E}_{\text{Rx}}(m',\ell')\Big|\mathcal{H}=1\right],
\end{align}
where the union is over all indices $(m',\ell')\in\{1,\ldots, \lfloor 2^{nR} \rfloor \}\times \{1,\ldots, \lfloor 2^{nR'} \rfloor \}$. Above probability is upper bounded by the sum of the probabilities of the following four events:
\begin{align}
\mathcal{B}_1 &\colon \{ \exists\; (m,\ell) \qquad\;\;\;\;\; \text{s.t.}\qquad \mathcal{E}_{\text{Tx}}(m,\ell)\qquad\text{and}\qquad \mathcal{E}_{\text{Rx}}(m,\ell)  \},\\
\mathcal{B}_2 &\colon \{\exists\; (m,m',\ell,\ell')\;\;\;\text{with}\qquad m\neq m'\qquad\text{and}\qquad \ell\neq \ell'\qquad \text{s.t.}\qquad \left(\mathcal{E}_{\text{Tx}}(m,\ell)\qquad\text{and}\qquad \mathcal{E}_{\text{Rx}}(m',\ell')\right) \},\\
\mathcal{B}_3 &\colon \{\exists\; (m,\ell,\ell')\qquad\;\;\text{with}\qquad \ell\neq \ell'\qquad \text{s.t.}\qquad \left(\mathcal{E}_{\text{Tx}}(m,\ell)\qquad\text{and}\qquad \mathcal{E}_{\text{Rx}}(m,\ell')\right) \},\\
{\mathcal{B}}_4 &\colon \{\forall\; (m,\ell) \;\;\;\;\mathcal{E}_{\text{Tx}}^c(m,\ell)\;\;\;\; \text{holds}\qquad \text{and}\qquad \exists\; (m',\ell')\;\;\; \text{s.t.}\;\;\; \mathcal{E}_{\text{Rx}}(m',\ell') \},
\end{align}
i.e., 
\begin{align}
\mathbb{E}_{\mathcal{C}}[\beta_n] \leq \sum_{i=1}^4 \Pr\big[\mathcal{B}_i \big| \mathcal{H}=1\big].\label{probfirst}
\end{align}
We will bound the four probabilities on the right-hand side of \eqref{probfirst} individually. To simplify notation, we introduce  the following sets of types
\begin{align}
\mathcal{P}_{\mu,\text{standard}} &= \{\pi_{SXY}\colon |\pi_{SX}-P_{SX}|<\mu/2,\qquad |\pi_{SY}-P_{SY}|<\mu \},\\
\mathcal{P}_{\mu,\text{decoding}} &= \{\pi_{SS'XY}\colon |\pi_{SX}-P_{SX}|<\mu/2,\qquad |\pi_{S'Y}-P_{SY}|<\mu,\qquad  H_{\pi}(S'|Y)\leq H_{\pi}(S|Y) \}.
\end{align}

Consider the probability of the first event $\mathcal{B}_1$:
	\begin{align}
		&\Pr \left[\mathcal{B}_1 | \mathcal{H}=1\right] \nonumber\\
		&\leq\sum_{m,\ell} \Pr \left[(S^n(m,\ell),X^n)\in \mathcal{T}_{\mu/2}^n(P_{SX}),\;\; (S^n(m,\ell),Y^n)\in \mathcal{T}_{\mu}^n(P_{SY}),\;\; (T^n,W^n(m),V^n)\in \mathcal{T}_{\mu}^n(P_{TWV})\; \Big|\; \mathcal{H}=1  \right]\nonumber\\
		&\leq\sum_{m,\ell} \Pr \left[(S^n(m,\ell),X^n)\in \mathcal{T}_{\mu/2}^n(P_{SX}),\;\; (S^n(m,\ell),Y^n)\in \mathcal{T}_{\mu}^n(P_{SY})\;  \Big|\; \mathcal{H}=1  \right]\nonumber\\
		&\stackrel{(a)}{\leq} 2^{n (R+R')} \cdot \max_{\substack{\pi:\\ |\pi_{SX}-P_{SX}|<\mu/2\\ |\pi_{SY}-P_{SY}|<\mu}} 2^{-n( D(\pi_{SXY}\|P_SQ_{XY}) -\mu)},\label{eq:prob2}
	\end{align}
	where  inequality $(a)$ follows by Sanov's theorem and by the way the source sequences, the codewords, and the channel outputs are generated. 
	Define now 
	\begin{align}
\tilde{\theta}_{\mu}^{\text{standard}} &:=\min_{\substack{\pi:\\ |\pi_{SX}-P_{SX}|<\mu/2\\ |\pi_{SY}-P_{SY}|<\mu}} D(\pi_{SXY}\|P_SQ_{XY})-R-R'-\mu,
		 \label{b1}
	\end{align}
	and observe that:
	\begin{IEEEeqnarray}{rCl}
		\tilde{\theta}_{\mu}^{\text{standard}}  &\stackrel{(\textnormal{eq.}\eqref{nois2}\&\eqref{nois0})}{=} &\min_{\substack{\pi_{SXY}:\\ |\pi_{SX}-P_{SX}|<\mu/2\\ |\pi_{SY}-P_{SY}|<\mu}} D(\pi_{SXY}\|P_SQ_{XY})-I(S;X)-2\mu\nonumber\\[1.2ex]
		&= & \min_{\pi_{SXY}\in\mathcal{P}_{\mu,\text{standard}}} D(\pi_{SXY}\|P_{S|X}Q_{XY})-\delta_1(\mu)\nonumber\\[1.2ex]
		&=: & \theta_{\mu}^{\text{standard}}-\delta_1(\mu),\label{t1final}
	\end{IEEEeqnarray}
	for a function $\delta_1(\mu)$ that goes to zero as $\mu\to 0$. 
	Combining \eqref{eq:prob2}--\eqref{t1final}, we obtain:
	\begin{align}
	\Pr\big[ \mathcal{B}_1 | \mathcal{H}=1\big] \leq 2^{-n\big(\theta_{\mu}^{\text{standard}}-\delta_1(\mu)\big)}.\label{stan-f}
	\end{align}
		
Consider next the probability of event $\mathcal{B}_2$:
	\begin{align}
		& \Pr \left[ \mathcal{B}_2 | \mathcal{H}=1\right]  
		\nonumber\\& \leq  \sum_{\substack{m,m':\\m \neq m'}}\;\;\sum_{\ell,\ell'} \Pr\Big[ (S^n(m,\ell),X^n)\in \mathcal{T}_{\mu/2}^n(P_{SX}),\;\; W^n(m) \textnormal{ is sent}, \;\; (S^n(m',\ell'),Y^n)\in \mathcal{T}_{\mu}^n(P_{SY}), \nonumber\\[-4ex]&\qquad \qquad \qquad\qquad H_{\text{tp}(S^n(m',\ell'),Y^n)}(S|Y) = \min_{\tilde{\ell}} H_{\text{tp}(S^n(m',\tilde{\ell}),Y^n)}(S|Y), \;\;  (T^n,W^n(m'),V^n)\in \mathcal{T}_{\mu}^n(P_{TWV})\;\; \Big|\; \mathcal{H}=1\Big]\\[2ex]
		& =  \sum_{\substack{m,m':\\m \neq m'}}\;\;\sum_{\ell,\ell'} \Pr\Big[ (S^n(m,\ell),X^n)\in \mathcal{T}_{\mu/2}^n(P_{SX}), \;\; (S^n(m',\ell'),Y^n)\in \mathcal{T}_{\mu}^n(P_{SY}), \nonumber\\[-4ex]
		&\hspace{7cm} H_{\text{tp}(S^n(m',\ell'),Y^n)}(S|Y) = \min_{\tilde{\ell}} H_{\text{tp}(S^n(m',\tilde{\ell}),Y^n)}(S|Y) \;\; \Big|\; \mathcal{H}=1\Big]\\[1ex]
	& \hspace{2.4cm} \cdot \Pr\left[ W^n(m) \textnormal{ is sent}, \;\;  (T^n,W^n(m'),V^n)\in \mathcal{T}_{\mu}^n(P_{TWV})\;\; \big|\; \mathcal{H}=1\right]\nonumber \\[2ex]
			& \leq   \sum_{\substack{m,m':\\m \neq m'}}\;\;\sum_{\ell,\ell'} \Pr\Big[ (S^n(m,\ell),X^n)\in \mathcal{T}_{\mu/2}^n(P_{SX}), \;\; (S^n(m',\ell'),Y^n)\in \mathcal{T}_{\mu}^n(P_{SY}), \nonumber\\[-4ex]
			&\hspace{7cm} H_{\text{tp}(S^n(m',\ell'),Y^n)}(S|Y) = \min_{\tilde{\ell}} H_{\text{tp}(S^n(m',\tilde{\ell}),Y^n)}(S|Y) \;\; \Big|\; \mathcal{H}=1\Big]\\[1ex]
			& \hspace{2.4cm} \cdot \Pr\left[  (T^n,W^n(m'),V^n)\in \mathcal{T}_{\mu}^n(P_{TWV})\;\; \big|\;W^n(m) \textnormal{ is sent}, \;\;  \mathcal{H}=1\right]\nonumber \\[2ex]
		& \leq 2^{n (2R+2R')}\cdot\max_{\substack{\pi_{SS'XY}:\\ |\pi_{SX}-P_{SX}|<\mu/2\\ |\pi_{S'Y}-P_{SY}|<\mu\\ H_{\pi}(S'|Y)\leq H_{\pi}(S|Y)}}\;   2^{-n\big(D\big(\pi_{SS'XY}\|P_SP_{S'}Q_{XY}\big)-\mu\big)} \;\; \cdot  \max_{ \substack{\pi_{TW'V}\colon \\ |\pi_{TW'V} - P_{TWV}| \leq \mu}}  2^{-n\big(D\big(\pi_{TW' V}\|P_{TV}P_{W'|T}\big)-\mu\big)},\label{b2first}
	\end{align}
	where the last inequality holds by Sanov's theorem and by the way the codebooks and the channel outputs are generated. 
	Define 
	\begin{align}
		 \tilde{\theta}_{\mu}^{\text{dec}}&:= \min_{\substack{\pi_{SS'XY}:\\ |\pi_{SX}-P_{SX}|<\mu/2\\ |\pi_{S'Y}-P_{SY}|<\mu\\  H_{\pi}(S'|Y)\leq H_{\pi}(S|Y)}}  D\big(\pi_{SS'XY}\|P_SP_{S'}Q_{XY}\big)+  \min_{ \substack{\pi_{TW'V}\colon \\ |\pi_{TW'V} - P_{TWV}| \leq \mu}}  D\big(\pi_{TW' V}\|P_{TV}P_{W'|T}\big)-2R-2R'-2\mu,\label{b2}
	\end{align}
	and observe that:
		\begin{IEEEeqnarray}{rCl}
		\tilde{\theta}_{\mu}^{\text{dec}} &\stackrel{(\textnormal{eq.}\eqref{nois2}\&\eqref{nois0})}{=}&\min_{\substack{\pi_{SS'XY}:\\ |\pi_{SX}-P_{SX}|<\mu/2\\ |\pi_{S'Y}-P_{SY}|<\mu\\ H_{\pi}(S'|Y)\leq H_{\pi}(S|Y)}}  D\big(\pi_{SS'XY }\|P_SP_{S'}Q_{XY}\big) +  \min_{ \substack{\pi_{TW'V}\colon \\ |\pi_{TW'V} - P_{TWV}| \leq \mu}}  D\big(\pi_{TW' V}\|P_{TV}P_{W'|T}\big)-2I(S;X)-4\mu
		\nonumber\\[2ex]
 	&\stackrel{(\textnormal{CR})}{=} & \min_{\substack{\pi_{SS'XY}:\\ |\pi_{SX}-P_{SX}|<\mu/2\\ |\pi_{S'Y}-P_{SY}|<\mu\\ H_{\pi}(S'|Y)\leq H_{\pi}(S|Y)}} \big[ D(\pi_{S XY}\|P_SQ_{XY}) +\mathbb{E}_{\pi_{SXY}}\left[D(\pi_{S'|SXY}\|P_{S'})\right] \big]  \nonumber \\
 	 &&\hspace{4cm}  + \min_{ \substack{\pi_{TW'V}\colon \\ |\pi_{TW'V} - P_{TWV}| \leq \mu}}  D\big(\pi_{TW' V}\|P_{TV}P_{W'|T}\big) -2I(S;X)-4\mu\nonumber\\[5ex]
		&\stackrel{(\textnormal{DP})}{\geq} & \min_{\substack{\pi_{SS'XY}:\\ |\pi_{SX}-P_{SX}|<\mu/2\\ |\pi_{S'Y}-P_{SY}|<\mu\\ H_{\pi}(S'|Y)\leq H_{\pi}(S|Y)}}\left[ D(\pi_{S XY}\|P_SQ_{XY})+\mathbb{E}_{\pi_Y}\left[D(\pi_{S'|Y}\|P_{S'})\right] \right] \nonumber \\
		&&\hspace{4cm}+   \min_{ \substack{\pi_{TW'V}\colon \\ |\pi_{TW'V} - P_{TWV}| \leq \mu}}  D\big(\pi_{TW' V}\|P_{TV}P_{W'|T}\big)-2I(S;X)-4\mu\nonumber\\[5ex]
		&\stackrel{(b)}{=} & \min_{\substack{\pi_{SS'XY}:\\ |\pi_{SX}-P_{SX}|<\mu/2\\ |\pi_{S'Y}-P_{SY}|<\mu\\ H_{\pi}(S'|Y)\leq H_{\pi}(S|Y)}} D(\pi_{S XY}\|P_{S|X}Q_{XY})+I(S;Y)-I(W;V|T)-2 I(S;X)-\delta_2(\mu)\nonumber\\[5ex]
		&= &\min_{\pi_{SS'XY}\in\mathcal{P}_{\mu,\text{decoding}}} D(\pi_{S XY}\|P_{S|X}Q_{XY})+I(S;Y) +I(W;V|T)-I(S;X)-\delta_2(\mu)\nonumber\\[2ex]
		&=:&\theta_{\mu}^{\text{dec}}-\delta_2(\mu),\label{t2final}
	\end{IEEEeqnarray}
	for a function $\delta_2(\mu)$ that goes to zero as $\mu\to 0$. Here, (CR) and (DP) refer to chain rule and data processing inequality arguments, 
	$(b)$ follows because $\pi_{TW'V}=P_{TWV}$ and $P_{W'|T}=P_{W|T}$ and because $\pi_{S'|Y}=P_{S|Y}$. (Notice that the DP-inequality can be shown to hold with equality.) Combining \eqref{b2first}, \eqref{b2}, and \eqref{t2final}, we have
	\begin{align}
 \Pr \big[ \mathcal{B}_2 | \mathcal{H}=1\big]   \leq 2^{-n \big( \theta_{\mu}^{\text{dec}}-\delta_2(\mu)\big)}.\label{dec-ff}
	\end{align}
	
Consider next the third event $\mathcal{B}_3$:
\begin{IEEEeqnarray}{rCl}
	\lefteqn{ \Pr \big[ \mathcal{B}_3 \big| \mathcal{H}=1\big] } 
\nonumber\\
& \leq & \sum_{m}\;\;\sum_{\ell,\ell'} \Pr\Big[ (S^n(m,\ell),X^n)\in \mathcal{T}_{\mu/2}^n(P_{SX}),\;\; W^n(m) \textnormal{ is sent}, \; \; (S^n(m,\ell'),Y^n)\in \mathcal{T}_{\mu}^n(P_{SY}),\nonumber\\[-1ex]
&& \hspace{2.5cm} \;\; (T^n,W^n(m),V^n)\in \mathcal{T}_{\mu}^n(P_{TWV}),\;\; H_{\text{tp}(S^n(m,\ell'),Y^n)}(S'|Y) = \min_{\tilde{\ell}} H_{\text{tp}(S^n(m,\tilde{\ell}),Y^n)}(S|Y) \;\; \Big| \;\; \mathcal{H}=1\Big]\nonumber\\
& \leq & \sum_{m}\;\;\sum_{\ell,\ell'} \Pr\Big[ (S^n(m,\ell),X^n)\in \mathcal{T}_{\mu/2}^n(P_{SX}),\;\; (S^n(m,\ell'),Y^n)\in \mathcal{T}_{\mu}^n(P_{SY}), \nonumber\\[-1ex]
&& \hspace{7.5cm} H_{\text{tp}(S^n(m,\ell'),Y^n)}(S'|Y) = \min_{\tilde{\ell}} H_{\text{tp}(S^n(m,\tilde{\ell}),Y^n)}(S|Y)\;\; \Big| \;\; \mathcal{H}=1\Big]\nonumber\\
&\leq &2^{n(R+2R')}\cdot \max_{\substack{\pi:\\ |\pi_{SX}-P_{SX}|<\mu/2\\ |\pi_{S'Y}-P_{SY}|<\mu\\ H_{\pi}(S'|Y)\leq H_{\pi}(S|Y)}}\;2^{-n\big(D(\pi_{SS'XY}\|P_SP_SQ_{XY})-\mu\big)},\label{b3event}
\end{IEEEeqnarray}
where the last inequality again holds by Sanov's theorem and the way the codebooks and the channel outputs are generated.

Define 
\begin{align}
\tilde{\theta}_{\mu}^{'\text{dec}}   &:= \min_{\substack{\pi_{SS'XY}\colon \\ |\pi_{SX}-P_{SX}|<\mu/2\\ |\pi_{S'Y}-P_{SY}|<\mu\\ H_{\pi}(S'|Y)\leq H_{\pi}(S|Y)}}  D(\pi_{SS'XY}\|P_SP_{S'}Q_{XY})-R-2R'-\mu,\label{b3}
\end{align}
and notice  that:
\begin{IEEEeqnarray}{rCl}
\tilde{\theta}_{\mu}^{'\text{dec}} &\stackrel{(\textnormal{eq.}\eqref{nois2}\&\eqref{nois0})}{=} & \min_{\substack{\pi_{SS'XY}\colon\\ |\pi_{SX}-P_{SX}|<\mu/2\\ |\pi_{S'Y}-P_{SY}|<\mu\\H_{\pi}(S'|Y)\leq H_{\pi}(S|Y)}}  D(\pi_{SS'XY}\|P_SP_{S'}Q_{XY})+I(W;V|T)-2 I(S;X) -4\mu\nonumber\\[0.5ex]
&\stackrel{(\textnormal{CR})\&(\textnormal{DP})}{\geq} & \min_{\substack{\pi_{SS'XY}\colon\\ |\pi_{SX}-P_{SX}|<\mu/2\\ |\pi_{S'Y}-P_{SY}|<\mu\\ H_{\pi}(S'|Y)\leq H_{\pi}(S|Y)}} 
\Big[D(\pi_{SXY}\|P_SQ_{XY})+\mathbb{E}_{\pi_Y}\big[D(\pi_{S'|Y}\|P_{S'})\big]\Big]+I(W;V|T)-2 I(S;X) -4\mu\nonumber\\[1ex]
&\stackrel{(c)}{=} & \min_{\pi_{SS'XY}\in\mathcal{P}_{\mu,\text{decoding}}} 
D(\pi_{SXY}\|P_{S|X}Q_{XY})+I(S;Y) +I(W;V|T)- I(S;X) -\delta_2'(\mu)\nonumber\\[1ex]
&=&\theta_{\mu}^{\text{dec}}-\delta_2'(\mu),\label{t3pfinal}
\end{IEEEeqnarray}
where $\delta_2'(\mu)$  is a function that goes to zero as $\mu\to 0$. Here, 
 $(c)$ holds because $\pi_{S'|Y}=P_{S|Y}$. (Notice that the DP-inequality can again be shown to hold with equality.) By \eqref{b3event}--\eqref{t3pfinal}, we conclude
\begin{align}
\Pr\big[\mathcal{B}_3|\mathcal{H}=1\big]\leq 2^{-n\big(\theta_{\mu}^{\text{dec}}-\delta_2'(\mu)\big)}.\label{dec-f}
\end{align}

Finally, consider the probability of the fourth event $\mathcal{B}_4$. By the union bound:
\begin{IEEEeqnarray}{rCl}
\lefteqn{ \Pr \big[ \mathcal{B}_4 \big| \mathcal{H}=1\big]  }
\nonumber\\[1.2ex]
& \leq & \sum_{m'}\;\;\sum_{\ell'} \Pr \bigg[ \bigg( \bigcap_{ (m,\ell)}  \mathcal{E}_{\text{Tx}}^c(m,\ell)\bigg)\;\;  \cap\; \;  \mathcal{E}_{\text{Rx}}(m',\ell')\bigg|\ \mathcal{H}=1\bigg]\nonumber\\[1.2ex] 
& \stackrel{(d)}{\leq} & \sum_{m'}\;\;\sum_{\ell'} \Pr \bigg[ (S^n(m',\ell'),Y^n)\in \mathcal{T}_{\mu}^n(P_{SY})  \bigg|\ \mathcal{H}=1\big] \cdot \Pr \bigg[ (T^n,W^n(m'),V^n)\in \mathcal{T}_{\mu}^n(P_{TWV}) \; \bigg| \;\bigg( \bigcap_{ (m,\ell)}  \mathcal{E}_{\text{Tx}}^c(m,\ell)\bigg), \; \mathcal{H}=1\bigg] \nonumber\\[2ex] 
& \stackrel{(e)}{\leq} & \sum_{m'}\sum_{\ell'}  \left( \sum_{\substack{\pi_{S'Y}:\\  |\pi_{S'Y}-P_{SY}|<\mu}}\;\;\; 2^{-nD(\pi_{S'Y}\| P_{S'}Q_{Y})} \right) \cdot\left( \sum_{\substack{\pi_{TW'V}: \\
|\pi_{TW'V}-P_{TWV}|<\mu}} \;\;\; 2^{-nD(\pi_{TW'V}\| P_{TW'} \Gamma_{V|W=T})}\right). \label{b4event} 
\end{IEEEeqnarray}	
where 
$(d)$ holds because the tuple $(T^n, W^n(m'), V^n)$ is generated independently of the pair $(S^n(m',\ell'), Y^n)$ and $(e)$ holds by Sanov's theorem and the way the codebooks and the source sequences are generated. 

Define now
\begin{align}
\tilde{\theta}_{\mu}^{\text{miss}} 
&:= \min_{\substack{\pi_{S'Y}:\\  |\pi_{S'Y}-P_{SY}|<\mu}} D(\pi_{S'Y}\| P_{S'}Q_{Y}) + \min_{\substack{\pi_{TW'V}: \\
|\pi_{TW'V}-P_{TWV}|<\mu}} D(\pi_{TW'V}\| P_{TW'}\Gamma_{V|W=T})-R-R'-\mu,\label{b4}
\end{align}
and notice that 
\begin{IEEEeqnarray}{rCl}
\tilde{\theta}_{\mu}^{\text{miss}}
& \stackrel{(\textnormal{eq.}\eqref{nois2}\&\eqref{nois0})}{=} &  \min_{\substack{\pi_{S'Y}:\\  |\pi_{S'Y}-P_{SY}|<\mu}} D(\pi_{S'Y}\| P_{S'}Q_{Y}) + \min_{\substack{\pi_{TW'V}: \\
|\pi_{TW'V}-P_{TWV}|<\mu}} D(\pi_{TW'V}\| P_{TW'}\Gamma_{V|W=T})-I(S;X)-2\mu \nonumber \\
&\stackrel{(f)}{=} &  D(P_{SY}\| P_{S}Q_{Y}) +  D(P_{TWV}\| P_{TW}\Gamma_{V|W=T})-I(S;X)-\delta_4(\mu) \nonumber \\
&\stackrel{(\textnormal{CR})}{=}&D(P_{ Y}\| Q_{Y})+I (S;Y) + D(P_{ TWV }\| P_{TW}\Gamma_{V|W=T})-I(S;X)-\delta_4(\mu)\nonumber\\
&:= &\theta_{\mu}^{\text{miss}}-\delta_4(\mu), \label{t4final}
\end{IEEEeqnarray}
for some function $\delta_4(\mu)$ that $\to 0$ as $\mu \to 0$. Here, $(f)$ holds because $\pi_{TW'V}=P_{TWV}$, $\pi_Y=P_Y$, and $\pi_{S'|Y}=P_{S|Y}$. By \eqref{b4event}--\eqref{t4final}, we have
\begin{align}
 \Pr \big[\mathcal{B}_4 \big| \mathcal{H}=1 \big] \leq 2^{-n \big( \theta_{\mu}^{\text{miss}}-\delta_4(\mu)\big)}.\label{mis-f}
\end{align}

		Combining \eqref{probfirst} with \eqref{stan-f}, \eqref{dec-ff}, \eqref{dec-ff} and \eqref{mis-f},
		proves that for sufficiently large blocklengths $n$, the average type-II error probability satisfies	
		\begin{align}
		\mathbb{E}_{\mathcal{C}}[\beta_n] \leq 4 \cdot \max\left\{  2^{-n \theta_{\mu}^{\text{standard}}}, \ 2^{-n \theta_{\mu}^{\text{dec}}}, \  2^{-n \theta_{\mu}^{\text{miss}}} \right\}.\label{averagec}
		\end{align}
		By standard arguments and successively eliminating the worst half of the codewords with respect to $\alpha_n$ and  the exponents $\theta_{\mu}^{\text{standard}}$, $\theta_{\mu}^{\text{dec}}$, and $\theta_{\mu}^{\text{miss}}$, it can be shown that there exists at least one codebook for which 
			\begin{IEEEeqnarray}{rCl}
			\alpha_n & \leq & \epsilon, \\
			\beta_n & \leq &64 \cdot \max\left\{  2^{-n \theta_{\mu}^{\text{standard}}}, \ 2^{-n \theta_{\mu}^{\text{dec}}}, \  2^{-n \theta_{\mu}^{\text{miss}}}  \right\}.
			\end{IEEEeqnarray}
		Letting $\mu\to 0$ and $n\to \infty$, we get $\theta_{\mu}^{\text{standard}}\to \theta^{\text{standard}}$, $\theta_{\mu}^{\text{dec}}\to \theta^{\text{dec}}$, $\theta_{\mu}^{\text{miss}}\to \theta^{\text{miss}}$.
		
		This proves the theorem for $I(S;X)\geq I(W;V|T)$. 
		When $I(S;X)<I(W;V|T)$, rates $R$ and $R'$ are chosen as in \eqref{nois0a} and \eqref{nois1a}. The analysis is similar to above, but since $R'=0$, event $\mathcal{B}_3$ can be omitted.

\section{Proof of Corollary~\ref{cor:extended_conditional}}\label{app:proof_of_corollary}
Let $f$ be a function satisfying the properties in the corollary.  In this case, $Q_{X|Y}=P_{X|f(Y)}$ and for the choice
\begin{equation}\label{eq:choice}
\pi_{SXY}=  P_{S|X} P_{X|f(Y)} P_Y
\end{equation}
the expectation in \eqref{eq:zero_condition} evaluates to 0. The proposed choice  in \eqref{eq:choice} Ωis a valid candidate for the minimization and in particular it satisfies the condition $H(S|Y) \leq H_{\pi}(S|Y)$. This can be seen by introducing the enhanced type
\begin{equation}
\pi_{SXYY'} =  \pi_{SXY}\cdot \pi_{Y'|Xf(Y)},
\end{equation}
with $\pi_{SXY}$ as chosen in \eqref{eq:choice} and $ \pi_{Y'|Xf(Y)}=P_{Y|Xf(Y)}$. Notice that under $\pi$ we have $f(Y)=f(Y')$ with probability $1$, and thus
\begin{IEEEeqnarray}{rCl}
H(S|Y) = H_\pi(S|Y') =   H_\pi(S|Y', f(Y)) = H_\pi(S|Y',f(Y),Y) \leq  H_\pi(S|Y).
\end{IEEEeqnarray}
We can thus conclude that we are in the case treated in  Remark~\ref{rem:no_red}.

We continue to evaluate the right-hand side of \eqref{eq:d2}. Let $P_{S|X}$ satisfy the stronger condition $I(S;X|f(Y)) \leq C$. Then, 
\begin{IEEEeqnarray}{rCl}
	\theta^{\textnormal{dec}}& \geq& D(P_Y\|Q_Y) + I(S;X|f(Y)) - I(S;X|Y) \nonumber \\
	& = &	D(P_Y\|Q_Y) + I(S;Y|f(Y)),
\end{IEEEeqnarray}
where the second inequality holds by the Markov chain $S \to X \to (Y,f(Y))$. Now, 
notice  that under the conditions of the corollary:
\begin{align}
		\sum_{x\in\mathcal{X}} 	P_{S|X}(s,x) Q_{XY}(x,y) 
	&=   \sum_{x\in\mathcal{X}}  \left(P_{S|X}(s|x) P_{X|f(Y)}(x|f(y))  \right)\cdot Q_Y(y) \nonumber \\[1ex]
	&= P_{S|f(Y)}(s|f(y))\cdot  Q_Y(y)
\end{align}
and thus by \eqref{eq:standard}:
\begin{IEEEeqnarray}{rCl}
	\theta^{\textnormal{standard}} & \geq & D(P_Y \| Q_Y) +  \min_{\substack{\pi_{SY}:\\ \pi_{SY}=P_{SY}}} D(\pi_{S|Y}\| P_{S|f(Y)}) \nonumber \\
	& = & D(P_Y\|Q_Y) + I(S;Y |f(Y)).
\end{IEEEeqnarray}

We now prove the converse direction. Defining $Z_i:=f(Y_i)$ and $\delta(\epsilon):=H(\epsilon)/n/(1-\epsilon)$ and following standard arguments \cite{Kim}, we obtain
\begin{IEEEeqnarray}{rCl}
	\theta &\leq & \frac{1}{(1-\epsilon)n} 	 D( P_{V^n Y^n | \mathcal{H}=0} \| P_{V^n Y^n | \mathcal{H}=1})+\delta(\epsilon)	\nonumber \\
	& \leq & \frac{1}{(1-\epsilon)n} 	 D( P_{V^n Y^n | \mathcal{H}=0} \| P_{V^n Y^n | \mathcal{H}=1})+\delta(\epsilon)	\nonumber \\
	& = & \frac{1}{(1-\epsilon)n} \mathbb{E}_{P_{Y^n}}	\left[ D( P_{V^n | Y^n, \mathcal{H}=0} \| P_{V^n | Z^n,  \mathcal{H}=1})\right]+ \frac{1}{(1-\epsilon)} \cdot D( P_Y \| Q_Y) +\delta(\epsilon) \nonumber \\
	& \leq & \frac{1}{(1-\epsilon)n} I(V^n; Y^n |Z^n) + \frac{1}{(1-\epsilon)} \cdot D( P_Y \| Q_Y) +\delta(\epsilon)	\nonumber \\
	& = & \frac{1}{(1-\epsilon)n} \sum_{i=1}^n I(V^n, Y^{i-1},  Z^{i-1}, Z_{i+1}^n ; Y_i | Z_i)+ \frac{1}{(1-\epsilon)}  \cdot D( P_Y \| Q_Y) +\delta(\epsilon) \nonumber \\
	& \leq & \frac{1}{(1-\epsilon)n} \sum_{i=1}^n I(V^n, X^{i-1},  Z^{i-1}, Z_{i+1}^n ; Y_i | Z_i) + \frac{1}{(1-\epsilon)}  \cdot D( P_Y \| Q_Y)+\delta(\epsilon) 	\nonumber \\	
	& \leq  & \frac{1}{(1-\epsilon)}\big( I(S; Y|f(Y)) + D( P_Y \| Q_Y)  \big)+\delta(\epsilon),
\end{IEEEeqnarray}
where the last inequality follows by introducing a time-sharing random variable $Q$ that is uniform over $\{1,\ldots, n\}$  and by defining $S:=(Q,V^n, X^{Q-1}, Z^{Q-1}, Z_{Q+1}^n)$ and $Y=Y_Q$.

We turn to the constraint on capacity:
\begin{IEEEeqnarray}{rCl}
	C & \geq & I(W^n;V^n) \nonumber \\
	& \geq & I(W^n;V^n|Z^n) \nonumber \\
	&  \geq & I(X^n; V^n|Z^n) 	\nonumber \\
	&  \geq &\sum_{i=1}^n I(X_i; V^n, X^{i-1}, Z^{i-1}, Z_{i+1}^n |Z_i ) 	\nonumber \\
	& \geq & I(X ; S| f(Y)),
\end{IEEEeqnarray}
where for the last inequality we defined $X=X_{Q}$.
The proof is established by noticing the Markov chain
\begin{equation}
S - X - Y.
\end{equation}

\section{Proof of Example~\ref{p2p-Gaussian}}\label{Gp2p-proof}
 We specialize Theorem~\ref{cor:extended_conditional} to the proposed Gaussian setup. Notice that $X$ and $Y$ are independent under $\mathcal{H}=1$. Moreover,  $Y$ (and $X$) has same marginal under both hypotheses. Therefore,  when applying Theorem~\ref{cor:extended_conditional}, the term   $D(P_Y \| Q_Y) =0$ and the function $f$ can be ignored.
 
  Let  now  $S=X+G$ with $G$  a zero-mean Gaussian  random variable of variance $\xi^2$ and  independent of $X$. For this choice:
\begin{align}
 I(S;Y) &= \frac{1}{2}\log \left(\frac{1}{1-\frac{\rho_0^2}{1+\xi^2}}\right),\label{g1}
 \end{align}
 and 
 \begin{align}
I(S;X) &= \frac{1}{2}\log \left( \frac{1+\xi^2}{\xi^2} \right).\label{g2}
\end{align}
Thus, by Theorem~\ref{cor:extended_conditional}, the optimal exponent for the presented Gaussian setup is lower bounded as:
\begin{align}
\theta^*& \geq \max_{\xi^2 \colon \frac{1}{2}\log \left( \frac{1+\xi^2}{\xi^2} \right)\leq C } \quad  \frac{1}{2}\log \left(\frac{1}{1-\frac{\rho_0^2}{1+\xi^2}}\right) 
\nonumber \\
& = 
\frac{1}{2}\log \left(\frac{1}{1-\rho_0^2+\rho_0^2\cdot 2^{-2C}}\right).\label{eq:last}
\end{align}
We now show that $\theta^*$ is also upper bounded by the right-hand side of \eqref{eq:last}. To this end, notice first that: 
\begin{align}
I(S;X)
=\frac{1}{2}\log(2\pi e)-h(X|S),
\end{align}
and thus constraint $C \geq I(S;X)$ is equivalent to:
\begin{align}
2^{2h(X|S)}\geq (2\pi e)\cdot 2^{-2C}.\label{gp2p-con}
\end{align}
Moreover,  (under $\mathcal{H}=0$) one can write $Y=\rho_0 X +F$, with $F$ zero-mean Gaussian of variance $1-\rho_0^2$ and independent of  $X$. This implies that for any $S$ forming the Markov chain $S - X - Y$, also the pair $(S,X)$ is independent of $F$. By the EPI and because $h(\rho_0 X)=\log|\rho_0| + h(X)$, we then have:
\begin{IEEEeqnarray}{rCl}
	h(Y|S)& \geq & \frac{1}{2} \log \left( 2 \pi e \left(\frac{1}{2\pi e} 2^{2h(\rho_0 X|S)} + (1- \rho_0^2)\right) \right) \nonumber \\
	&=  &\frac{1}{2} \log \left( 2 \pi e \left(\frac{\rho_0^2}{2\pi e} 2^{2h( X|S)} + (1- \rho_0^2)\right)\right).\label{eq:EPI}
	\end{IEEEeqnarray}
By Theorem~\ref{cor:extended_conditional}, the optimal error exponent is upper bounded as:
\begin{align}
\theta^*&= \max_{\substack{S:\\\text{s.t.}\; \eqref{gp2p-con}}} I(S;Y)\nonumber\\&=h(Y)-\min_{\substack{S:\\\text{s.t.}\; \eqref{gp2p-con}}}h(Y|S)\nonumber\\
&\stackrel{(a)}{\leq}\frac{1}{2}\log (2\pi e)-\min_{\substack{S:\\\text{s.t.}\; \eqref{gp2p-con}}}\frac{1}{2}\log \left( 2\pi e\left(\frac{\rho_0^2}{2\pi e}2^{2h( X|S)}+(1-\rho_0^2)\right)\right)\nonumber\\
&\stackrel{(b)}{\leq} \frac{1}{2}\log (2\pi e)-\frac{1}{2}\log \left( 2\pi e \left(\rho_0^2\cdot 2^{-2C}+(1-\rho_0^2)\right)\right)\nonumber\\
&=\frac{1}{2}\log \left(\frac{1}{1-\rho_0^2+\rho_0^2\cdot 2^{-2C}}\right),
	\end{align}
where $(a)$ holds by  \eqref{eq:EPI} and $(b)$ by \eqref{gp2p-con}. Combining this upper bound with the lower bound in \eqref{eq:last}, completes the proof.

\section{Proof of Theorem \ref{macthm}}\label{macsecproof}

The proof is based on the scheme of Section \ref{sec:MACcoding}. Fix a choice of the blocklength $n$, the small positive $\mu$,  the (conditional) pmfs $P_{T_1T_2}$, $P_{S_1|X_1T_1T_2}$ and $P_{S_2|X_2T_1T_2}$, and the functions $f_1$ and $f_2$ so that \eqref{eq:Iconditions} holds.  Define the set $\mathcal{P}_{\mu,\text{type-I}}^n$ to be the subset of types $\pi_{S_1S'_1S_2S'_2VYT_1T_2}\in\mathcal{P}^n$ such that  for all $(s_1,s'_1,s_2,s'_2,v,y,t_1,t_2)\in\mathcal{S}_1\times \mathcal{S}_1\times \mathcal{S}_2\times \mathcal{S}_2\times \mathcal{V}\times \mathcal{Y}\times \mathcal{W}_1\times \mathcal{W}_2$
\begin{subequations}
	\begin{align}
	|\pi_{S_iX_iT_1T_2}(s_i,x_i,t_1,t_2) - P_{S_iX_iT_1T_2}(s_i,x_i,t_1,t_2)| &\leq \mu/2,\qquad\qquad\;\;\;\;\;\;\;\;i\in\{1,2\},\label{eq:types5mac-typ1}\\
	|\pi_{S_1S_2YVT_1T_2}(s_1,s_2,y,v,t_1,t_2)-P_{S_1S_2YVT_1T_2}(s_1,s_2,y,v,t_1,t_2)| &\leq \mu,\label{eq:types2mac-typ1}\\
	H_{\pi_{S'_1S'_2YVT_1T_2}}(S_1',S_2'|Y,V,T_1,T_2) &\leq H_{\pi_{S_1S_2YVT_1T_2}}(S_1,S_2|Y,V,T_1,T_2),\label{eq:types4mac-typ1}\\
	|\pi_{S'_1S'_2T_1T_2}(s'_1,s'_2,t_1,t_2)-P_{S_1S_2T_1T_2}(s'_1,s'_2,t_1,t_2)| &\leq \mu,
	\end{align}
\end{subequations}
Also, set for convenience of notation:
\begin{align}
P_{S_1'|T_1T_2}(s_1|t_1,t_2) &= P_{S_1|T_1T_2}(s_1|t_1,t_2),\quad \forall (s_1,t_1,t_2)\in \mathcal{S}_1\times  \mathcal{T}_1\times \mathcal{T}_2,\\
P_{S_2'|T_1T_2}(s_2|t_1,t_2) &= P_{S_2|T_1T_2}(s_2|t_1,t_2),\quad \forall (s_2,t_1,t_2)\in \mathcal{S}_2\times \mathcal{T}_1\times \mathcal{T}_2.
\end{align}
In the following, for simplicity of presentation, we abbreviate the pair $(T_1^n,T_2^n)$ by $\mathbf{T}^n$ and its realization  $(t_1^n,t_2^n)$ by $\mathbf{t}^n$.

We first analyze the type-I error probability averaged over the random code construction.  Let $(M_1,M_2)$ be the indices of the chosen codewords at the transmitters, if they exist, and define the following events:
\begin{align}
\mathcal{E}_{\text{Tx}_i}&\colon \left\{ \nexists\; m_i \colon (S_i^n(m_i),X_i^n,\mathbf{T}^n)\in\mathcal{T}_{\mu/2}^n(P_{S_iX_i\mathbf{T}})\right\},\qquad\qquad\qquad i\in\{1,2\},\\
\mathcal{E}_{\text{Rx}}^{(1)}&\colon \left\{  (S_1^n(M_1),S_2^n(M_2),Y^n,V^n,\mathbf{T}^n)\notin\mathcal{T}_{\mu}^n(P_{S_1S_2YV\mathbf{T}}) \right\},\\ 
\mathcal{E}_{\text{Rx}}^{(2)}&\colon  \Big\{ \exists\; m'_1\neq M_1, m'_2\neq M_2\colon  H_{\text{tp}(s_1^n(m'_1),s_2^n(m'_2),y^n,v^n,\mathbf{t}^n)}(S_1,S_2|Y,V,\mathbf{T})=\nonumber\\&\qquad\qquad\qquad\qquad\qquad\qquad\qquad\qquad\qquad\qquad\qquad\qquad\qquad \min_{\tilde{m}_1,\tilde{m}_2 } H_{\text{tp}(s_1^n(\tilde{m}_1),s_2^n(\tilde{m}_2),y^n,v^n,\mathbf{t}^n)}(S_1,S_2|Y,V,\mathbf{T})\Big\},\\
\mathcal{E}_{\text{Rx}}^{(3)}&\colon \left\{\exists \; m'_2\neq M_2\colon  H_{\text{tp}(s_1^n(M_1),s_2^n(m'_2),y^n,v^n,\mathbf{t}^n)}(S_1,S_2|Y,V,\mathbf{T})= \min_{\tilde{m}_2 } H_{\text{tp}(s_1^n(M_1),s_2^n(\tilde{m}_2),y^n,v^n,\mathbf{t}^n)}(S_1,S_2|Y,V,\mathbf{T})\right\},\\
\mathcal{E}_{\text{Rx}}^{(4)}&\colon \left\{\exists \; m'_1\neq M_1\colon  H_{\text{tp}(s_1^n(m'_1),s_2^n(M_2),y^n,v^n,\mathbf{t}^n)}(S_1,S_2|Y,V,\mathbf{T})= \min_{\tilde{m}_1 } H_{\text{tp}(s_1^n(\tilde{m}_1),s_2^n(M_2),y^n,v^n,\mathbf{t}^n)}(S_1,S_2|Y,V,\mathbf{T})\right\}.
\end{align}
Notice that the event $\Big(\mathcal{E}_{\text{Tx}_1} \cup \mathcal{E}_{\text{Tx}_2} \cup \mathcal{E}_{\text{Rx}}^{(1)} \cup \mathcal{E}_{\text{Rx}}^{(2)}\cup \mathcal{E}_{\text{Rx}}^{(3)} \cup \mathcal{E}_{\text{Rx}}^{(4)}\Big)^c$ implies that the receiver decides on $\hat{\mathcal{H}}=0$. Thus, we obtain
\begin{align}
\mathbb{E}_{\mathcal{C}}[\alpha_n ]
&\leq \Pr\left[\mathcal{E}_{\text{Tx}_1}\right]+\Pr\left[\mathcal{E}_{\text{Tx}_2}\right]+\Pr\left[\mathcal{E}_{\text{Rx}}^{(1)}|\mathcal{E}_{\text{Tx}_1}^{c},\mathcal{E}_{\text{Tx}_2}^{c}\right] +\Pr\left[\mathcal{E}_{\text{Rx}}^{(2)}|\mathcal{E}_{\text{Tx}_1}^{c},\mathcal{E}_{\text{Tx}_2}^{c},\mathcal{E}_{\text{Rx}}^{(1)c}\right]+\Pr\left[\mathcal{E}_{\text{Rx}}^{(3)}|\mathcal{E}_{\text{Tx}_1}^{c},\mathcal{E}_{\text{Tx}_2}^{c},\mathcal{E}_{\text{Rx}}^{(1)c}\right]\nonumber\\&\;\;\;+\Pr\left[\mathcal{E}_{\text{Rx}}^{(4)}|\mathcal{E}_{\text{Tx}_1}^{c},\mathcal{E}_{\text{Tx}_2}^{c},\mathcal{E}_{\text{Rx}}^{(1)c}\right]\label{eq:dmac}\\
&\leq  \epsilon/6 + \epsilon/6 + \epsilon/6 +  \epsilon/6 + \epsilon/6 + \epsilon/6 \nonumber\\&=\epsilon,
\end{align}
where the second inequality holds  for all sufficiently small values of $\mu$ and sufficiently large blocklengths $n$ and can be proved as follows. The first and second summands of \eqref{eq:dmac} can be upper bounded by means of the covering lemma \cite{ElGamal} and the rate constraint \eqref{nois1mac};  the third by means of the Markov lemma \cite{ElGamal}. To prove the upper bound on the fourth term,  consider the following set of inequalities
\begin{align}
&\Pr\left[\mathcal{E}_{\text{Rx}}^{(2)} \Big|\mathcal{E}_{\text{Tx}_1}^{c},\mathcal{E}_{\text{Tx}_2}^{c},\mathcal{E}_{\text{Rx}}^{(1)c}, \;\mathcal{H}=0\right] 
\nonumber\\&=\Pr \bigg[H_{\text{tp}(s_1^n(M_1),s_2^n(M_2),y^n,v^n,\mathbf{t}^n)}(S_1',S_2'|Y,V,\mathbf{T})\geq \min_{\substack{\tilde{m}_1\neq M_1\\\tilde{m}_2\neq M_2} } H_{\text{tp}(s_1^n(\tilde{m}_1),s_2^n(\tilde{m}_2),y^n,v^n,\mathbf{t}^n)}(S_1,S_2|Y,V,\mathbf{T})\bigg| \nonumber\\&\qquad \;\;\; \qquad (S_i^n(M_i),X_i^n,\mathbf{T}^n)\in\mathcal{T}_{\mu/2}^n(P_{S_iX_i\mathbf{T}}),\; i\in\{1,2\},\;\;  (S_1^n(M_1),S_2^n(M_2),Y^n,V^n,\mathbf{T}^n)\in\mathcal{T}_{\mu}^n(P_{S_1S_2YV\mathbf{T}}), \;\; \mathcal{H}=0\bigg]\nonumber\\[2ex]
&\stackrel{(a)}{=}\Pr \bigg[H_{\text{tp}(s_1^n(1),s_2^n(1),y^n,v^n,\mathbf{t}^n)}(S_1',S_2'|Y,V,\mathbf{T})\geq \min_{\substack{\tilde{m}_1\neq 1\\\tilde{m}_2\neq 1} } H_{\text{tp}(s_1^n(\tilde{m}_1),s_2^n(\tilde{m}_2)|y^n,v^n,\mathbf{t}^n)}(S_1,S_2|Y,V,\mathbf{T})\bigg| \nonumber\\&\hspace{6cm} (S_i^n(1),X_i^n,\mathbf{T}^n)\in\mathcal{T}_{\mu/2}^n(P_{S_iX_i\mathbf{T}}),\; i\in\{1,2\}, \;\; \nonumber \\
&\hspace{6cm} (S_1^n(1),S_2^n(1),Y^n,V^n,\mathbf{T}^n)\in\mathcal{T}_{\mu}^n(P_{S_1S_2YV\mathbf{T}}), \; M_1=M_2=1, \; \;\; \mathcal{H}=0\bigg]\nonumber\\[2ex]
&=\Pr \bigg[\bigcup_{\substack{\tilde{m}_1\neq 1\\\tilde{m}_2\neq 1}} \Big\{ H_{\text{tp}(s_1^n(1),s_2^n(1),y^n,v^n,\mathbf{t}^n)}(S_1',S_2'|Y,V,\mathbf{T}) \geq H_{\text{tp}(s_1^n(\tilde{m}_1),s_2^n(\tilde{m}_2),y^n,v^n,\mathbf{t}^n)}(S_1,S_2|Y,V,\mathbf{T}) \Big\} \bigg| \nonumber\\&\hspace{6cm} (S_i^n(1),X_i^n,\mathbf{T}^n)\in\mathcal{T}_{\mu/2}^n(P_{S_iX_i\mathbf{T}}),\; i\in\{1,2\},\nonumber\\&\hspace{6cm} (S_1^n(1),S_2^n(1),Y^n,V^n,\mathbf{T}^n)\in\mathcal{T}_{\mu}^n(P_{S_1S_2YV\mathbf{T}}), \; M_1=M_2=1, \; \mathcal{H}=0\bigg]\nonumber\\[2ex]
&\stackrel{(b)}{\leq} \sum_{\substack{\pi_{S_1S'_1S_2S'_2VY\mathbf{T}}\in \mathcal{P}_{\mu,\text{type-I}}^n}}\;\;\sum_{\tilde{m}_1=2}^{2^{nR_1}}\;\;\sum_{\tilde{m}_2=2}^{2^{nR_2}}\sum_{\substack{s_1^n,s'^{n}_1,s_2^n,s'^{n}_2,v^n,y^n,\mathbf{t}^n:\\\text{tp}(s_1^n,s'^{n}_1,s_2^n,s'^{n}_2,v^n,y^n,\mathbf{t}^n)=\pi_{S_1S'_1S_2S'_2VY\mathbf{T}}}}\nonumber\\&\qquad  \Pr \bigg[S_1^n(1)=s_1^n,S_2^n(1)=s_2^n,V^n=v^n,Y^n=y^n,\mathbf{T}^n=\mathbf{t}^n \bigg| (S_i^n(1),X_i^n,\mathbf{T}^n)\in\mathcal{T}_{\mu/2}^n(P_{S_iX_i\mathbf{T}}), i\in\{1,2\},\nonumber\\&\hspace{7cm} (S_1^n(1),S_2^n(1),Y^n,V^n,\mathbf{T}^n)\in\mathcal{T}_{\mu}^n(P_{S_1S_2YV\mathbf{T}}), M_1=M_2=1, \; \mathcal{H}=0\bigg]\nonumber\\&\qquad\qquad\cdot \Pr\big[S_1^n(\tilde{m}_1)=s'^{n}_1|\mathbf{T}^n=\mathbf{t}^n,M_1=M_2=1, \; \mathcal{H}=0\big] \cdot \Pr\big[S_2^n(\tilde{m}_2)=s'^{n}_2|\mathbf{T}^n=\mathbf{t}^n,M_1=M_2=1,\; \mathcal{H}=0\big]\nonumber\\[3ex]
&\stackrel{(c)}{\leq} \sum_{\substack{\pi_{S_1S'_1S_2S'_2VY\mathbf{T}}\in \mathcal{P}_{\mu,\text{type-I}}^n}}\;\;2^{nR_1}\cdot 2^{nR_2}\cdot 2^{nH_{\pi}(S_1,S'_1,S_2,S'_2,V,Y,\mathbf{T})} \cdot 2^{-nH_{\pi}(S_1,S_2,V,Y,\mathbf{T})}\cdot 2^{-nH_{\pi}(S'_1|\mathbf{T})}\cdot 2^{-nH_{\pi}(S'_2|\mathbf{T})}\nonumber\\[3ex]
&=\sum_{\substack{\pi_{S_1S'_1S_2S'_2VY\mathbf{T}}\in \mathcal{P}_{\mu,\text{type-I}}^n}}\!\!\!\!\! 2^{n(R_1+R_2-I_{\pi}(S'_1,S'_2;S_1,S_2,V,Y|\mathbf{T})-I_{\pi}(S'_1;S'_2|\mathbf{T}))}\nonumber\\[3ex]
&\leq\sum_{\substack{\pi_{S_1S'_1S_2S'_2VY\mathbf{T}}\in \mathcal{P}_{\mu,\text{type-I}}^n}} 2^{n(R_1+R_2-I_{\pi}(S'_1,S'_2;V,Y|\mathbf{T})-I_{\pi}(S'_1;S'_2|\mathbf{T}))}\nonumber\\[3ex]
&\stackrel{(d)}{\leq} (n+1)^{|\mathcal{S}_1|^2.|\mathcal{S}_2|^2.|\mathcal{V}|.|\mathcal{Y}|}\cdot \max_{\substack{\pi_{S_1S'_1S_2S'_2VY\mathbf{T}}\in \mathcal{P}_{\mu,\text{type-I}}^n}} \;\; 2^{n(R_1+R_2-I_{\pi}(S_1,S_2;V,Y|\mathbf{T})-I(S_1;S_2|\mathbf{T})+\delta(\mu))}\nonumber\\[1ex]
&\stackrel{(e)}{\leq}\epsilon/6,
\end{align}
where 
\begin{itemize}
	\item $(a)$ holds by the symmetry of the code construction and the encoding;
	\item $(b)$ holds by the union bound and  because conditioned on $\mathbf{T}^n$ and $M_1=M_2=1$, the sequences $S_1^n(\tilde{m}_1)$ and $S_2^n(\tilde{m}_2)$ are generated independently of each other and of all other sequences;
	\item $(c)$ holds because all $2^{nH_{\pi}(S_1,S_2,V,Y,\mathbf{T})}$ tuples $(s_1^n,s_2^n,v^n,y^n,\mathbf{t}^n)$ of the same type $\pi$ have same conditional probability and similarly all $2^{nH_{\pi}(S'_i|\mathbf{T})}$ sequences $s_i^n$, for $ i\in\{1,2\}$, of same joint type with $\mathbf{t}^n$ have same conditional probability;
	\item $(d)$ holds because for all $\pi$ in $\mathcal{P}_{\mu,\text{type-I}}^n$, $H_{\pi}(S'_1,S'_2|V,Y,\mathbf{T})\leq H_{\pi}(S_1,S_2|V,Y,\mathbf{T})$ and $|\pi_{S'_1S'_2T}-P_{S_1S_2T}|\leq\mu$ and $|\pi_{S_1S_2VYT}-P_{S_1S_2VYT}|\leq \mu$; and
	\item $(e)$ holds by the rate constraint in \eqref{nois4mac}.
\end{itemize}
That also the fifth and sixth summands of \eqref{eq:dmac} are upper bounded by $\epsilon/6$, can be shown in a similar way. 
\medskip

Next, we analyze the type-II error probability averaged over the random code construction. Define   events:
\begin{IEEEeqnarray}{rCl}
\mathcal{E}_{\text{Tx}_i}(m_i) &\colon &\big\{\big(S_i^n(m_i),X_i^n,T_1^n,T_2^n\big) \in \mathcal{T}_{\mu/2}^n(P_{S_iX_iT_1T_2})\;\; \textnormal{ and } \;\; W_i^n=f_i(S_i^n(m_i),X_i^n)\; \text{is sent}\big\}\\
 \mathcal{E}_{\text{Rx}}(m'_1,m'_2) &\colon& \Big\{ \big( S_1^n(m'_1),S_2^n(m'_2),Y^n,T_1^n,T_2^n,V^n \big)\in \mathcal{T}_{\mu}^n(P_{S_1S_2YT_1T_2V}) \hspace{0.2cm}\textnormal{ and}\nonumber\\\;\;\; &&\hspace{0.7cm} H_{\text{tp}(S_1^n(m'_1),S_2^n(m'_2),Y^n,T_1^n,T_2^n,V^n)}(S_1',S_2'|Y,T_1,T_2,V)  \nonumber \\ && \hspace{5cm}= \min_{\substack{\tilde{m}_1,\tilde{m}_2}}   H_{\text{tp}(S_1^n(\tilde{m}_1),S_2^n(\tilde{m}_2),Y^n,T_1^n,T_2^n,V^n)}(S_1,S_2|Y,T_1,T_2,V)\Big\},
\end{IEEEeqnarray}
and notice that
\begin{IEEEeqnarray}{rCl}
	\mathbb{E}_{\mathcal{C}}[\beta_n] &= &\Pr\left[\hat{\mathcal{H}}=0|\mathcal{H}=1\right] \leq\Pr \left[ \displaystyle \bigcup_{m'_1,m'_2}\mathcal{E}_{\text{Rx}}(m'_1,m'_2)\Big|\mathcal{H}=1 \right],\nonumber
\end{IEEEeqnarray}
where the union is over indices $(m_1',m_2')\in \{1,\ldots, \big \lfloor 2^{nR_1} \big \rfloor \}\times \{1,\ldots, \big \lfloor 2^{nR_2} \big \rfloor \}$. Notice further that the above probability is upper bounded by the sum of the probabilities of the following nine events:

\begin{enumerate}
	\item[$\mathcal{B}_1$:] $\big\{ \exists (m_1,m_2)\;\;$    s.t. $\;\;\big(\mathcal{E}_{\text{Tx}_1}(m_1)\quad \textnormal{and}\quad \mathcal{E}_{\text{Tx}_2}(m_2) \quad\textnormal{and} \quad\mathcal{E}_{\text{Rx}}(m_1,m_2)\big)\big\}$\\
	\item[$\mathcal{B}_2$:] $\big\{\exists (m_1,m_1',m_2)\;\;$ with $\;\;m_1\neq m_1'\;\;$ s.t. $\;\;\big(\mathcal{E}_{\text{Tx}_1}(m_1)\quad \textnormal{and}\quad \mathcal{E}_{\text{Tx}_2}(m_2) \quad\textnormal{and} \quad\mathcal{E}_{\text{Rx}}(m_1',m_2)\big)\big\}$\\
	\item[$\mathcal{B}_3$:] $\big\{\exists (m_1,m_2,m_2')\;\;$ with $\;\;m_2\neq m_2'\;\;$ s.t. $\;\;\big(\mathcal{E}_{\text{Tx}_1}(m_1)\quad \textnormal{and}\quad \mathcal{E}_{\text{Tx}_2}(m_2) \quad\textnormal{and} \quad\mathcal{E}_{\text{Rx}}(m_1,m_2')\big)\big\}$\\
	\item[$\mathcal{B}_4$:] $\big\{\exists (m_1,m_1',m_2,m_2')\;\;$ with $\;\;m_1\neq m_1'\;\;$ and  $m_2\neq m_2'\;\;$ s.t. $\;\;\big(\mathcal{E}_{\text{Tx}_1}(m_1)\quad \textnormal{and}\quad \mathcal{E}_{\text{Tx}_2}(m_2) \quad\textnormal{and} \quad\mathcal{E}_{\text{Rx}}(m_1',m_2')\big)\big\}$\\
	\item[$\mathcal{B}_5$:] $\big\{\forall m_1\quad \mathcal{E}_{\text{Tx}_1}^c(m_1)\quad \text{holds}  \quad \textnormal{and}\quad \exists (m_1',m_2,m_2')\;\;$ with $\;\;m_2\neq m_2' \;\;\;$ s.t. $\;\;\;\;\mathcal{E}_{\text{Tx}_2}(m_2) \quad\textnormal{and} \qquad\mathcal{E}_{\text{Rx}}(m_1',m_2')\big\}$\\
	\item[$\mathcal{B}_6$:] $\big\{\forall m_1\quad \mathcal{E}_{\text{Tx}_1}^c(m_1)\quad \text{holds} \big\} \; \cup \; \big\{ \exists (m_1',m_2)\;\;$ s.t. $\;\;\;\mathcal{E}_{\text{Tx}_2}(m_2) \quad\textnormal{and} \quad\mathcal{E}_{\text{Rx}}(m_1',m_2)\big\}$\\
	\item[$\mathcal{B}_7$:] $\big\{\forall m_2\quad \mathcal{E}_{\text{Tx}_2}^c(m_2)\quad \text{holds}  \big\} \; \cup \; \big\{ \exists (m_1,m_1',m_2')\;\;$ with $\;\;m_1\neq m_1'\;\;$ s.t. $ \;\;\big(\mathcal{E}_{\text{Tx}_1}(m_1) \quad \textnormal{and}\quad \mathcal{E}_{\text{Rx}}(m_1',m_2')\big)\big\}$\\
	\item[$\mathcal{B}_8$:] $\big\{\forall m_2\quad \mathcal{E}_{\text{Tx}_2}^c(m_2)\quad \text{holds}  \big\} \; \cup \big\{\exists (m_1,m_2')\;\;$ s.t. $\;\;\big(\mathcal{E}_{\text{Tx}_1}(m_1) \quad \textnormal{and}\quad \mathcal{E}_{\text{Rx}}(m_1,m_2')\big)\big\}$\\
	\item[$\mathcal{B}_9$:] $\big\{\forall (m_1,m_2)\quad \big(\mathcal{E}_{\text{Tx}_1}^c(m_1)\; \cup \; \mathcal{E}_{\text{Tx}_2}^c(m_2) \big)\quad \text{hold} \big\} \;  \cup \;  \big\{ \exists (m_1',m_2')\;\;$ s.t. $\;\;\mathcal{E}_{\text{Rx}}(m_1',m_2')\big\}$
\end{enumerate}So, we have
\begin{equation}
\mathbb{E}_{\mathcal{C}}[\beta_n] \leq \sum_{\ell=1}^9 \Pr\big[ \mathcal{B}_\ell \big| \mathcal{H}=1\big]. \label{mac-events} \end{equation}
We will bound the nine probabilities on the right-hand side of \eqref{mac-events} individually. To simplify the notation, we introduce the following set of types:
\begin{IEEEeqnarray}{rCl}
\mathcal{P}_{\mu,\text{standard}} &:=& \{\pi_{S_1S_2X_1X_2YT_1T_2V}\colon \;\; |\pi_{S_iX_iT_1T_2}-P_{S_iX_iT_1T_2}|<\mu/2, \;\; i\in\{1,2\},\; |\pi_{S_1S_2YT_1T_2V}-P_{S_1S_2YT_1T_2V}|< \mu\},\nonumber\\\\
\mathcal{P}_{\mu,\text{dec,1}} &:=& \big\{\pi_{S_1S'_1S_2X_1X_2YT_1T_2V}\colon  |\pi_{S_iX_iT_1T_2}-P_{S_iX_iT_1T_2}|<\mu/2,\;\;i\in \{1,2\}, \;\;  |\pi_{S'_1S_2YT_1T_2V}-P_{S_1S_2YT_1T_2V}|< \mu,\nonumber\\&&\hspace{6.5cm}\qquad H_{\pi}(S'_1|S_2,Y,T_1,T_2,V)\leq H_{\pi}(S_1|S_2,Y,T_1,T_2,V)\big\},\\[1ex]
\mathcal{P}_{\mu,\text{dec,2}} &:=& \big\{\pi_{S_1S_2S'_2X_1X_2YT_1T_2V}\colon  |\pi_{S_iX_iT_1T_2}-P_{S_iX_iT_1T_2}|<\mu/2,\;\;i\in \{1,2\}, \;\; |\pi_{S_1S'_2YT_1T_2V}-P_{S_1S_2YT_1T_2V}|< \mu,\nonumber\\&&\hspace{6.5cm}  \qquad H_{\pi}(S'_2|S_1,Y,T_1,T_2,V)\leq H_{\pi}(S_2|S_1,Y,T_1,T_2,V)\big\},\\[1ex]
\mathcal{P}_{\mu,\text{dec,12}} &:=& \big\{\pi_{S_1S'_1S_2S'_2X_1X_2YT_1T_2V}\colon  |\pi_{S_iX_iT_1T_2}-P_{S_iX_iT_1T_2}|<\mu/2,\;\;i\in \{1,2\}, \;\; |\pi_{S'_1S'_2YT_1T_2V}-P_{S_1S_2YT_1T_2V}|< \mu, \nonumber\\&&\hspace{5.8cm}  \qquad\qquad H_{\pi}(S'_1,S'_2|Y,T_1,T_2,V)\leq H_{\pi}(S_1,S_2|Y,T_1,T_2,V)\big\},\\[1ex]
\mathcal{P}_{\mu,\text{miss,1a}} &:=& \big\{\pi_{S'_1S_2S'_2X_2YT_1T_2V}\colon  |\pi_{S_2X_2T_1T_2}-P_{S_2X_2T_1T_2}|<\mu/2, \;\; |\pi_{S'_1S'_2YT_1T_2V}-P_{S_1S_2YT_1T_2V}|< \mu,\nonumber\\&&\hspace{6.cm}  \qquad\qquad H(S'_1,S'_2|Y,T_1,T_2,V)\leq H_{\pi}(S'_1,S_2|Y,T_1,T_2,V)\big\},\\[1ex]
\mathcal{P}_{\mu,\text{miss,1b}} &:=& \big\{\pi_{S'_1S_2X_2YT_1T_2V}\colon  |\pi_{S_2X_2T_1T_2}-P_{S_2X_2T_1T_2}|<\mu/2, \;\; |\pi_{S'_1S_2YT_1T_2V}-P_{S_1S_2YT_1T_2V}|< \mu\big\},\\[1ex]
\mathcal{P}_{\mu,\text{miss,2a}} &:=& \big\{\pi_{S_1S'_1S'_2X_1YT_1T_2V}\colon  |\pi_{S_1X_1T_1T_2}-P_{S_1X_1T_1T_2}|<\mu/2, \;\; |\pi_{S'_1S'_2YT_1T_2V}-P_{S'_1S'_2YT_1T_2V}|< \mu,\nonumber\\&&\hspace{6.1cm}  \qquad\qquad H(S'_1,S'_2|Y,T_1,T_2,V)\leq H_{\pi}(S_1,S'_2|Y,T_1,T_2,V)\big\},\\[1ex]
\mathcal{P}_{\mu,\text{miss,1b}} &:=& \big\{\pi_{S_1S'_2X_1YT_1T_2V}\colon  |\pi_{S_1X_1T_1T_2}-P_{S_1X_1T_1T_2}|<\mu/2, \;\; |\pi_{S_1S'_2YT_1T_2V}-P_{S_1S_2YT_1T_2V}|< \mu\big\}.
	\end{IEEEeqnarray}
Consider the probability of event $\mathcal{B}_1$.
By Sanov's theorem \cite{Cover} and the way the source sequences and the codebooks are generated, we have
\begin{IEEEeqnarray}{rCl}
\Pr\big[ \mathcal{B}_1\big| \mathcal{H}=1\big]& \leq & \sum_{m_1,m_2} 
\Pr \Big[ (S_i^n(m_i),X_i^n,T_1^n,T_2^n)\in \mathcal{T}_{\mu/2}^n(P_{S_iX_iT_1T_2}) \; \textnormal{and}\;W_i^n=f_i(S_i^n(m_i),X_i^n)\; \text{is sent for } i\in\{1,2\}, \nonumber\\[-2ex]&&\hspace{4.5cm} \textnormal{and} \quad (S_1^n(m_1),S_2^n(m_2),Y^n,T_1^n,T_2^n,V^n)\in\mathcal{T}_{\mu}^n(P_{S_1S_2YT_1T_2V})\Big|\mathcal{H}=1\Big]\nonumber\\[1ex]
&\leq & 2^{n(R_1+R_2)}\cdot \max_{\substack{\pi\in\mathcal{P}_{\mu,\text{standard}}}} 2^{-n(D(\pi_{S_1S_2X_1X_2YT_1T_2V}\|P_{S_1|T_1T_2}P_{S_2|T_1T_2}Q_{X_1X_2Y}P_{T_1T_2}\Gamma_{V|S_1S_2X_1X_2})-\mu)}.\label{b1event}
\end{IEEEeqnarray}
Define now:
\begin{align}
\tilde{\theta}_{\mu}^{\text{standard}} := \min_{\substack{\pi\in\mathcal{P}_{\mu,\text{standard}}}} D(\pi_{S_1S_2X_1X_2YT_1T_2V}\|P_{S_1|T_1T_2} P_{S_2|T_1T_2}Q_{X_1X_2Y}P_{T_1T_2}\Gamma_{V|S_1S_2X_1X_2})
	-R_1-R_2-\mu,
\end{align}
and observe that:
\begin{IEEEeqnarray}{rCl}
 \tilde{\theta}_{\mu}^{\text{standard}}
&\stackrel{(\text{eq.}\;\eqref{nois1mac})}{=}&\min_{\substack{\pi\in\mathcal{P}_{\mu,\text{standard}}}} D(\pi_{S_1S_2X_1X_2YT_1T_2V}\|P_{S_1|T_1T_2} P_{S_2|T_1T_2}Q_{X_1X_2Y}P_{T_1T_2}\Gamma_{V|S_1S_2X_1X_2})  \nonumber\\
&&\quad-I(S_1;X_1|T_1,T_2)-I(X_2;S_2|T_1,T_2)-3\mu\nonumber\\
&\stackrel{(a)}{=}&\min_{\substack{\pi\in\mathcal{P}_{\mu,\text{standard}}}} D(\pi_{S_1S_2X_1X_2YT_1T_2V}\|P_{S_1|X_1T_1T_2} P_{S_2|X_2T_1T_2}Q_{X_1X_2Y}P_{T_1T_2}\Gamma_{V|S_1S_2X_1X_2})-\delta_1(\mu)\nonumber\\
&:=& \theta_{\mu}^{\text{standard}}-\delta_1(\mu),\label{stanmac}
\end{IEEEeqnarray}
where $\delta_1(\mu)$ is a function that goes to zero as $\mu\to 0$. Here, $(a)$ follows by re-arranging  terms and by the continuity of KL-divergence.
Combining \eqref{b1event}--\eqref{stanmac}, we have:
\begin{align}
 \Pr\left[ \mathcal{B}_1\big| \mathcal{H}=1\right] \leq 2^{-n\left(\theta_{\mu}^{\text{standard}}-\delta_1(\mu)\right)}.
\end{align}

Consider next event $\mathcal{B}_2$. Its probability can be upper bounded as:
\begin{IEEEeqnarray}{rCl}
&&\Pr\big[ \mathcal{B}_2\big| \mathcal{H}=1\big]\nonumber\\&&\leq \sum_{m_1,m'_1,m_2}\Pr \Big[ \big( \mathcal{E}_{\text{Tx}_1}(m_1)\cap \mathcal{E}_{\text{Tx}_2}(m_2)\cap \mathcal{E}_{\text{Rx}}(m'_1,m_2) \big) \Big|\mathcal{H}=1\Big]\nonumber\\
&&\leq 2^{n(2R_1+R_2)} \cdot \nonumber\\&&\Pr \Big[ (S_i^n(m_i),X_i^n,T_1^n,T_2^n)\in \mathcal{T}_{\mu/2}^n(P_{S_iX_iT_1T_2}) \; \textnormal{ and }\;W_i^n=f_i(S_i^n(m_i),X_i^n)\; \text{is sent for }i\in\{1,2\},  \;\;\nonumber\\&&\hspace{0.7cm} (S_1^n(m'_1),S_2^n(m_2),Y^n,T_1^n,T_2^n,V^n)\in\mathcal{T}_{\mu}^n(P_{S_1S_2YT_1T_2V}),\nonumber\\&&\hspace{0.7cm} H_{\text{tp}(S_1^n(m'_1),S_2^n(m_2),Y^n,T_1^n,T_2^n,V^n)}(S'_1|S_2,Y,T_1,T_2,V)\leq H_{\text{tp}(S_1^n(m_1),S_2^n(m_2),Y^n,T_1^n,T_2^n,V^n)}(S_1|S_2,Y,T_1,T_2,V) \Big|\mathcal{H}=1\Big]\label{b2event}\nonumber \\[2ex]
&& = \sum_{m_1,m'_1,m_2}\;\; \sum_{\substack{\pi_{S_1S'_1S_2X_1X_2YT_1T_2V}:\\|\pi_{S_iX_iT_1T_2}-P_{S_iX_iT_1T_2}|<\mu/2,\\|\pi_{S'_1S_2YT_1T_2V}-P_{S_1S_2YT_1T_2V}|<\mu,\\H_{\pi}(S'_1|S_2,Y,T_1,T_2,V)\leq H_{\pi}(S_1|S_2,Y,T_1,T_2,V)}} \nonumber\\[1ex]&&\hspace{0.8cm}\Pr\left[W_i^n=f_i(S_i^n(m_i),X_i^n),  \;\; \text{tp}\left(S_1^n(m_1),S_1^n(m'_1),S_2^n(m_2),X_1^n,X_2^n,Y^n,T_1^n,T_2^n,V^n\right)=\pi_{S_1S'_1S_2X_1X_2YT_1T_2V} \big| \mathcal{H}=1 \right]\nonumber\\[2ex]
&& \leq \sum_{m_1,m'_1,m_2}\;\; \sum_{\substack{\pi_{S_1S'_1S_2X_1X_2YT_1T_2V}:\\|\pi_{S_iX_iT_1T_2}-P_{S_iX_iT_1T_2}|<\mu/2,\\|\pi_{S'_1S_2YT_1T_2V}-P_{S_1S_2YT_1T_2V}|<\mu,\\H_{\pi}(S'_1|S_2,Y,T_1,T_2,V)\leq H_{\pi}(S_1|S_2,Y,T_1,T_2,V)}} \nonumber\\[1.5ex]&&\hspace{1cm}\Pr\left[ \;\; \text{tp}\left(S_1^n(m_1),S_1^n(m'_1),S_2^n(m_2),X_1^n,X_2^n,Y^n,T_1^n,T_2^n\right)=\pi_{S_1S'_1S_2X_1X_2YT_1T_2} \big| \mathcal{H}=1 \right]\nonumber\\[1.5ex]
&&\hspace{0.8cm}\cdot \Pr\Big[ \text{cond-tp}\left(V^n|S_1^n(m_1),S_1^n(m'_1),S_2^n(m_2),X_1^n,X_2^n,Y^n,T_1^n,T_2^n\right)=\pi_{V|S_1S'_1S_2X_1X_2YT_1T_2} \Big|\nonumber\\&&\hspace{2cm}\text{tp}\left(S_1^n(m_1),S_1^n(m'_1),S_2^n(m_2),X_1^n,X_2^n,Y^n,T_1^n,T_2^n\right)=\pi_{S_1S'_1S_2X_1X_2YT_1T_2},\;\; W_i^n=f_i(S_i^n(m_i),X_i^n),  \;\;\mathcal{H}=1 \Big]\nonumber\\
&&\leq \sum_{m_1,m'_1,m_2}\;\; \sum_{\substack{\pi_{S_1S'_1S_2X_1X_2YT_1T_2V}:\\|\pi_{S_iX_iT_1T_2}-P_{S_iX_iT_1T_2}|<\mu/2,\\|\pi_{S'_1S_2YT_1T_2V}-P_{S_1S_2YT_1T_2V}|<\mu,\\H_{\pi}(S'_1|S_2,Y,T_1,T_2,V)\leq H_{\pi}(S_1|S_2,Y,T_1,T_2,V)}} \nonumber\\
&&\hspace{0.8cm} 2^{-nD(\pi_{S_1S'_1S_2X_1X_2YT_1T_2}|| P_{S_1|T_1T_2}P_{S'_1|T_1T_2}P_{S_2|T_1T_2}Q_{X_1X_2Y}P_{T_1T_2})}\cdot 2^{-nD(\pi_{V|S_1S'_1S_2X_1X_2YT_1T_2}||\Gamma_{V|S_1S_2X_1X_2})}\nonumber\\[2ex]
&& \leq 2^{-n\tilde{\theta}_{\mu}^{\text{dec,1}}}, 
\end{IEEEeqnarray}
where we define:
\begin{IEEEeqnarray}{rCl}
\tilde{\theta}_{\mu}^{\text{dec,1}}&:= & \min_{\substack{\pi:\\|\pi_{S_iX_iT_1T_2}-P_{S_iX_iT_1T_2}|<\mu/2\\ |\pi_{S'_1S_2YT_1T_2V}-P_{S_1S_2YT_1T_2V}|< \mu\\ H_{\pi}(S'_1|S_2,Y,T_1,T_2,V)\leq H_{\pi}(S_1|S_2,Y,T_1,T_2,V)}}\!\!\!\!\!\!\!\!\! D\big(\pi_{S_1S'_1S_2X_1X_2YT_1T_2V}\|P_{S_1|T_1T_2}P_{S_1'|T_1T_2}P_{S_2|T_1T_2} Q_{X_1X_2Y}P_{T_1T_2}\Gamma_{V|S_1S_2X_1X_2}\big)\nonumber\\[-4ex]&&\hspace{9cm}-2R_1-R_2-\mu.\nonumber\\[2ex]\label{tt1final}
\end{IEEEeqnarray}
Notice the following set of inequalities:
\begin{IEEEeqnarray}{rCl}
\tilde{\theta}_{\mu}^{\text{dec,1}} &\stackrel{(\text{eq.}\;\eqref{nois1mac})}{=}&\hspace{-0.5cm}\min_{\substack{\pi:\\|\pi_{S_iX_iT_1T_2}-P_{S_iX_iT_1T_2}|<\mu/2\\ |\pi_{S'_1S_2YT_1T_2V}-P_{S_1S_2YT_1T_2V}|< \mu\\ H_{\pi}(S'_1|S_2,Y,T_1,T_2,V)\leq H_{\pi}(S_1|S_2,Y,T_1,T_2,V)}}\!\!\!\!\!\!\!\!\! D\big(\pi_{S_1S'_1S_2X_1X_2YT_1T_2V}\|P_{S_1|T_1T_2}P_{S_1'|T_1T_2}P_{S_2|T_1T_2}  Q_{X_1X_2Y}P_{T_1T_2}\Gamma_{V|S_1S_2X_1X_2}\big)\nonumber\\[-4ex]
&&\hspace{9cm}-2I(S_1;X_1|T_1,T_2)-I(S_2;X_2|T_1,T_2)-4\mu\nonumber\\[2ex]
&\stackrel{\text{(CR)}}{=}&\min_{\substack{\pi:\\|\pi_{S_iX_iT_1T_2}-P_{S_iX_iT_1T_2}|<\mu/2\\ |\pi_{S'_1S_2YT_1T_2V}-P_{S_1S_2YT_1T_2V}|< \mu\\ H_{\pi}(S'_1|S_2,Y,T_1,T_2,V)\leq H_{\pi}(S_1|S_2,Y,T_1,T_2,V)}} \Big[D\big(\pi_{S_1S_2X_1X_2YT_1T_2V}\|P_{S_1|T_1T_2}P_{S_2|T_1T_2}Q_{X_1X_2Y}P_{T_1T_2} \Gamma_{V|S_1S_2X_1X_2}\big)\nonumber\\[-6ex]
&&\hspace{8.5cm}+ \mathbb{E}_{\pi_{S_2X_1X_2YT_1T_2V}}\big[ D(\pi_{S_1'|S_2X_1X_2YT_1T_2V} \| P_{S_1'|T_1T_2}) \big] \Big]\nonumber\\[1ex]
&&\hspace{8.5cm}-2I(S_1;X_1|T_1,T_2)-I(S_2;X_2|T_1,T_2)-4\mu\nonumber\\[1ex]
&\stackrel{\text{(DP)}}{\geq}&\min_{\substack{\pi:\\|\pi_{S_iX_iT_1T_2}-P_{S_iX_iT_1T_2}|<\mu/2\\ |\pi_{S'_1S_2YT_1T_2V}-P_{S_1S_2YT_1T_2V}|< \mu\\ H_{\pi}(S'_1|S_2,Y,T_1,T_2,V)\leq H_{\pi}(S_1|S_2,Y,T_1,T_2,V)}} \Big[D\big(\pi_{S_1S_2X_1X_2YT_1T_2V}\|P_{S_1|T_1T_2}P_{S_2|T_1T_2}Q_{X_1X_2Y}P_{T_1T_2} \Gamma_{V|S_1S_2X_1X_2}\big)\nonumber\\[-6ex]
&&\hspace{7cm}+ \mathbb{E}_{\pi_{S_2YT_1T_2V}}\big[ D(\pi_{S_1'|S_2YT_1T_2V} \| P_{S_1'|T_1T_2}) \big] \Big]\nonumber\\[1ex]
&&\hspace{7cm}-2I(S_1;X_1|T_1,T_2)-I(S_2;X_2|T_1,T_2)-4\mu\nonumber\\
&\stackrel{(c)}{=} &\min_{\substack{\pi\in\mathcal{P}_{\mu,\text{dec,1}}}} D(\pi_{S_1S_2X_1X_2YT_1T_2V}\|P_{S_1|X_1T_1T_2}P_{S_2|X_2T_1T_2}Q_{X_1X_2Y}P_{T_1T_2} \Gamma_{V|S_1S_2X_1X_2})\nonumber\\
&&\qquad\qquad\qquad+I(S_1;S_2,Y,V|T_1,T_2)-I(S_1;X_1|T_1,T_2)-\delta_2(\mu)\nonumber\\
&\stackrel{(d)}{=}&\min_{\substack{\pi\in\mathcal{P}_{\mu,\text{dec,1}}}} D(\pi_{S_1S_2X_1X_2YT_1T_2V}\|P_{S_1|X_1T_1T_2}P_{S_2|X_2T_1T_2}Q_{X_1X_2Y}P_{T_1T_2} \Gamma_{V|S_1S_2X_1X_2})\nonumber\\
&&\qquad\qquad\qquad+I(S_1;Y,V|S_2,T_1,T_2)-I(S_1;X_1|S_2,T_1,T_2)-\delta_2(\mu)\nonumber\\
&:=& \theta_{\mu}^{\text{dec,1}}-\delta_2(\mu),\label{tdec1final}
\end{IEEEeqnarray}
where $\delta_2(\mu)$ is a function that goes to zero as $\mu\to 0$;  $(c)$ holds because $\pi_{S_1'|S_2YT_1T_2V}=P_{S_1|S_2YT_1T_2V}$; and $(d)$ holds by the Markov chain $S_2\to (X_1,T_1,T_2)\to S_1$. Combining \eqref{b2event}--\eqref{tdec1final}, one then obtains:
\begin{align}
 \Pr\big[ \mathcal{B}_2\big| \mathcal{H}=1\big] \leq 2^{-n\left(\theta_{\mu}^{\text{dec,1}}-\delta_2(\mu)\right)}.
\end{align}
In a similar way, one can also derive the upper bound
\begin{align}
& \Pr\big[ \mathcal{B}_3\big| \mathcal{H}=1\big]
\leq 2^{-n\left(\theta_{\mu}^{\text{dec,2}}-\delta_3(\mu)\right)},
\end{align} 
where 
\begin{align}
\theta_{\mu}^{\text{dec,2}}&:=\min_{\substack{\pi\in\mathcal{P}_{\mu,\text{dec,2}}}} D(\pi_{S_1S_2X_1X_2YT_1T_2V}\|P_{S_1|X_1T_1T_2}P_{S_2|X_2T_1T_2} Q_{X_1X_2Y}P_{T_1T_2}\Gamma_{V|S_1S_2X_1X_2})\nonumber\\
&\qquad\qquad\qquad\qquad+I(S_2;Y,V|S_1,T_1,T_2)-I(S_2;X_2|S_1,T_1,T_2),
\end{align}
and  $\delta_3(\mu)$ is a function that goes to zero as $\mu\to 0$.

Next, consider event $\mathcal{B}_4$. Its probability is upper bounded as
\begin{align}
\Pr\big[ \mathcal{B}_4\big| \mathcal{H}=1\big] \leq 2^{-n\tilde{\theta}_{\mu}^{\text{dec,12}}},\label{t12event}
\end{align}
where 
\begin{IEEEeqnarray}{rCl}
 \tilde{\theta}_{\mu}^{\text{dec,12}}&:=\hspace{-1cm}\min_{\substack{\pi:\\ |\pi_{S_iX_iT_1T_2}-P_{S_iX_iT_1T_2}|<\mu/2,\\ |\pi_{S'_1S'_2YVT_1T_2}-P_{S_1S_2YVT_1T_2}|< \mu \\ H_{\pi}(S'_1,S'_2|Y,V,T_1,T_2)\leq H_{\pi}(S_1,S_2|Y,V,T_1,T_2)}}\hspace{-1cm} D(\pi_{S_1S'_1S_2S'_2X_1X_2YVT_1T_2}\|P_{S_1|T_1T_2}P_{S_1'|T_1T_2}P_{S_2|T_1T_2}	 P_{S_2'|T_1T_2} Q_{X_1X_2Y}P_{T_1T_2} \Gamma_{V|S_1S_2X_1X_2})\nonumber\\[-6ex]&\hspace{6cm} -2R_1-2R_2-\mu.\nonumber\\[2ex]\label{t12}\end{IEEEeqnarray}
 Notice the following set of inequalities:
 \begin{IEEEeqnarray}{rCl}
\tilde{\theta}_{\mu}^{\text{dec,12}}&\stackrel{(\text{eq.}\;\eqref{nois1mac})}{=}&\hspace{-1.7cm}\min_{\substack{\pi:\\ |\pi_{S_iX_iT_1T_2}-P_{S_iX_iT_1T_2}|<\mu/2,\\ |\pi_{S'_1S'_2YT_1T_2V}-P_{S_1S_2YT_1T_2V}|< \mu \\ H_{\pi}(S'_1,S'_2|Y,T_1,T_2,V)\leq H_{\pi}(S_1,S_2|Y,T_1,T_2,V)}}\hspace{-1cm} D(\pi_{S_1S'_1S_2S'_2X_1X_2YT_1T_2V}\|P_{S_1|T_1T_2}P_{S_1'|T_1T_2}P_{S_2|T_1T_2}	 P_{S_2'|T_1T_2}Q_{X_1X_2Y}P_{T_1T_2}  \Gamma_{V|S_1S_2X_1X_2})\nonumber\\[-6ex]&&\hspace{6cm} -2I(S_1;X_1|T_1,T_2)-2I(S_2;X_2|T_1,T_2)-5\mu\nonumber\\[5ex]
&\stackrel{\text{(CR) \& (DP)}}{\geq}&\hspace{-1.5cm}\min_{\substack{\pi:\\ |\pi_{S_iX_iT_1T_2}-P_{S_iX_iT_1T_2}|<\mu/2,\\ |\pi_{S'_1S'_2YT_1T_2V}-P_{S_1S_2YT_1T_2V}|< \mu \\ H_{\pi}(S'_1,S'_2|Y,T_1,T_2,V)\leq H_{\pi}(S_1,S_2|Y,T_1,T_2,V)}} \Big[D(\pi_{S_1S_2X_1X_2YT_1T_2V}\|P_{S_1|T_1T_2}	 P_{S_2|T_1T_2}Q_{X_1X_2Y}P_{T_1T_2}\Gamma_{V|S_1S_2X_1X_2} )\nonumber\\[-6ex]
&&\hspace{8cm}+\mathbb{E}_{\pi_{YT_1T_2V}}\big[ D(\pi_{S'_1S'_2|YT_1T_2V}\|P_{S_1'|T_1T_2}	 P_{S_2'|T_1T_2})\big]\Big]\nonumber\\[1ex]
&&\hspace{8cm} -2I(S_1;X_1|T_1,T_2)-2I(S_2;X_2|T_1,T_2)-5\mu\nonumber\\[1ex]
&\stackrel{(e)}{=} &\min_{\substack{\pi\in\mathcal{P}_{\mu,\text{dec,12}}}} D(\pi_{S_1S_2X_1X_2YT_1T_2V}\|P_{S_1|X_1T_1T_2}P_{S_2|X_2T_1T_2} Q_{X_1X_2Y}P_{T_1T_2}\Gamma_{V|S_1S_2X_1X_2})\nonumber\\
&&\hspace{2.5cm}+I(S_1,S_2;Y,V|T_1,T_2)+I(S_1;S_2|T_1,T_2)-I(S_1;X_1|T_1,T_2)-I(S_2;X_2|T_1,T_2)-\delta_4(\mu)\nonumber\\[1ex]
&\stackrel{(f)}{=}&\min_{\substack{\pi\in\mathcal{P}_{\mu,\text{dec,12}}}} D(\pi_{S_1S_2X_1X_2YT_1T_2V}\|P_{S_1|X_1T_1T_2}P_{S_2|X_2T_1T_2} Q_{X_1X_2Y}P_{T_1T_2}\Gamma_{V|S_1S_2X_1X_2})\nonumber\\
&&\hspace{2.5cm}+I(S_1,S_2;Y,V|T_1,T_2)-I(S_1,S_2;X_1,X_2|T_1,T_2)-\delta_4(\mu)\nonumber\\[1ex]
&:=&\theta_{\mu}^{\text{dec,12}}-\delta_4(\mu),\label{t12final}
\end{IEEEeqnarray}
where  $\delta_4(\mu)$ is a function that goes to zero as $\mu\to 0$; $(e)$ holds by $\pi_{S'_1S'_2|YT_1T_2V}=P_{S_1S_2|YT_1T_2V}$, by re-arranging terms, and by the continuity of KL-divergence; and $(f)$ holds by the Markov chains $(S_2,X_2)\to (X_1,T_1,T_2)\to S_1$ and $(S_1,X_1)\to (X_2,T_1,T_2)\to S_2$. Combining \eqref{t12event}--\eqref{t12final}, one then obtains:
\begin{align}
\Pr\big[ \mathcal{B}_4\big| \mathcal{H}=1\big] \leq 2^{-n\left(\theta_{\mu}^{\text{dec,12}}-\delta_4(\mu)\right)}.
\end{align}

We upper bound the probability of event $\mathcal{B}_{5}$. Recall that cond\_tp$(a^n|b^n)$ denotes the conditional type of sequence $a^n$ given $b^n$. We have:
\begin{IEEEeqnarray}{rCl}
	&& \Pr\big[ \mathcal{B}_5\big| \mathcal{H}=1\big]  \nonumber\\
	&&\leq\sum_{m'_1,m_2,m_2'}\Pr \Big[  W_1^n=T_1^n,\quad (S_2^n(m_2),X_2^n,T_1^n,T_2^n)\in\mathcal{T}_{\mu/2}^n(P_{S_2X_2T_1T_2}),\nonumber\\
	&&\hspace{2.3cm}(S_1^n(m'_1),S_2^n(m'_2),Y^n,T_1^n,T_2^n,V^n)\in\mathcal{T}_{\mu}^n(P_{S_2Y T_1T_2V})\nonumber\\&&\hspace{1cm} H_{\text{tp}(S_1^n(m'_1),S_2^n(m'_2),Y^n,T_1^n,T_2^n,V^n)}(S_1',S'_2|Y,T_1,T_2,V)\leq H_{\text{tp}(S_1^n(m'_1),S_2^n(m_2),Y^n,T_1^n,T_2^n,V^n)}(S_1,S_2|Y,T_1,T_2,V)\Big|\mathcal{H}=1 \Big]\nonumber\\[2ex]
		&&=\sum_{m'_1,m_2,m_2'}\sum_{\substack{\pi_{S_1'S_2S_2'XYT_1T_2V}:\\ |\pi_{S_2X_2T_1T_2}-P_{S_2X_2T_1T_2}|<\mu/2,\\ |\pi_{S'_1S'_2YT_1T_2V}-P_{S_1S_2YT_1T_2V}|< \mu \\ H_{\pi}(S'_1,S'_2|Y,T_1,T_2,V)\leq H_{\pi}(S_1',S_2|Y,T_1,T_2,V)}}   
		\nonumber\\[2ex]
		&&\hspace{1cm}
		\Pr \Big[  W_1^n=T_1^n,\quad \text{tp}(S_1^n(m'_1),S_2^n(m_2),S_2^n(m_2'), X_2^n,Y^n,T_1^n,T_2^n,V^n) =\pi_{S_1'S_2S_2'XYT_1T_2V} \Big|\mathcal{H}=1 \Big]\nonumber\\[2ex] 
				&&\leq\sum_{m'_1,m_2,m_2'}\sum_{\substack{\pi_{S_1'S_2S_2'XYT_1T_2V}:\\ |\pi_{S_2X_2T_1T_2}-P_{S_2X_2T_1T_2}|<\mu/2,\\ |\pi_{S'_1S'_2YT_1T_2V}-P_{S_1S_2YT_1T_2V}|< \mu \\ H_{\pi}(S'_1,S'_2|Y,T_1,T_2,V)\leq H_{\pi}(S_1',S_2|Y,T_1,T_2,V)}}   
				\nonumber\\[2ex]
				&&\hspace{.6cm}
				\Pr \Big[ \text{tp}(S_1^n(m'_1),S_2^n(m_2), S_2^n(m_2'), X_2^n,Y^n,T_1^n,T_2^n) =\pi_{S_1'S_2S_2'XYT_1T_2} \Big| \mathcal{H}=1 \Big]		\nonumber \\
				& & \hspace{.6cm} \cdot	\Pr \Big[ \text{cond\_tp}(V^n|S_1^n(m_1'), S_2^n(m_2), S_2^n(m_2'),X_2^n, Y^n,T_1^n, T_2^n) =\pi_{V|S_1'S_2S_2'XYT_1T_2} \Big| \nonumber \\
				& & \hspace{5.4cm} \text{tp}(S_1^n(m_1'),S_2^n(m_2),S_2^n(m_2'), X_2^n,Y^n,T_1^n,T_2^n) =\pi_{S_1'S_2S_2'XYT_1T_2}, \; W_1^n=T_1^n,\;\mathcal{H}=1 \Big]\nonumber\\[2ex] 
						&&\leq\sum_{m'_1,m_2,m_2'}\sum_{\substack{\pi_{S_1'S_2S_2'XYT_1T_2V}:\\ |\pi_{S_2X_2T_1T_2}-P_{S_2X_2T_1T_2}|<\mu/2,\\ |\pi_{S'_1S'_2YT_1T_2V}-P_{S_1S_2YT_1T_2V}|< \mu \\ H_{\pi}(S'_1,S'_2|Y,T_1,T_2,V)\leq H_{\pi}(S_1',S_2|Y,T_1,T_2,V)}}   
						\nonumber\\[2ex]
						&&\hspace{1cm}
					2^{- n D( \pi_{S_1'S_2S_2'X_2YT_1T_2} \| P_{S_1'|T_1T_2}P_{S_2|T_1T_2} P_{S_2'|T_1T_2} Q_{X_2Y} P_{T_1T_2})} \cdot 	2^{- n D( \pi_{V|S_1'S_2S_2'X_2YT_1T_2} \| \Gamma^{(1)}_{V|T_1S_2X_2})} 	
					 \nonumber\\[2ex] 						
&&{\leq}  2^{-n\tilde{\theta}_{\mu}^{\text{miss,1a}}},\label{tm1bevent}
 \end{IEEEeqnarray}
where 
\begin{IEEEeqnarray}{rCl}
\tilde{\theta}_{\mu}^{\text{miss,1a}}&:=&\!\!\!\!\!\!\!\!\!\!\!\!\!\min_{\substack{\pi:\\ |\pi_{S_2X_2T_1T_2}-P_{S_2X_2T_1T_2}|<\mu/2,\\ |\pi_{S'_1S'_2YT_1T_2V}-P_{S_1S_2YT_1T_2V}|< \mu \\ H_{\pi}(S'_1,S'_2|Y,T_1,T_2,V)\leq H_{\pi}(S_1',S_2|Y,T_1,T_2,V)}}\!\!\!\!\!\!\!\!\!\!\!\!\! D(\pi_{S'_1S_2S'_2X_1X_2YT_1T_2V}\|P_{S'_1|T_1T_2}P_{S_2|T_1T_2}P_{S'_2|T_1T_2} Q_{X_2Y}P_{T_1T_2}\Gamma_{V|T_1S_2X_2}^{(1)})\nonumber\\[-6ex]
&&\hspace{9cm}-R_1-2R_2-\mu.\nonumber\\[1ex]\end{IEEEeqnarray}
We have the following set of inequalities:
\begin{IEEEeqnarray}{rCl}
\tilde{\theta}_{\mu}^{\text{miss,1a}} &\stackrel{(\text{eq.}\;\eqref{nois1mac})}{=}&\!\!\!\!\!\!\!\!\!\!\!\min_{\substack{\pi:\\ |\pi_{S_2X_2T_1T_2}-P_{S_2X_2T_1T_2}|<\mu/2,\\ |\pi_{S'_1S'_2YT_1T_2V}-P_{S_1S_2YT_1T_2V}|< \mu \\ H_{\pi}(S'_1,S'_2|Y,T_1,T_2,V)\leq H_{\pi}(S_1,S_2|Y,T_1,T_2,V)}}\!\!\!\!\!\!\!\!\!\!\!\!\! D(\pi_{S'_1S_2S'_2X_1X_2YT_1T_2V}\|P_{S'_1|T_1T_2}P_{S_2|T_1T_2}P_{S'_2|T_1T_2} Q_{X_2Y}P_{T_1T_2}\Gamma_{V|T_1S_2X_2}^{(1)})\nonumber\\[-6ex]
&&\hspace{7cm}-I(S_1;X_1|T_1,T_2)-2I(S_2;X_2|T_1,T_2)-4\mu\nonumber\\[3ex]
&\stackrel{\text{(CR) \& (DP)}}{\geq}&\hspace{-1cm}\min_{\substack{\pi:\\ |\pi_{S_2X_2T_1T_2}-P_{S_2X_2T_1T_2}|<\mu/2,\\ |\pi_{S'_1S'_2YT_1T_2V}-P_{S_1S_2YT_1T_2V}|< \mu \\ H_{\pi}(S'_1,S'_2|Y,T_1,T_2,V)\leq H_{\pi}(S_1',S_2|Y,T_1,T_2,V)}} \Big[D(\pi_{S_2X_1X_2YT_1T_2V}\|P_{S_2|T_1T_2}Q_{X_2Y}P_{T_1T_2} \Gamma_{V|T_1S_2X_2}^{(1)})\nonumber\\[-6ex]
&&\hspace{7cm} +\mathbb{E}_{\pi_{YVT_1T_2}}\big[D(\pi_{S'_1S'_2|YVT_1T_2}\|P_{S'_1|T_1T_2}P_{S'_2|T_1T_2}) \big]\Big]\nonumber\\
&&\hspace{7cm}-I(S_1;X_1|T_1,T_2)-2I(S_2;X_2|T_1,T_2)-4\mu\nonumber\\[1ex]
&= &\min_{\substack{\pi\in\mathcal{P}_{\mu,\text{miss,1a}}}}  D(\pi_{S_2X_2YT_1T_2V}\|P_{S_2|X_2T_1T_2} Q_{X_2Y}P_{T_1T_2} \Gamma_{V|T_1S_2X_2}^{(1)})\nonumber\\
&&\hspace{2cm} +I(S_1,S_2;Y,V|T_1,T_2)+I(S_1;S_2|T_1,T_2)-I(S_1;X_1|T_1,T_2)-I(S_2;X_2|T_1,T_2)-\delta_5(\mu)\nonumber\\[1ex]
&=&\min_{\substack{\pi\in\mathcal{P}_{\mu,\text{miss,1a}}}} D(\pi_{S_2X_2YT_1T_2V}\|P_{S_2|X_2T_1T_2} Q_{X_2Y}P_{T_1T_2} \Gamma_{V|T_1S_2X_2}^{(1)})\nonumber\\
&&\hspace{2cm}+I(S_1,S_2;Y,V|T_1,T_2)-I(S_1,S_2;X_1,X_2|T_1,T_2)-\delta_5(\mu)\nonumber\\[1ex]
&:=&\theta_{\mu}^{\text{miss,1a}}-\delta_5(\mu), \label{eq:mise1a}
\end{IEEEeqnarray}where $\delta_5(\mu)$ is a function that goes to zero as $\mu\to 0$.
Combining~\eqref{tm1bevent}--\eqref{eq:mise1a} leads to:
\begin{align}
 \Pr\big[ \mathcal{B}_5\big| \mathcal{H}=1\big] \leq 2^{-n\left(\theta_{\mu}^{\text{miss,1a}}-\delta_5(\mu)\right)}.
\end{align}
The probability of  event $\mathcal{B}_6$ can be upper bounded in a similar way to obtain:
\begin{align}
&\Pr\big[ \mathcal{B}_6\big| \mathcal{H}=1\big] 
\leq  2^{-n\tilde{\theta}_{\mu}^{\text{miss,1b}}},
\end{align}
where 
\begin{IEEEeqnarray}{rCl}
	\tilde{\theta}_{\mu}^{\text{miss,1b}}&:=&
	\min_{\substack{\pi:\\ |\pi_{S_2X_2T_1T_2}-P_{S_2X_2T_1T_2}|<\mu/2,\\ |\pi_{S'_1S_2YT_1T_2V}-P_{S_1S_2YT_1T_2V}|< \mu }}\!\!\!\!\!\!\!\!\!\!\!\! D(\pi_{S'_1S_2X_2YT_1T_2V}\|P_{S'_1|T_1T_2}P_{S_2|T_1T_2} Q_{X_2Y}P_{T_1T_2} \Gamma_{V|T_1S_2X_2}^{(1)})-R_1-R_2-\mu.\nonumber\\[1ex]\label{t1mbt}\end{IEEEeqnarray}
We have the following set of inequalities:
\begin{IEEEeqnarray}{rCl}
	\tilde{\theta}_{\mu}^{\text{miss,1b}}&\stackrel{(\text{eq.}\;\eqref{nois1mac})}{=}&\min_{\substack{\pi_{S_1'S_2X_2YT_1T_2V}:\\ |\pi_{S_2X_2T_1T_2}-P_{S_2X_2T_1T_2}|<\mu/2,\\ |\pi_{S'_1S_2YT_1T_2V}-P_{S_1S_2YT_1T_2V}|< \mu  }} D(\pi_{S'_1S_2X_2YT_1T_2V}\|P_{S'_1|T_1T_2}P_{S_2|T_1T_2} Q_{X_2Y}P_{T_1T_2} \Gamma_{V|T_1S_2X_2}^{(1)})\nonumber\\[-4.5ex]
	&&\hspace{8cm}-I(S_1;X_1|T_1,T_2)-I(S_2;X_2|T_1,T_2)-3\mu\nonumber\\[3ex]&\stackrel{\text{(CR)}}{=}&\min_{\substack{\pi_{S_1'S_2X_2YT_1T_2V}:\\ |\pi_{S_2X_2T_1T_2}-P_{S_2X_2T_1T_2}|<\mu/2,\\ |\pi_{S'_1S_2YT_1T_2V}-P_{S_1S_2YT_1T_2V}|< \mu }} \Big[D(\pi_{S_2X_2YT_1T_2V}\|P_{S_2|T_1T_2} Q_{X_2Y}P_{T_1T_2} \Gamma_{V|T_1S_2X_2}^{(1)})\nonumber\\[-5.5ex]
	&&\hspace{7cm}+\mathbb{E}_{\pi_{S_2X_2YT_1T_2V}}\big[ D(\pi_{S'_1|S_2X_2YT_1T_2V}\|P_{S'_1|T_1T_2}) \big]\Big]\nonumber\\
	&&\hspace{7cm}-I(S_1;X_1|T_1,T_2)-I(S_2;X_2|T_1,T_2)-3\mu\nonumber\\[2ex]
	&\stackrel{\text{(DP)}}{\geq}&\min_{\substack{\pi_{S_1'S_2X_2YT_1T_2V}:\\ |\pi_{S_2X_2T_1T_2}-P_{S_2X_2T_1T_2}|<\mu/2,\\ |\pi_{S'_1S_2YT_1T_2V}-P_{S_1S_2YT_1T_2V}|< \mu }} \Big[D(\pi_{S_2X_2YT_1T_2V}\|P_{S_2|T_1T_2} Q_{X_2Y}P_{T_1T_2} \Gamma_{V|T_1S_2X_2}^{(1)})\nonumber\\[-5.5ex]
	&&\hspace{7cm}+\mathbb{E}_{\pi_{S_2YT_1T_2V}}\big[ D(\pi_{S'_1|S_2YT_1T_2V}\|P_{S_1'|T_1T_2}) \big]\Big]\nonumber\\
	&&\hspace{7cm}-I(S_1;X_1|T_1,T_2)-I(S_2;X_2|T_1,T_2)-3\mu\nonumber\\[2ex]
	&\stackrel{(h)}{=} &\min_{\substack{\pi_{S_1'S_2X_2YT_1T_2V}\in\mathcal{P}_{\mu,\text{miss,1b}}}}  D(\pi_{S_2X_2YT_1T_2V}\|P_{S_2|T_1T_2}  Q_{X_2Y}P_{T_1T_2}\Gamma_{V|T_1S_2X_2}^{(1)})\nonumber\\
	&&\hspace{5cm}+I(S_1;S_2,Y,V|T_1,T_2)-I(S_1;X_1|T_1,T_2)-I(S_2;X_2|T_1,T_2)-\delta_6(\mu)\nonumber\\[1ex]
	&=&\min_{\substack{\pi\in\mathcal{P}_{\mu,\text{miss,1b}}}}  D(\pi_{S_2X_2YT_1T_2V}\|P_{S_2|X_2T_1T_2}  Q_{X_2Y}P_{T_1T_2}\Gamma_{V|T_1S_2X_2}^{(1)})\nonumber\\
	&&\hspace{5cm}+I(S_1;S_2,Y,V|T_1,T_2)-I(S_1;X_1|T_1,T_2)-\delta_6(\mu)\nonumber\\[1ex]
	&\stackrel{(i)}{=}&\min_{\substack{\pi\in\mathcal{P}_{\mu,\text{miss,1b}}}}  D(\pi_{S_2X_2YT_1T_2V}\|P_{S_2|X_2T_1T_2}  Q_{X_2Y}P_{T_1T_2}\Gamma_{V|T_1S_2X_2}^{(1)})\nonumber\\
	&&\hspace{5cm}+I(S_1;Y,V|S_2,T_1,T_2)-I(S_1;X_1|S_2,T_1,T_2)-\delta_6(\mu)\nonumber\\[1ex]
	&:=&\theta_{\mu}^{\text{miss,1b}}-\delta_6(\mu),
	\label{tm1b}
\end{IEEEeqnarray}
where $\delta_6(\mu)$ is a function that goes to zero as $\mu\to 0$.
Here, $(h)$ holds because $\pi_{S'_1|YVT_1T_2}=P_{S_1|YVT_1T_2}$ and $(i)$ holds because of the Markov chain $S_1\to (X_1,T_1,T_2)\to S_2$. From \eqref{tm1bevent}--\eqref{tm1b}, we obtain
\begin{align}
 \Pr \big[ \mathcal{B}_6\big| \mathcal{H}=1\big] \leq 2^{-n\left(\theta_{\mu}^{\text{miss,1b}}-\delta_6(\mu)\right)}.
\end{align}

Following similar steps to above, one can show that 
\begin{align}
\Pr\big[ \mathcal{B}_7\big| \mathcal{H}=1\big]  &\leq 2^{-n\left(\theta_{\mu}^{\text{miss,2a}}-\delta_7(\mu)\right)},\\
 \Pr\big[ \mathcal{B}_8\big| \mathcal{H}=1\big] &\leq  2^{-n\left(\theta_{\mu}^{\text{miss,2b}}-\delta_8(\mu)\right)},
\end{align}
where $\delta_7(\mu)$ and $\delta_8(\mu)$ are functions that go to zero as $\mu\to 0$ and 
\begin{align}
\theta_{\mu}^{\text{miss,2a}} &:= \min_{\substack{\pi\in\mathcal{P}_{\mu,\text{miss,2a}}}}  D(\pi_{S_1X_1YT_1T_2V}\|P_{S_1|X_1T_1T_2}  Q_{X_1Y}P_{T_1T_2}\Gamma_{V|S_1X_1T_2}^{(2)})\nonumber\\
&\hspace{6cm}+I(S_1,S_2;Y,V|T_1,T_2)-I(S_1,S_2;X_1,X_2|T_1,T_2),\\
\theta_{\mu}^{\text{miss,2b}} &:= \min_{\substack{\pi\in\mathcal{P}_{\text{miss,2b}}}}  D(\pi_{S_1X_1YT_1T_2V}\|P_{S_1|X_1T_1T_2} Q_{X_1Y}P_{T_1T_2} \Gamma_{V|S_1X_1T_2}^{(2)})\nonumber\\&\hspace{6cm}+I(S_2;Y,V|S_1,T_1,T_2)-I(S_2;X_2|S_1,T_1,T_2).
\end{align}

Finally, the probability of event $\mathcal{B}_9$ can be upper bounded as:
\begin{align}
 \Pr\big[ \mathcal{B}_9\big| \mathcal{H}=1\big] \leq  2^{-n\tilde{\theta}_{\mu}^{\text{miss,12}}},\label{eq:bevent9}
\end{align}
where 
\begin{IEEEeqnarray}{rCl}
\tilde{\theta}_{\mu}^{\text{miss,12}}&:=& \hspace{-0.8cm}\min_{\substack{\pi:\\ |\pi_{S'_1S'_2YT_1T_2V}-P_{S_1S_2YT_1T_2V}|< \mu }} D(\pi_{S'_1S'_2YT_1T_2V}\|P_{S_1'|T_1T_2}P_{S_2'|T_1T_2} Q_{Y}P_{T_1T_2}\Gamma_{V|T_1T_2}^{(12)})-R_1-R_2-\mu,
\end{IEEEeqnarray}
We have the following set of inequalities:
\begin{IEEEeqnarray}{rCl}
\tilde{\theta}_{\mu}^{\text{miss,12}} &\stackrel{(\text{eq.}\;\eqref{nois1mac})}{=}&\min_{\substack{\pi:\\ |\pi_{S'_1S'_2YT_1T_2V}-P_{S_1S_2YT_1T_2V}|< \mu }} \hspace{-0.2cm} D(\pi_{S'_1S'_2YT_1T_2V}\|P_{S_1'|T_1T_2}P_{S_2'|T_1T_2}Q_{Y}P_{T_1T_2} \Gamma_{V|T_1T_2}^{(12)})\nonumber\\[-2ex]
&&\hspace{7cm}-I(S_1;X_1|T_1,T_2)-I(S_2;X_2|T_1,T_2)-3\mu\nonumber\\[1.5ex]
&=& \min_{\substack{\pi:\\ |\pi_{S'_1S'_2YT_1T_2V}-P_{S_1S_2YT_1T_2V}|< \mu }} \Big[D(\pi_{YT_1T_2V}\| Q_{Y}P_{T_1T_2}\Gamma_{V|T_1T_2}^{(12)})\nonumber\\[-3ex]
&&\hspace{7cm}+\mathbb{E}_{\pi_{YT_1T_2V}}\big[ D(\pi_{S'_1S'_2|YT_1T_2V}\|P_{S_1'|T_1T_2}P_{S_2'|T_1T_2})\big]\Big]\nonumber\\
&&\hspace{7cm}-I(S_1;X_1|T_1,T_2)-I(S_2;X_2|T_1,T_2)-3\mu\nonumber\\[1ex]
&\stackrel{(j)}{=}&\mathbb{E}_{P_{T_1T_2}} \big[D(P_{YV|T_1T_2}\| Q_{Y}\Gamma_{V|T_1T_2}^{(12)})\big]+I(S_1,S_2;Y,V|T_1,T_2)-I(S_1,S_2;X_1,X_2|T_1,T_2)-\delta_9(\mu)\nonumber\\
&=: &\theta_{\mu}^{\text{miss,12}}-\delta_9(\mu),\label{eq:ms12}
\end{IEEEeqnarray}
where $\delta_9(\mu)$ is a function that goes to zero as $\mu\to 0$. Here, $(j)$ holds because $\pi_{YT_1T_2V}=P_{YT_1T_2V}$, $\pi_{S'_1S'_2|YT_1T_2V}=P_{S_1S_2|YT_1T_2V}$ and by  the Markov chains $S_1\to (X_1,T_1,T_2)\to S_2$ and $S_2\to (X_2,T_1,T_2)\to S_1$.
Combining \eqref{eq:bevent9}--\eqref{eq:ms12} yields:
\begin{align}
 \Pr\big[ \mathcal{B}_9\big| \mathcal{H}=1\big] \leq 2^{-n\left(\theta_{\mu}^{\text{miss,12}}-\delta_9(\mu)\right)}.
\end{align}
Therefore, the average type-II error probability satisfies:
\begin{align}
\mathbb{E}_{\mathcal{C}}[\beta_n] \leq \max \left\{ 2^{-n\theta_{\mu}^{\text{standard}}}, 2^{-n\theta_{\mu}^{\text{dec,1}}}, 2^{-n\theta_{\mu}^{\text{dec,2}}},2^{-n\theta_{\mu}^{\text{dec,12}}},2^{-n\theta_{\mu}^{\text{miss,1a}}},2^{-n\theta_{\mu}^{\text{miss,1b}}},2^{-n\theta_{\mu}^{\text{miss,2a}}},2^{-n\theta_{\mu}^{\text{miss,2b}}},2^{-n\theta_{\mu}^{\text{miss,12}}} \right\}.
\end{align}
By standard arguments and successively eliminating the worst half of the codebooks, it can be shown that there exists at least one codebook for which:
\begin{align}
\alpha_n &\leq \epsilon,\\
\beta_n &\leq 1024\cdot  \max \left\{ 2^{-n\theta_{\mu}^{\text{standard}}}, 2^{-n\theta_{\mu}^{\text{dec,1}}}, 2^{-n\theta_{\mu}^{\text{dec,2}}},2^{-n\theta_{\mu}^{\text{dec,12}}},2^{-n\theta_{\mu}^{\text{miss,1a}}},2^{-n\theta_{\mu}^{\text{miss,1b}}},2^{-n\theta_{\mu}^{\text{miss,2a}}},2^{-n\theta_{\mu}^{\text{miss,2b}}},2^{-n\theta_{\mu}^{\text{miss,12}}} \right\}.
\end{align}
By letting $\mu\to 0$ and $n\to\infty$ for $i\in\{1,2\}$, we get $\theta_{\mu}^{\text{standard}}\to \theta^{\text{standard}}$, $\theta_{\mu}^{\text{dec,i}}\to \theta^{\text{dec,i}}$, $\theta_{\mu}^{\text{dec,12}}\to \theta^{\text{dec,12}}$, $\theta_{\mu}^{\text{miss,ia}}\to \theta^{\text{miss,ia}}$, $\theta_{\mu}^{\text{miss,ib}}\to \theta^{\text{miss,ib}}$ and $\theta_{\mu}^{\text{miss,12}}\to \theta^{\text{miss,12}}$, which concludes the proof of the theorem.

\section{Proof of Corollary~\ref{corr-mac-dif-marginal}}\label{remark-dif-marginal}

For the described setup and for any choice of the (conditional) pmfs $P_{T_1T_2}, P_{S_1|X_1T_1T_2}, P_{S_2|X_2T_1T_2}$ and  functions $f_1, f_2$, the error exponents in \eqref{eq:condmac} simplify as follows.
For the  decoding-error exponents, we have:
\begin{IEEEeqnarray}{rCl}
	\theta^{\text{dec,1}} &:= &\hspace{-1cm} \min_{\substack{\tilde{P}_{S_1S_2X_1X_2YT_1T_2V}:\\\tilde{P}_{S_iX_iT_1T_2}=P_{S_iX_iT_1T_2}, \; i\in\{1,2\},\\\tilde{P}_{S_2\bar{Y}ZT_1T_2V}=P_{S_2\bar{Y}ZT_1T_2V}\\H(S_1|S_2,\bar Y, Z, T_1,T_2,V)\leq H_{\tilde{P}}(S_1|S_2,\bar{Y},Z,T_1,T_2,V)}} \hspace{-1cm} D\big(\tilde{P}_{S_1S_2X_1X_2 \bar{Y}ZT_1T_2V}\|P_{S_1|X_1T_1T_2} P_{S_2|X_2T_1T_2}   P_{X_1X_2Z} Q_{\bar{Y}|Z} P_{T_1T_2} \Gamma_{V|S_1S_2X_1X_2}  \big)\nonumber\\[-4ex]
	& &\hspace{8cm} +I(S_1;\bar{Y},Z,V|S_2,T_1,T_2)-I(S_1;X_1|S_2,T_1,T_2)\\[4.5ex]
	&\stackrel{(\textnormal{CR})}{=}&\hspace{-10mm} \min_{\substack{\tilde{P}_{S_1S_2X_1X_2T_1T_2}:\\\tilde{P}_{S_iX_iT_1T_2}=P_{S_iX_iT_1T_2}, \; i\in\{1,2\},\\\tilde{P}_{S_2\bar{Y}ZT_1T_2V}=P_{S_2\bar{Y}ZT_1T_2V}\\H(S_1|S_2,\bar{Y},Z,T_1,T_2,V)\leq H_{\tilde{P}}(S_1|S_2,\bar{Y},Z,T_1,T_2,V)}} \hspace{-7mm} \Big[ D\big(\tilde{P}_{S_1S_2X_1X_2 ZT_1T_2}\|P_{S_1|X_1T_1T_2} P_{S_2|X_2T_1T_2}    P_{X_1X_2Z} P_{T_1T_2} \big)  \nonumber\\[-7ex]
	&&\hspace{7cm} + \mathbb{E}_{\tilde{P}_{S_1S_2X_1X_2 ZT_1T_2}} [ D( \tilde{P}_{\bar{Y}V|S_1S_2X_1X_2 ZT_1T_2} \|  Q_{\bar{Y}|Z}  \Gamma_{V|S_1S_2X_1X_2}) \big] \Big]\nonumber\\[3ex]
	& &\hspace{7cm} +I(S_1;\bar{Y},Z,V|S_2,T_1,T_2)-I(S_1;X_1|S_2,T_1,T_2)\\[4.5ex]
	&\stackrel{(\textnormal{DP})}{\geq}&\hspace{-10mm} \min_{\substack{\tilde{P}_{S_1S_2X_1X_2,T_1,T_2}:\\\tilde{P}_{S_iX_i,T_1,T_2}=P_{S_iX_i,T_1,T_2}, \; i\in\{1,2\},\\\tilde{P}_{S_2\bar{Y}ZT_1T_2V}=P_{S_2\bar{Y}ZT_1T_2}\\H(S_1|S_2,\bar{Y},Z,T_1,T_2,V)\leq H_{\tilde{P}}(S_1|S_2,\bar{Y},Z,T_1,T_2,V)}} \hspace{-7mm} \Big[ D\big(\tilde{P}_{S_1S_2X_1X_2 ZT_1T_2}\|P_{S_1|X_1T_1T_2} P_{S_2|X_2T_1T_2}    P_{X_1X_2Z}P_{T_1T_2}  \big)\nonumber\\[-7ex]
	&&\hspace{8cm} +   \mathbb{E}_{\tilde{P}_{S_2 ZT_1T_2}} [ D( \tilde{P}_{\bar{Y}V|S_2 ZT_1T_2} \|  Q_{\bar{Y}|Z}  \Gamma_{V|S_2ZT_1T_2}) \big] \Big]\nonumber\\[3ex]
	& &\hspace{8cm} +I(S_1;\bar{Y},Z,V|S_2,T_1,T_2)-I(S_1;X_1|S_2,T_1,T_2) \nonumber\\[4.5ex]
	&\stackrel{(a)}{=}&\hspace{-10mm} \min_{\substack{\tilde{P}_{S_1S_2X_1X_2}:\\\tilde{P}_{S_iX_iT_1T_2}=P_{S_iX_iT_1T_2}, \; i\in\{1,2\},\\\tilde{P}_{S_2,\bar{Y},ZT_1T_2V}=P_{S_2\bar{Y}ZT_1T_2V}\\H(S_1|S_2,\bar{Y},Z,T_1,T_2,V)\leq H_{\tilde{P}}(S_1|S_2,\bar{Y},Z,T_1,T_2,V)}} \hspace{-7mm} D\big(\tilde{P}_{S_1S_2X_1X_2 ZT_1T_2}\|P_{S_1|X_1T_1T_2} P_{S_2|X_2T_1T_2}   P_{X_1X_2Z} P_{T_1T_2}  \big)
	\nonumber\\[-7ex]
	&&\hspace{8cm} + \mathbb{E}_{{P}_{S_2 ZT_1T_2}} [ D({P}_{\bar{Y}|S_2 ZT_1T_2V} \|  Q_{\bar{Y}|Z} ) \big] \nonumber\\[3ex]
	& &\hspace{8cm} +I(S_1;\bar{Y},Z,V|S_2,T_1,T_2)-I(S_1;X_1|S_2,T_1,T_2)\\[4.5ex]
	&\stackrel{(b)}{=}& \mathbb{E}_{{P}_{S_2 ZT_1T_2}} [ D({P}_{\bar{Y}|S_2 ZT_1T_2V} \|  Q_{\bar{Y}|Z}  \big) +I(S_1;\bar{Y},Z,V|S_2,T_1,T_2)-I(S_1;X_1|S_2,T_1,T_2), \label{dec1_final}
\end{IEEEeqnarray}
where $(a)$ holds by the second constraint in the minimization and $(b)$ holds because KL-divergence is nonnegative and $\tilde{P}_{S_1S_2X_2ZT_1T_2}= P_{S_1|X_1T_1T_2} P_{S_2|X_2T_1T_2}  P_{X_1X_2Z}P_{T_1T_2}$ is a valid choice in the minimization, and because in this example, $P_{V|S_2ZT_1T_2}=\Gamma_{V|S_2ZT_1T_2}$. 

Moreover, above inequality $\stackrel{(\textnormal{DP})}{\geq}$ holds with equality, because evaluating   $D\big(\tilde{P}_{S_1S_2X_1X_2 \bar{Y}ZT_1T_2V}\|P_{S_1|X_1T_1T_2} P_{S_2|X_2T_1T_2}   P_{X_1X_2Z}  $ $Q_{\bar{Y}|Z} P_{T_1T_2}\Gamma_{V|S_1S_2X_1X_2}  \big)$ for the choice 
\begin{align}
\tilde{P}_{S_1S_2X_1X_2 \bar{Y}ZT_1T_2V} =P_{S_1|X_1T_1T_2}P_{S_2|X_2T_1T_2}  P_{X_1X_2Z}  P_{T_1T_2} P_{\bar{Y}V|S_2ZT_1T_2}  
\end{align}
(which is a valid candidate for the minimization) results in the KL-divergence on the right-hand side of \eqref{dec1_final}. So, we conclude that 
\begin{subequations}\label{eq:decoding_final}
\begin{IEEEeqnarray}{rCl}
\theta^{\text{dec,1}} & = & \mathbb{E}_{{P}_{S_2 ZT_1T_2V}} [ D({P}_{\bar{Y}|S_2 ZT_1T_2V} \|  Q_{\bar{Y}|Z}  )\big] +I(S_1;\bar{Y},Z,V|S_2,T_1,T_2)-I(S_1;X_1|S_2,T_1,T_2),\label{eq:dec1_final}
\end{IEEEeqnarray}
and in an analogous way it can be shown that also
\begin{IEEEeqnarray}{rCl}
\theta^{\text{dec,2}} & = & \mathbb{E}_{{P}_{S_1 ZT_1T_2V}} [ D({P}_{\bar{Y}|S_1 ZT_1T_2V} \|  Q_{\bar{Y}|Z} \big) +I(S_2;\bar{Y},Z,V|S_1,T_1,T_2)-I(S_2;X_1|S_1,T_1,T_2),\label{eq:dec2_final}
\end{IEEEeqnarray}
and
\begin{IEEEeqnarray}{rCl}
\theta^{\text{dec,12}} & = & \mathbb{E}_{{P}_{ZT_1T_2V}} [ D({P}_{\bar{Y}|ZT_1T_2V} \|  Q_{\bar{Y}|Z} \big) +I(S_1,S_2;\bar{Y},Z,V|T_1,T_2)-I(S_1,S_2;X_1,X_2|T_1,T_2).\label{eq:dec12_final}
\end{IEEEeqnarray}
\end{subequations}

Moreover, following similar steps, we obtain for the miss-0 error exponents:
\begin{IEEEeqnarray}{rCl}
\theta^{\text{miss},1\textnormal{a}} &= &\min_{\substack{\tilde{P}_{S_2X_2\bar{Y}ZT_1T_2V}:\\\tilde{P}_{S_2X_2T_1T_2}=P_{S_2X_2T_1T_2}\\\tilde{P}_{\bar{Y}ZT_1T_2V}=P_{\bar{Y}ZT_1T_2V}\\H(S_2|\bar{Y},Z,T_1,T_2,V)\leq H_{\tilde{P}}(S_2|\bar{Y},Z,T_1,T_2,V)}}D\big(\tilde{P}_{S_2X_2\bar{Y}ZT_1T_2V}\|P_{S_2|X_2T_1T_2}     P_{X_2Z} Q_{\bar{Y}|Z} P_{T_1T_2} \Gamma^{(1)}_{V|T_1S_2X_2}  \big)\nonumber\\[-6ex]
&&\hspace{6cm}+
I(S_1,S_2;V,\bar{Y},Z|T_1,T_2)-I(S_1,S_2;X_1,X_2|T_1,T_2)\nonumber
\\[6ex]
&\stackrel{(\textnormal{CR})\& (\textnormal{DP})}{\geq} &\hspace{-1cm}  \min_{\substack{\tilde{P}_{S_2X_2\bar{Y}ZT_1T_2V}:\\\tilde{P}_{S_2X_2T_1T_2}=P_{S_2X_2T_1T_2}\\\tilde{P}_{\bar{Y}ZT_1T_2V}=P_{\bar{Y}ZT_1T_2V}\\H(S_2|\bar{Y},Z,T_1,T_2,V)\leq H_{\tilde{P}}(S_2|\bar{Y},Z,T_1,T_2,V)}} \hspace{-1cm} \Big[ D\big(\tilde{P}_{S_2X_2ZT_1T_2}\|P_{S_2|X_2T_1T_2}     P_{X_2Z}  P_{T_1T_2}\big)+ \mathbb{E}_{\tilde{P}_{ZT_1T_2}}\big[D\big(\tilde{P}_{\bar{Y}V|ZT_1T_2}\|     Q_{\bar{Y}|Z} \Gamma^{(1)}_{V|ZT_1T_2}\big)\big]\Big]\nonumber\\[-6ex]
&&\hspace{6cm}+
I(S_1,S_2;V,\bar{Y},Z|T_1,T_2)-I(S_1,S_2;X_1,X_2|T_1,T_2)\Big]\nonumber\\[2ex]
&= &\mathbb{E}_{P_{ZT_1T_2}}\big[ D(P_{\bar{Y}V|ZT_1T_2} \|   Q_{\bar{Y}|Z} \Gamma^{(1)}_{V|ZT_1T_2}  ) \big]+
I(S_1,S_2;V,\bar{Y},Z|T_1,T_2)-I(S_1,S_2;X_1,X_2|T_1,T_2),\label{miss1a_final}
\end{IEEEeqnarray}
 and 
\begin{IEEEeqnarray}{rCl}
\theta^{\text{miss},1\textnormal{b}}  &= & \min_{\substack{\tilde{P}_{S_2X_2\bar{Y}ZT_1T_2V}:\\\tilde{P}_{S_2X_2T_1T_2}=P_{S_2X_2T_1T_2}\\\tilde{P}_{S_2\bar{Y}ZT_1T_2V}=P_{S_2\bar{Y}ZT_1T_2V}}} D\big(\tilde{P}_{S_2X_2\bar{Y}ZT_1T_2V}\|P_{S_2|X_2T_1T_2}     P_{X_2Z}  Q_{\bar{Y}|Z} P_{T_1T_2}\Gamma^{(1)}_{V|T_1S_2X_2} \big) \nonumber\\[-4ex]
& &\hspace{5cm} +
I(S_1;V,\bar{Y},Z|S_2,T_1,T_2)-I(S_1;X_1|S_2,T_1,T_2)\nonumber\\[3ex]
&\stackrel{(\textnormal{CR})\&(\textnormal{DP})}{\geq} &\hspace{-0.9cm} \min_{\substack{\tilde{P}_{S_2X_2\bar{Y}ZT_1T_2V}:\\\tilde{P}_{S_2X_2T_1T_2}=P_{S_2X_2T_1T_2}\\\tilde{P}_{S_2\bar{Y}ZT_1T_2V}=P_{S_2\bar{Y}ZT_1T_2V}}}\hspace{-0.5cm} \Big[ D\big(\tilde{P}_{S_2X_2ZT_1T_2}\|P_{S_2|X_2T_1T_2}     P_{X_2Z} P_{T_1T_2} \big)+ \mathbb{E}_{\tilde{P}_{S_2ZT_1T_2}}\big[D\big(\tilde{P}_{\bar{Y}V|S_2ZT_1T_2}\|   Q_{\bar{Y}|Z} \Gamma^{(1)}_{V|S_2ZT_1T_2}  \big)\big]\Big]\nonumber\\[-5ex]
& &\hspace{5cm} +
I(S_1;V,\bar{Y},Z|S_2,T_1,T_2)-I(S_1;X_1|S_2,T_1,T_2)\nonumber\\[3ex]
&=&\mathbb{E}_{P_{S_2ZT_1T_2}}\big[ D(P_{\bar{Y}V|S_2ZT_1T_2} \| Q_{\bar{Y}|Z}\Gamma^{(1)}_{V|S_2ZT_1T_2}) \big]+I(S_1;V,\bar{Y},Z|S_2,T_1,T_2)-I(S_1;X_1|S_2,T_1,T_2).
\IEEEeqnarraynumspace\label{miss1b_final}
\end{IEEEeqnarray}
Moreover, above two inequalities can be shown to hold with equality, and thus
\begin{subequations}
\begin{IEEEeqnarray}{rCl}
\theta^{\text{miss,1\textnormal{a}}}
&= &\mathbb{E}_{P_{ZT_1T_2}}\big[ D(P_{\bar{Y}V|ZT_1T_2} \|    Q_{\bar{Y}|Z} \Gamma^{(1)}_{V|ZT_1T_2} ) \big]+
I(S_1,S_2;V,\bar{Y},Z|T_1,T_2)-I(S_1,S_2;X_1,X_2|T_1,T_2) \label{eq:miss1a_final}\\
\theta^{\text{miss,1\textnormal{b}}} &= &  \mathbb{E}_{P_{S_2ZT_1T_2}}\big[ D(P_{\bar{Y}V|S_2ZT_1T_2} \| Q_{\bar{Y}|Z}\Gamma^{(1)}_{V|S_2ZT_1T_2}) \big]+I(S_1;V,\bar{Y},Z|S_2,T_1,T_2)-I(S_1;X_1|S_2,T_1,T_2)\label{eq:miss1b_final}.
\end{IEEEeqnarray}
By similar arguments, also
\begin{IEEEeqnarray}{rCl}
\theta^{\text{miss,2\textnormal{a}}}
&= &\mathbb{E}_{P_{ZT_1T_2}}\big[ D(P_{\bar{Y}V|ZT_1T_2} \|      Q_{\bar{Y}|Z}\Gamma^{(2)}_{V|ZT_1T_2}) \big]+
I(S_1,S_2;V,\bar{Y},Z|T_1,T_2)-I(S_1,S_2;X_1,X_2|T_1,T_2)\label{eq:miss2a_final}\\
\theta^{\text{miss,2\textnormal{b}}} &=&  \mathbb{E}_{P_{S_1ZT_1T_2}}\big[ D(P_{\bar{Y}V|S_1ZT_1T_2} \| Q_{\bar{Y}|Z}\Gamma^{(2)}_{V|S_2ZT_1T_2}) \big]+I(S_2;V,\bar{Y},Z|S_1,T_1,T_2)-I(S_2;X_1|S_1,T_1,T_2).\label{eq:miss2b_final}
\end{IEEEeqnarray}
Finally, it is straightforward to see:
\begin{IEEEeqnarray}{rCl}
\theta^{\text{miss,12}} &= &\mathbb{E}_{P_{ZT_1T_2}}\big[ D(P_{\bar{Y}V|ZT_1T_2} \|   Q_{\bar{Y}|Z}  \Gamma^{(12)}_{V|ZT_1T_2} ) \big]+
I(S_1,S_2;V,\bar{Y},Z|T_1,T_2)-I(S_1,S_2;X_1,X_2|T_1,T_2).\label{eq:miss12_final}
\end{IEEEeqnarray}
\end{subequations}

Notice now that 
\begin{IEEEeqnarray}{rCl}
\lefteqn{\mathbb{E}_{P_{S_2ZT_1T_2}}\big[ D(P_{\bar{Y}V|S_2ZT_1T_2} \|   Q_{\bar{Y}|Z} \Gamma^{(1)}_{V|S_2ZT_1T_2}  ) \big]} \quad \nonumber \\
& = & \sum_{s_2,z, \bar{y},t_1,t_2,v} P_{S_2Z\bar{Y}T_1T_2V} (s_2,z, \bar{y},t_1,t_2,v) \log  \left( \frac{ P_{\bar{Y}V|S_2ZT_1T_2} (\bar{y},v |s_2,z,t_1,t_2)}{  \Gamma^{(1)}_{V|S_2ZT_1T_2} (v|s_2,z,t_1,t_2) Q_{\bar{Y}|Z}(\bar{y}|z)} \cdot \frac{ \Gamma_{V|S_2ZT_1T_2} (v|s_2,z,t_1,t_2) }{\Gamma_{V|S_2ZT_1T_2} (v|s_2,z,t_1,t_2)   } \right)\nonumber \\[1.2ex]
&\stackrel{(a)}{=} & \mathbb{E}_{P_{S_2ZT_1T_2V}}\big[ D(  P_{\bar{Y}|S_2ZT_1T_2V} \| Q_{\bar{Y}|Z} )\big] + \mathbb{E}_{P_{S_2ZT_1T_2}}\big[ D(  \Gamma_{V|S_2ZT_1T_2} \| \Gamma^{(1)}_{V|S_2ZT_1T_2})\big] \nonumber \\
& \stackrel{(b)}{\geq } &  \mathbb{E}_{P_{S_2ZT_1T_2V}}\big[ D(  P_{\bar{Y}|S_2ZT_1T_2V} \| Q_{\bar{Y}|Z} )\big] , \label{eq:channel_KLplus}
\end{IEEEeqnarray}
where $(a)$ holds because for the present example with $Q_{X_1X_2Z}=P_{X_1X_2Z}$ we have $P_{V|S_2ZT_1T_2} =\Gamma_{V|S_2ZT_1T_2}$ and $(b)$ holds because KL-divergence is nonnegative. 

Comparing \eqref{eq:dec1_final} with \eqref{eq:miss1b_final}, in view of \eqref{eq:channel_KLplus}, we see that exponent $\theta^{\text{miss},1\textnormal{b}}$ is redundant in view of exponent $\theta^{\text{dec},1}$. In the same way, it can be shown that $\theta^{\text{miss},2\textnormal{b}}$ is redundant in view of $\theta^{\text{dec},2}$ and the three exponents $\theta^{\text{miss},1\textnormal{a}}, \theta^{\text{miss},2\textnormal{a}},\theta^{\text{miss},12}$ are redundant in view of $\theta^{\text{dec},12}$.

We thus conclude that in this example, any error exponent $\theta$ satisfying
\begin{equation} \label{eq:theta2}
\theta \leq \max \min \{\theta^{\text{standard}},\theta^{\text{dec},1}, \theta^{\text{dec},2},\theta^{\text{dec},12}\}
\end{equation}
is achievable, where  $\theta^{\text{dec},1}, \theta^{\text{dec},2},\theta^{\text{dec},12}$ are given in \eqref{eq:decoding_final} and $\theta^{\textnormal{standard}}$ can be simplified to:
\begin{IEEEeqnarray}{rCl}
\theta^{\text{standard}} &= & \mathbb{E}_{{P}_{S_1S_2ZT_1T_2V}} \big[ D({P}_{\bar{Y}|S_1S_2ZT_1T_2V} \|  Q_{\bar{Y}|Z} ) \big].
 \nonumber \\
& = &  \mathbb{E}_{{P}_{T_1T_2Z}} \big[ D({P}_{\bar{Y}|ZT_1T_2V} \|  Q_{\bar{Y}|Z} ) \big] + I(S_1,S_2;\bar{Y}|Z,T_1,T_2,V) 
\label{eq:standard_final}
\end{IEEEeqnarray}

The proof of the corollary is finally concluded by showing that if the pmfs $P_{S_1|X_1T_1T_2}$ and $P_{S_2|X_2T_1T_2}$ and the functions $f_1$ and $f_2$ are chosen to satisfy inequalities \eqref{eq:cond_ind}, then the minimum in \eqref{eq:theta2} is attained by $\theta^{\text{standard}}$. In fact, by the Markov chain $S_1 - X_1 - X_2- S_2$ and by expanding KL-divergences, one can show that:
\begin{IEEEeqnarray}{rCl}
\theta^{\text{dec,1}} & \geq  & \mathbb{E}_{{P}_{S_2 ZT_1T_2V}} \big[ D({P}_{\bar{Y}|S_2 ZT_1T_2V} \|  Q_{\bar{Y}|Z}  )\big] +I(S_1;\bar{Y}|S_2,Z,T_1,T_2,V)\nonumber  \\
& = & \mathbb{E}_{{P}_{ZT_1T_2V}} [ D({P}_{\bar{Y}| ZT_1T_2V} \|  Q_{\bar{Y}|Z}  )\big] +I(S_1,S_2;\bar{Y}|Z,V,T_1,T_2) \nonumber \\
& = & \theta^{\text{standard}} 
\end{IEEEeqnarray}
and by similar arguments also 
\begin{IEEEeqnarray}{rCl}
\theta^{\text{dec,2}} & \geq & \theta^{\text{standard}} \\
\theta^{\text{dec,12}} &  \geq &\theta^{\text{standard}} .
\end{IEEEeqnarray}

\section{Proof of Converse for Theorem \ref{thm2opt}}\label{app:conv}
All mutual informations are calculated with respect to the pmfs under $\mathcal{H}=0$. Define $\bar{S}_{1,t} := (V_1^n,X_1^{t-1})$ and $\bar{S}_{2,t} := (V_2^n,X_2^{t-1})$. Notice that the Markov chains $\bar{S}_{1,t}\to X_{1,t}\to \bar{S}_{2,t}$ and $\bar{S}_{2,t}\to X_{2,t}\to S_{1,t}$ hold. Define $\delta(\epsilon) := H(\epsilon)/n/(1-\epsilon)$ as in \cite{Kim}. Then:
\begin{align}
\theta &\leq \frac{1}{n(1-\epsilon)}D(P_{V^nY^n|\mathcal{H}=0}\|P_{V^nY^n|\mathcal{H}=1})+\delta(\epsilon)\nonumber\\
&=\frac{1}{n(1-\epsilon)}\mathbb{E}_{P_{Y^n}}\big[D(P_{V^n|Y^n,\mathcal{H}=0}\|P_{V^n|Y^n,\mathcal{H}=1})\big]+\frac{1}{1-\epsilon}\cdot D(P_Y\| Q_Y)+\delta(\epsilon)\nonumber\\
&=\frac{1}{n(1-\epsilon)}\mathbb{E}_{P_{Y^n}}\big[D(P_{V^n|Y^n,\mathcal{H}=0}\|P_{V^n|\mathcal{H}=1})\big]+\frac{1}{1-\epsilon}\cdot D(P_Y\| Q_Y)+\delta(\epsilon)\nonumber\\
&=\frac{1}{n(1-\epsilon)}I(V^n;Y^n)+\frac{1}{1-\epsilon}\cdot D(P_Y\| Q_Y)+\delta(\epsilon)\nonumber\\
&=\frac{1}{n(1-\epsilon)}\sum_{t=1}^n I(V^n,Y^{t-1};Y_t)+\frac{1}{1-\epsilon}\cdot D(P_Y\| Q_Y)+\delta(\epsilon)\nonumber\\
&=\frac{1}{n(1-\epsilon)}\sum_{t=1}^n I(V^n,Y^{t-1};Y_t)+\frac{1}{1-\epsilon}\cdot D(P_Y\| Q_Y)+\delta(\epsilon)\nonumber\\
&\stackrel{(a)}{\leq} \frac{1}{n(1-\epsilon)}\sum_{t=1}^n I(V^n,X_1^{t-1},X_2^{t-1};Y_t)+\frac{1}{1-\epsilon}\cdot D(P_Y\| Q_Y)+\delta(\epsilon)\nonumber\\
&=\frac{1}{n(1-\epsilon)}\sum_{t=1}^n I(\bar{S}_{1,t},\bar{S}_{2,t};Y_t)+\frac{1}{1-\epsilon}\cdot D(P_Y\| Q_Y)+\delta(\epsilon),\nonumber\\
&=\frac{1}{1-\epsilon}I(\bar{S}_1,\bar{S}_2;Y)+\frac{1}{1-\epsilon}\cdot D(P_Y\| Q_Y)+\delta(\epsilon),
\end{align}\color{black}
where $(a)$ follows from the Markov chain $Y^{t-1}\to (V^n,X_1^{t-1},X_2^{t-1})\to Y_t$. The last equality holds by defining a time-sharing random variable  $Q$ that is uniform over $\{1,\ldots, n\}$ and $\bar{S}_i:=(Q,V_i^n, X_i^{Q-1})$, for $i\in\{1,2\}$,  and $Y:=Y_Q$.

Next, consider the following term,
\begin{align}
&I(X_1^n;V_1^n|V_2^n) \nonumber\\&= \sum_{t=1}^n I(X_{1,t};V_1^n|X_1^{t-1},V_2^n)\nonumber\\
&\stackrel{(b)}{=} \sum_{t=1}^n I(X_{1,t};X_1^{t-1},V_1^n|V_2^n)\nonumber\\
&\stackrel{(c)}{=} \sum_{t=1}^n I(X_{1,t};X_1^{t-1},V_1^n,X_2^{t-1}|V_2^n)\nonumber\\
&\geq \sum_{t=1}^n I(X_{1,t};X_1^{t-1},V_1^n|X_2^{t-1},V_2^n)\nonumber\\
&=\sum_{t=1}^n I(X_{1,t};\bar{S}_{1,t}|\bar{S}_{2,t})\nonumber\\
&=nI(X_1;\bar{S}_1|\bar{S}_2)
\end{align}
where $(b)$ and $(c)$ follow from the Markov chains $X_{1,t}\to V_2^n\to X_1^{t-1}$ and $X_{1,t}\to (V_1^n,V_2^n,X_1^{t-1})\to X_2^{t-1}$, respectively. Both Markov chains hold because $X_{1}^n$ and $X_2^n$ are independent under both hypotheses and by the orthogonality of the MAC. The last equality holds by defining $X_i:=(Q,X_{i,Q})$,  for $i\in\{1,2\}$. Notice that $\bar{S}_i \to X_i \to S_i$.

Similarly, we get
\begin{align}
I(X_2^n;V_2^n|V_1^n) &\geq nI(X_2;\bar{S}_2|\bar{S}_1),\\
I(X_1^n,X_2^n;V_1^n,V_2^n) &\geq nI(X_1,X_2;\bar{S}_1,\bar{S}_2).
\end{align}
On the other hand, we have
\begin{align}
&I(X_1^n;V_1^n|V_2^n) \nonumber\\&\leq I(W_1^n;V_1^n|V_2^n)\nonumber\\
&=H(V_1^n|V_2^n)-H(V_1^n|W_1^n,V_2^n)\nonumber\\
&\leq H(V_1^n)-H(V_1^n|W_1^n,V_2^n)\nonumber\\
&\stackrel{(d)}{=}H(V_1^n)-H(V_1^n|W_1^n)\nonumber\\
&=I(W_1^n;V_1^n)\nonumber\\
&\leq \sum_{t=1}^n I(W_{1,t};V_{1,t})\nonumber\\
&=nI(W_1;V_1)\nonumber\\
&\leq nC_1,
\end{align}
where $(d)$ follows from the Markov chain $V_1^n\to W_1^n\to V_2^n$ and the orthogonality assumption. The last equality holds  by defining $W_i:=(Q,W_{i,Q})$ and $V_i=V_{i,Q}$ for $i\in\{1,2\}$.
Similarly, we have
\begin{align}
I(X_2^n;V_2^n|V_1^n) &\leq nC_2,\\
I(X_2^n,X_1^n;V_1^n,V_2^n) &\leq nC_1+nC_2.
\end{align}
Appropriately combining  the derived inequalities concludes the proof of the converse.

\section{Converse  Proof for Proposition~\ref{separation-ortho}}\label{separation-proof}
The proof follows similar steps to \cite{Luo}. Define $\delta(\epsilon) := H(\epsilon)/n/(1-\epsilon)$ as in \cite{Kim}. First, consider the error exponent:
\begin{IEEEeqnarray}{rCl}
\theta &\leq & \frac{1}{n(1-\epsilon)}D(P_{V^nY^n|\mathcal{H}=0}\|P_{V^nY^n|\mathcal{H}=1})+\delta(\epsilon)\nonumber\\
&=&\frac{1}{n(1-\epsilon)}\mathbb{E}_{P_{Y^n}}\big[D(P_{V^n|Y^n,\mathcal{H}=0}\|P_{V^n|Y^n,\mathcal{H}=1})\big]+\frac{1}{1-\epsilon}\cdot D(P_Y\| Q_Y)+\delta(\epsilon)\nonumber\\
&=&\frac{1}{n(1-\epsilon)}\mathbb{E}_{P_{Y^n}}\big[D(P_{V^n|Y^n,\mathcal{H}=0}\|P_{V^n|\mathcal{H}=1})\big]+\frac{1}{1-\epsilon}\cdot D(P_Y\| Q_Y)+\delta(\epsilon)\nonumber\\
&=&\frac{1}{n(1-\epsilon)}I(V^n;Y^n)+\frac{1}{1-\epsilon}\cdot D(P_Y\| Q_Y)+\delta(\epsilon)\nonumber\\
&=&\frac{1}{n(1-\epsilon)}I(V_1^n,V_2^n;Y^n)+\frac{1}{1-\epsilon}\cdot D(P_Y\| Q_Y)+\delta(\epsilon).
\end{IEEEeqnarray}
Next, consider the following set of inequalities:

\begin{IEEEeqnarray}{rCl}
I(X_1^n;V_1^n|V_2^n) &\leq &I(W_1^n;V_1^n|V_2^n)\nonumber\\
&=&H(V_1^n|V_2^n)-H(V_1^n|W_1^n,V_2^n)\nonumber\\
&\leq& H(V_1^n)-H(V_1^n|W_1^n,V_2^n)\nonumber\\
&\stackrel{(a)}{=}&H(V_1^n)-H(V_1^n|W_1^n)\nonumber\\
&=&I(W_1^n;V_1^n)\nonumber\\
&\leq &\sum_{t=1}^n I(W_{1,t};V_{1,t})\nonumber\\
&=&nI(W_1;V_1)\nonumber\\
&\leq & nC_1,
\end{IEEEeqnarray}
where $(a)$ follows from the Markov chain $V_2^n\to W_1^n\to V_1^n$. Similarly, we have
\begin{IEEEeqnarray}{rCl}
	I(X_2^n;V_2^n|V_1^n) &\leq & nC_2,\\
	I(X_1^n,X_2^n;V_1^n,V_2^n) &\leq & n(C_1+C_2).
	\end{IEEEeqnarray}
Defining the auxiliaries $S_1^n$ and $S_2^n$ to be $V_1^n$ and $V_2^n$, respectively, considering the Markov chains $V_1^n\to X_1^n\to V_2^n$, $V_2^n\to X_2^n\to V_1^n$ and letting $\epsilon\to 0$ completes the proof of the theorem.

\section{Proof of Example~\ref{ex:Gauss_MAC}}\label{gcap-proof}
We simplify the result of Theorem \ref{thm2opt} for the proposed Gaussian setup. Notice  first that since $X_1$ and $X_2$ are independent and because of the Markov chains $\bar S_1 \to X_1 \to X_2$ and $\bar{S}_2 \to X_2 \to X_1$, the pair $(X_1,\bar{S}_1)$ is independent of $(X_2,\bar{S}_2)$. As a consequence,
\begin{align}
I(\bar{S}_1;X_1|\bar{S}_2)&=I(\bar{S}_1;X_1)\\
I(\bar{S}_2;X_2|\bar{S}_1)&=I(\bar{S}_2;X_2)\\
I(\bar{S}_1,\bar{S}_2;X_1,X_2)&=I(\bar{S}_1;X_1)+I(\bar{S}_2;X_2),
\end{align}
and the three constraints in the maximization of \eqref{eq:exp_opt} simplify to the two constraints:
\begin{subequations}\label{cons-or}
\begin{align}
&I(\bar{S}_1;X_1)\leq C_1,\label{cons1-or}\\
&I(\bar{S}_2;X_2)\leq C_2.\label{cons2-or}
\end{align}
\end{subequations}

Choose now the auxiliary random variables $\bar{S}_1$ and $\bar{S}_2$ as 
\begin{align}
\bar{S}_i=X_i+F_i,\qquad F_i\sim \mathcal{N}(0,\xi_i^2),\qquad i\in\{1,2\},
\end{align}
where
\begin{align}
\xi_i^2 := \frac{1}{2^{2C_i}-1},\qquad i\in\{1,2\}.
\end{align}
 It is easily checked that this choice satisfies  constraints  \eqref{cons-or}.  
Moreover, the mutual information term in the achievable error exponent evaluates to:
\begin{align}
I(\bar{S}_1,\bar{S}_2;Y) = \frac{1}{2}\log \frac{2+\sigma_0^2}{\sigma_0^2+\frac{\xi_1^2}{1+\xi_1^2}+\frac{\xi_2^2}{1+\xi_2^2}}= \frac{1}{2} \log \frac{2+\sigma_0^2}{ 2^{-2C_1}+2^{-2C_2}+\sigma_0^2},\label{MI-or}
\end{align}
and the KL-divergence term to:
\begin{align}
D(P_{Y}\|Q_Y) &= -h(Y)+\mathbb{E}_{P_Y}\left[\log \frac{1}{Q_Y}\right]\nonumber\\
&=-h(Y)+\mathbb{E}_{P_Y}\left[\log \left(\sqrt{2\pi \sigma_y^2}e^{\frac{y^2}{2\sigma_y^2}}\right)\right]\nonumber\\
&=-h(Y)+\frac{1}{2}\log \left(2\pi \sigma_y^2\right)+\mathbb{E}_{P_Y}\left[\frac{Y^2}{2\sigma_y^2}\right]\cdot\log e\nonumber\\
&=-h(Y)+\frac{1}{2}\log \left(2\pi \sigma_y^2\right)+\frac{2+\sigma_0^2}{2\sigma_y^2}\cdot \log e\nonumber\\
&= -\frac{1}{2}\log \left(2\pi e (2+\sigma_0^2) \right)+\frac{1}{2}\log \left(2\pi \sigma_y^2\right)+\frac{2+\sigma_0^2}{2\sigma_y^2}\cdot \log e\nonumber\\
&=\frac{1}{2}\log \left(\frac{\sigma_y^2}{2+\sigma_0^2} \right)+\left(\frac{2+\sigma_0^2}{2\sigma_y^2}-\frac{1}{2}\right)\cdot \log e.\label{KL-or}
\end{align}
Combining \eqref{MI-or} and \eqref{KL-or}, by Theorem~\ref{thm2opt}, any error exponent $\theta\geq 0$ is achievable if it satisfies:
\begin{align}
\theta \leq \frac{1}{2} \log \frac{\sigma_y^2}{ 2^{-2C_1}+2^{-2C_2}+\sigma_0^2}+\left(\frac{2+\sigma_0^2}{2\sigma_y^2}-\frac{1}{2}\right)\cdot \log e.
\end{align}

We now show that by Theorem~\ref{thm2opt} no larger exponent is achievable.  Notice first that  since each $X_i$ is standard Gaussian, constraints  \eqref{cons-or} are equivalent to
\begin{align}
2^{2h(X_i|\bar{S}_i)} \geq 2\pi e \cdot 2^{-2C_i}, \quad  i\in\{1,2\}.\label{ineq1}
\end{align}
Then, by Theorem~\ref{thm2opt}, any exponent has to satisfy:
\begin{align}
\theta &\leq D(P_Y\|Q_Y)+  \max_{\substack{\bar{S}_1,\bar{S}_2\\ \textnormal{ s.t. } \eqref{ineq1}}} I(\bar{S}_1,\bar{S}_2;Y)\nonumber\\
&=D(P_Y\|Q_Y)+h(Y)-\min_{\substack{\bar{S}_1,\bar{S}_2\\ \textnormal{ s.t. } \eqref{ineq1}}} h(Y|\bar{S}_1,\bar{S}_2) \nonumber\\
&=D(P_Y\|Q_Y)+h(Y)- \min_{\substack{\bar{S}_1,\bar{S}_2\\ \textnormal{ s.t. } \eqref{ineq1}}} h(Y|\bar{S}_1,\bar{S}_2)\nonumber\\
&\stackrel{(a)}{\leq}D(P_Y\|Q_Y)+h(Y)-\min_{\substack{\bar{S}_1,\bar{S}_2\\ \textnormal{ s.t. } \eqref{ineq1}}} \frac{1}{2} \log \left( 2\pi e \left(\frac{1}{2\pi e}2^{2h(X_1|\bar{S}_1,\bar{S}_2)}+\frac{1}{2\pi e}2^{2h(X_2|\bar{S}_1,\bar{S}_2)}+\frac{1}{2\pi e}2^{2h(N_0|\bar{S}_1,\bar{S}_2)}\right) \right)\nonumber\\
&\stackrel{(b)}{=}D(P_Y\|Q_Y)+h(Y)-\min_{\substack{\bar{S}_1,\bar{S}_2\\ \textnormal{ s.t. } \eqref{ineq1}}}\frac{1}{2} \log \left( 2\pi e \left(\frac{1}{2\pi e}2^{2h(X_1|\bar{S}_1)}+\frac{1}{2\pi e}2^{2h(X_2|\bar{S}_2)}+\sigma_0^2\right) \right)\nonumber\\
&\stackrel{(c)}{\leq}D(P_Y\|Q_Y)+h(Y)-\frac{1}{2} \log \Big( 2\pi e \Big(2^{-2C_1}+2^{-2C_2}+\sigma_0^2 \Big)\Big)\nonumber\\
&=\frac{1}{2}\log \left( \frac{\sigma_y^2}{ 2^{-2C_1}+2^{-2C_2}+\sigma_0^2} \right)+\left(\frac{2+\sigma_0^2}{2\sigma_y^2}-\frac{1}{2}\right)\cdot \log e,
\end{align}
where $(a)$ follows the from conditional EPI and  the fact that given $(\bar{S}_1,\bar{S}_2)$, the three random variables $X_1$, $X_2$, and $N_0$ are independent; $(b)$ follows because $X_1$ is independent of $\bar{S}_2$, $X_2$ is independent of $\bar{S}_1$ and $N_0$ is  independent of both $(\bar{S}_1,\bar{S}_2)$; and $(c)$ follows by \eqref{ineq1}. This concludes the proof.

\section{Proof of Corollary~\ref{cor-gaussian-separate}}\label{app:hybrid}
We evaluate the exponent in Corollary~\ref{corr-mac-dif-marginal} for the following choice of \emph{Gaussian}  auxiliary random variables. Let
$F_1,F_2,G_1,G_2$ be independent zero-mean Gaussian random variables of variances $\xi^2, \xi^2, \gamma^2, \gamma^2$ and independent of the source variables $(X_1,X_2,Y)$. Then, define
\begin{IEEEeqnarray}{rCl}
	\bar{S}_i :=X_i +G_i,\qquad i\in\{1,2\},
\end{IEEEeqnarray}
and
\begin{align}
S_i = (\bar{S}_i,F_i),\qquad i\in\{1,2\}.
\end{align}
We apply hybrid coding with channel inputs:
\begin{align}
W_i = \alpha X_i+\beta G_i+ F_i,\label{hybaux}
\end{align}
for some real numbers $\alpha$ and $\beta$ such that
\begin{align}
\gamma^2+\alpha^2 + \beta^2\cdot \xi^2 = P.
\end{align}

We first investigate for which  parameters $\alpha, \beta, \gamma, \xi$, the presented choice of random variables satisfies the three constraints in the corollary. Notice first that:
\begin{subequations}
	\label{eq:SV}
\begin{align}
I(S_1;V|S_2)
&=\frac{1}{2}\log \left(\sigma^2+ 2P-\gamma^2+2\alpha^2\rho-\frac{(\alpha\cdot (1+\rho)+\beta \cdot \xi^2)^2}{1+\xi^2}\right)-\frac{1}{2} \log \Big( \sigma^2+\frac{2(\alpha-\beta)^2\cdot (1+\rho)\xi^2}{1+\rho+\xi^2}\Big),\label{gh1-new}\\[1.2ex]
I(S_2;V|S_1)&=  \frac{1}{2}\log \left(\sigma^2+ 2P+2\alpha^2\rho-\frac{(\alpha\cdot (1+\rho)+\beta \cdot \xi^2)^2}{1+\xi^2}\right)-\frac{1}{2} \log \left( \sigma^2+\frac{2(\alpha-\beta)^2\cdot (1+\rho)\xi^2}{1+\rho+\xi^2}\right),\label{gh2-new}
\end{align}
and
\begin{align}
I(S_1,S_2;V) &= \frac{1}{2}\log\left( \frac{\sigma^2+2P+2\alpha^2\rho}{\sigma^2+\frac{2(\alpha-\beta)^2\cdot (1+\rho)\xi^2}{1+\rho+\xi^2}} \right).\label{gh3-new}
\end{align}
\end{subequations}
Moreover,
\begin{subequations}
	\label{eq:SX}
\begin{align}
I(\bar{S}_1;X_1|\bar{S}_2) &= \frac{1}{2}\log \left( \frac{(1+\xi^2)^2-\rho^2}{(1+\xi^2)\xi^2}\right),\label{gh4-new}\\
I(\bar{S}_2;X_2|\bar{S}_1) &= \frac{1}{2}\log \left( \frac{(1+\xi^2)^2-\rho^2}{(1+\xi^2)\xi^2}\right),\label{gh5-new}
\end{align}and
\begin{align}
I(\bar{S}_1,\bar{S}_2;X_1,X_2) &=\frac{1}{2}\log \left( \frac{(1+\xi^2)^2-\rho^2}{\xi^4}\right).\label{gh6-new}
\end{align}
\end{subequations}
Combining \eqref{eq:SV} and \eqref{eq:SX}, shows  that the presented choice of auxiliaries satisfies the three constraints \eqref{con-ind1}--\eqref{con-ind3} in Corollary~\ref{corr-mac-dif-marginal}, whenever the two conditions~\eqref{eq:condhyb} are satisfied.

We now evaluate the error exponent  \eqref{ghyb-ach-theta} for the proposed choice of auxiliaries. To this end, notice that 
\begin{IEEEeqnarray}{rCl}
	\mathbb{E}_{P_V}\left[D(P_{Y|V}\| Q_Y)\right] +I(S_1,S_2;Y|V) & =& D(P_Y\| Q_Y) +I(V;Y) +I(S_1,S_2;Y) \nonumber \\
	& = &   D(P_Y\| Q_Y)+I(S_1,S_2,V;Y). \label{gaus-eq1}	
	\end{IEEEeqnarray}
Moreover, 
\begin{align}
I(S_1,S_2,V;Y) 
&=\frac{1}{2}\log \left( \sigma_0^2+2+2\rho\right)-\frac{1}{2}
\log \left( \sigma_0^2+\frac{2\xi^2(1+\rho)\sigma^2}{2\xi^2(\alpha-\beta)^2\cdot (1+\rho)+\sigma^2(1+\rho+\xi^2)} \right)\label{gh7-new}
\end{align}
and (by similar steps as in \eqref{KL-or}):
\begin{align}
D(P_{Y}\|Q_Y) &= -h(Y)+\mathbb{E}_{P_Y}\left[\log \frac{1}{Q_Y}\right]\nonumber\\
&=\frac{1}{2}\log \left(\frac{\sigma_y^2}{2+2\rho+\sigma_0^2} \right)+\left(\frac{2+2\rho+\sigma_0^2}{2\sigma_y^2}-\frac{1}{2}\right)\cdot \log e.\label{gh8-new}
\end{align}
Combining 
\eqref{gh7-new} and \eqref{gh8-new} yields the error exponent  in \eqref{ghyb-ach-theta}. This concludes the proof.

\section{Proof of Theorem \ref{gout-thm}}\label{gout-proof}

Fix a blocklength $n$ and encoding and decoding/testing functions. Then, notice that by Witsenhausen's max-correlation argument \cite{Witsenhausen}, see also \cite{LapidothTinguely}, 
\begin{IEEEeqnarray}{rCl}
	\frac{1}{2}\log \left( 1+\frac{2P(1+\rho)}{\sigma^2} \right)&\geq & \frac{1}{n}I(W_1^n,W_2^n;V^n)\nonumber\\
	& \stackrel{(a)}{\geq} &  \frac{1}{n}I(X_1^n,X_2^n;V^n)\nonumber\\
	& = & \frac{1}{n}h(X_1^n,X_2^n)-\frac{1}{n}h(X_1^n,X_2^n|V^n)\nonumber\\
	&\stackrel{(b)}{=}&\frac{1}{n}h(X_1^n,X_2^n) -\frac{1}{n}h(X_1^n+X_2^n, X_1^n - X_2^n|V^n) +1 \nonumber\\
		&=&\frac{1}{n}h(X_1^n,X_2^n) -\frac{1}{n}h(X_1^n+X_2^n |V^n)  -\frac{1}{n}h(X_1^n-X_2^n |X_1^n+X_2^n, V^n) +1 \nonumber\\
				&\stackrel{(c)}{\geq} &\frac{1}{n}h(X_1^n,X_2^n) -\frac{1}{n}h(X_1^n+X_2^n |V^n)  -\frac{1}{n}h(X_1^n-X_2^n) +1 \nonumber\\
	&=& \frac{1}{2} \log \Big( (2\pi e)\cdot (2+2\rho) \Big)-\frac{1}{n}h(X_1^n+X_2^n|V^n),\label{eq:condjoint}
	\end{IEEEeqnarray}
where $(a)$ holds by  the Markov chain $(X_1^n,X_2^n)\to (W_1^n,W_2^n)\to V^n$; $(b)$ holds because for each $t$ the vector $(X_{1,t}+ X_{2,t}, X_{1,t}- X_{2,t})$ is obtained from $(X_1,X_2)$ by  rotating it with the matrix 
\begin{equation}
A=\begin{pmatrix} 1 & 1 \\ 1 & -1\end{pmatrix},
\end{equation} and because for any bivariate vector $\mathbf{X}$ differential entropy satisfies $h(A \mathbf{X})= h(\mathbf{X})+\log|A| =h(\mathbf{X})+1$; and $(c)$ holds because conditioning cannot increase differential entropy.  
Inequality~\eqref{eq:condjoint} is equivalent to:
\begin{align}
2^{\frac{2}{n}h(X_1^n+X_2^n|V^n)} \geq 2\pi e\cdot \frac{ 2(1+\rho) \sigma^2}{2P(1+\rho)+\sigma^2}.\label{C3}
\end{align}

We proceed to upper bound the error exponent. Define $\delta(\epsilon) := H(\epsilon)/n/(1-\epsilon)$. Then, 
\begin{IEEEeqnarray}{rCl}
\theta &\leq & \frac{1}{n(1-\epsilon)}D(P_{V^nY^n|\mathcal{H}=0}\|P_{V^nY^n|\mathcal{H}=1})+\delta(\epsilon)\nonumber\\
&=&\frac{1}{1-\epsilon}\cdot D(P_Y\| Q_Y)+\frac{1}{n(1-\epsilon)}\mathbb{E}_{P_{Y^n}}\big[D(P_{V^n|Y^n,\mathcal{H}=0}\|P_{V^n|Y^n,\mathcal{H}=1})\big]+\delta(\epsilon)\nonumber\\
&=&\frac{1}{1-\epsilon}\cdot D(P_Y\| Q_Y)+\frac{1}{n(1-\epsilon)}\mathbb{E}_{P_{Y^n}}\big[D(P_{V^n|Y^n,\mathcal{H}=0}\|P_{V^n|\mathcal{H}=1})\big]+\delta(\epsilon)\nonumber\\
&=&\frac{1}{1-\epsilon}\cdot D(P_Y\| Q_Y)+\frac{1}{n(1-\epsilon)}I(V^n;Y^n)+\delta(\epsilon)\nonumber\\
&= &\frac{1}{1-\epsilon}\cdot D(P_Y\| Q_Y)+\frac{1}{n(1-\epsilon)}\big[ h(Y^n)-h(Y^n|V^n) \big]+\delta(\epsilon)\nonumber\\
&=&\frac{1}{1-\epsilon}\cdot \Big[ D(P_Y\|Q_Y)+h(Y)\Big]- \frac{1}{n(1-\epsilon)}h(Y^n|V^n)+\delta(\epsilon)\nonumber\\
&\stackrel{(d)}{=}&\frac{1}{1-\epsilon}\cdot \Big[ D(P_Y\|Q_Y)+h(Y)\Big]- \frac{1}{n(1-\epsilon)}h(X_1^n+X_2^n+N_0^n|V^n)+\delta(\epsilon)\nonumber\\
&\stackrel{(e)}{\leq} &\frac{1}{1-\epsilon}\cdot \Big[ D(P_Y\|Q_Y)+h(Y)\Big]- \frac{1}{2(1-\epsilon)}\log\left( 2\pi e \left(\frac{1}{2\pi e}2^{\frac{2}{n}{h(X_1^n+X_2^n|V^n)}}+\sigma_0^2 \right)\right)+\delta(\epsilon)\nonumber\\
&\stackrel{(f)}{\leq} & \frac{1}{1-\epsilon}\cdot \left[ D(P_Y\|Q_Y)+h(Y)\right]- \frac{1}{2(1-\epsilon)}\log\left( 2\pi e \left(\frac{ 2(1+\rho) \sigma^2}{2P(1+\rho)+\sigma^2}+\sigma_0^2\right) \right)+\delta(\epsilon)\nonumber\\
&\stackrel{(g)}{=}&\frac{1}{2(1-\epsilon)}\cdot \left[\log \left(\frac{\sigma_y^2}{\frac{ 2(1+\rho) \sigma^2}{2P(1+\rho)+\sigma^2}+\sigma_0^2 } \right)+\left(\frac{2+2\rho+\sigma_0^2}{\sigma_y^2}-1\right)\cdot \log e\right]+\delta(\epsilon),
\end{IEEEeqnarray}
where $(d)$ follows from the definition of $Y^n$ in \eqref{gdef2}; $(e)$ follows from the conditional EPI and noting that for given $V^n$, the two random variables $N_0^n$ and $X_1^n+X_2^n$ are independent; $(f)$ follows from \eqref{C3}; $(g)$ follows from \eqref{gh8-new} The proof is concluded by letting $\epsilon\to 0$.

\section{Proof of Theorem \ref{thm-BC}}\label{BC-proof}

The proof is based on the scheme of Section \ref{BC-feedback}. Fix a choice of blocklength $n$, the small positive $\mu$ and the (conditional) pmfs $p_T$, $p^1_{T_1|T}$, $p^2_{T_2|T}$, $p^1_{W|TT_1}$, $p^2_{W|TT_2}$, $p^1_{S|X}$ and $p^2_{S|X}$ so that \eqref{constraint-BC} holds. Assume that $I_{p^1}(S;X)\geq I_{p^1}(W;V_1|T,T_1)$ and $I_{p^2}(S;X)\geq I_{p^2}(W;V_2|T,T_2)$ in which case $R_1, R_2, R_1', R_2'$ are given by \eqref{rate-BC1} and \eqref{rate-BC2}.  Also, set for convenience of notation:
\begin{IEEEeqnarray}{rCl}
	p^i_{S'}(s) &=&p^i_{S}(s), \qquad \forall s \in \mathcal{S}, \\
	p^i_{W'|TT_i}(w|t,t_i) & =& p^i_{W|TT_i}(w|t,t_i), \qquad \forall t,t_i,w \in \mathcal{W}.
\end{IEEEeqnarray}

The analysis of type-I error probability is similar as in the previous Appendices. The main novelty is that because $p_X^1(x)\neq p_X^2(x)$ for some $x \in\mathcal{X}$, for sufficiently small values of $\mu>0$, the source sequence cannot lie in both $\mathcal{T}_{\mu/2}(p_X^1)$  and $\mathcal{T}_{\mu/2}(p_X^2)$. Details are omitted.

Consider the type-II error probability at Receiver 1  averaged over all random codebooks. Define the following events for $i\in\{1,2\}$:
\begin{align}
&\mathcal{E}_{\text{Tx},i}(m,\ell)\colon \qquad \{(S^n(i;m,\ell),X^n)\in \mathcal{T}_{\mu/2}^n(p^i_{SX}), \qquad (T^n,T_i^n,W^n(i;m))\in \mathcal{T}_{\mu/2}^n(p^i_{TT_iW}) , \qquad W^n(i;m)) \textnormal{ is sent}\},\\
&\mathcal{E}_{\text{Rx},i}(m',\ell')\colon \;\;\;\;\; \{(S^n(i;m',\ell'),Y^n)\in \mathcal{T}_{\mu}^n(p^i_{SY_i}), \qquad (T^n,T_i^n,W^n(i;m'),V_i^n)\in \mathcal{T}_{\mu}^n(p^i_{TT_iW V_i}), \nonumber\\&\hspace{8cm} H_{\text{tp}(S^n(i;m',\ell'),Y_1^n)}(S|Y_i) = \min_{\tilde{l}} H_{\text{tp}(S^n(i;m',\tilde{\ell}),Y_i^n)}(S|Y_i)\}.
\end{align}

Notice that 
\begin{align}
\mathbb{E}_{\mathcal{C}}[\beta_{1,n}] = \Pr[\hat{\mathcal{H}}_1=0|\mathcal{H}=\h_1] \leq \Pr\left[\bigcup_{m',\ell'}\mathcal{E}_{\text{Rx},1}(m',\ell')\Bigg| \mathcal{H}=\h_1\right].
\end{align}
Above probability is upper bounded as:
\begin{IEEEeqnarray}{rCl}
\Pr\left[\bigcup_{m',\ell'}\mathcal{E}_{\text{Rx},1}(m',\ell')\big| \mathcal{H}=\h_1\right]&\leq & \Pr\left[\left(\bigcup_{m',\ell'}\mathcal{E}_{\text{Rx},1}(m',\ell')\right)\cap  \left(\bigcup_{m,\ell}\mathcal{E}_{\text{Tx},1}(m,\ell)\right)\Bigg| \mathcal{H}=\h_1\right] \nonumber\\
&+&\Pr\left[\left(\bigcup_{m',\ell'}\mathcal{E}_{\text{Rx},1}(m',\ell')\right)\cap \left(\bigcap_{m,\ell}\mathcal{E}_{\text{Tx},1}^c(m,\ell)\right)\cap \left(\bigcup_{m,\ell}\mathcal{E}_{\text{Tx},2}(m,\ell)\right)\Bigg| \mathcal{H}=\h_1\right]\nonumber\\
&+&\Pr\left[\left(\bigcup_{m',\ell'}\mathcal{E}_{\text{Rx},1}(m',\ell')\right)\cap \left(\bigcap_{m,\ell}\mathcal{E}_{\text{Tx},1}^c(m,\ell)\right)\cap \left(\bigcap_{m,\ell}\mathcal{E}^c_{\text{Tx},2}(m,\ell)\right)\Bigg| \mathcal{H}=\h_1\right].\nonumber
\end{IEEEeqnarray}
The sum of above probabilities can be upper bounded by the sum of the probabilities of the following events:
\begin{IEEEeqnarray}{rCl}
\mathcal{B}_1&\colon & \left\{\exists (m,\ell)\qquad\qquad\;\; \text{s.t.}\qquad \left(\mathcal{E}_{\text{Tx},1}(m,\ell)\;\;\;\; \text{and}\;\;\;\; \mathcal{E}_{\text{Rx},1}(m,\ell)\right) \right\},\\
\mathcal{B}_2 &\colon & \left\{\exists (m,\ell,\ell')\qquad\;\;\;\;\;\text{with}\qquad \ell\neq \ell'\qquad \text{s.t.}\qquad \left(\mathcal{E}_{\text{Tx},1}(m,\ell)\;\;\;\; \text{and}\;\;\;\; \mathcal{E}_{\text{Rx},1}(m,\ell')\right) \right\},\\
\mathcal{B}_3 &\colon & \left\{\exists (m,m',\ell,\ell')\qquad\text{with}\qquad \ell\neq \ell'\qquad\text{and}\qquad m\neq m'\qquad \text{s.t.}\qquad \left( \mathcal{E}_{\text{Tx},1}(m,\ell)\;\;\;\; \text{and}\;\;\;\; \mathcal{E}_{\text{Rx},1}(m',\ell') \right)\right\},\\
\mathcal{B}_4 &\colon & \left\{\forall (m,\ell) \qquad  \mathcal{E}_{\text{Tx},1}^c(m,\ell)\right\}\;\; \cap\;\; \left\{\exists (m,m',\ell,\ell')\qquad \text{s.t.}\qquad \mathcal{E}_{\text{Tx},2}(m,\ell)\;\; \cap\;\;   \mathcal{E}_{\text{Rx},1}(m',\ell') \right\},\\
\mathcal{B}_5 &\colon & \left\{\forall (m,\ell) \qquad  \mathcal{E}_{\text{Tx},1}^c(m,\ell)\qquad \text{and}\qquad \mathcal{E}_{\text{Tx},2}^c(m,\ell) \right\}\;\;\; \cap\;\; \left\{\exists (m',\ell')\qquad\;\; \text{s.t.}\qquad \mathcal{E}_{\text{Rx},1}(m',\ell')\right\}.
\end{IEEEeqnarray}
Thus, we have
\begin{align}
\mathbb{E}_{\mathcal{C}}\big[ \beta_{1,n}\big] \leq \sum_{i=1}^5 \Pr \big[ \mathcal{B}_i \big| \mathcal{H}=\h_1\big].
\end{align}
The probabilities of events $\mathcal{B}_1$, $\mathcal{B}_2$, $\mathcal{B}_3$ and $\mathcal{B}_5$ can be bounded following similar steps to Appendix~\ref{sec:proof}. This yields:
\begin{align}
 \Pr \big[ \mathcal{B}_1 \big| \mathcal{H}=\h_1 \big]& \leq 2^{-n\left(\theta_{\mu,\text{standard},1}-\delta_1(\mu)\right)},\\
 \Pr\big[\mathcal{B}_2\big|\mathcal{H}=\h_1\big]&\leq 2^{-n\left(\theta_{\mu,\text{dec},1}-\delta_2(\mu)\right)},\\
 \Pr \big[ \mathcal{B}_3 \big| \mathcal{H}=\h_1\big]   &\leq 2^{-n\left(\theta_{\mu,\text{dec},1}-\delta'_2(\mu)\right)},\\
\Pr \big[ \mathcal{B}_5 \big| \mathcal{H}=\h_1 \big]&\leq 2^{-n\left(\theta_{\mu,\text{miss},1}-\delta_4(\mu)\right)},
\end{align}
for some functions $\delta_1(\mu)$, $\delta_2(\mu)$, $\delta'_2(\mu)$ and $\delta_4(\mu)$ that go to zero as $n$ goes to infinity and  $\mu \to 0$, and where we define:
\begin{IEEEeqnarray}{rCl}
\theta_{\text{standard},i} &:= &\min_{\substack{\tilde{P}_{SXY_i}:\\ |\pi_{SX}-p^i_{SX}|< \mu/2 \\ |\pi_{SY_i}-{p}^{i}_{SY_i} | <\mu}} D(\pi_{SXY_i}\| p^i_{S|X}q^i_{XY_i}) ,\\[2ex]
\theta_{\text{dec},i} &:=&\!\!\!\!\!\!\!\! \min_{\substack{\tilde{P}_{SXY_i}:\\|\pi_{SX}-p^i_{SX} |< \mu/2 \\ |\pi_{Y_i}-p^i_{Y_i}| <\mu\\ H_{p^i}(S|Y_i)\leq H_{\pi}(S|Y_i)}}\!\!\!\!\!\!\! D(\pi_{SXY_i}\| p^i_{S|X}q^i_{XY_i})-I_{p^i}(S;X|Y_i)+ I_{p^i}(W;V_i|T,T_i),\\[2ex]
\theta_{\text{miss},i} &:= &D(p^i_{Y_i}\|q^i_{Y_i})+\mathbb{E}_{p_T}\left[ D\big(p^{i}_{V_i|T}\| \Gamma_{V_i|W=T}\big)\right]-I_{{p^i}}(S;X|Y_i)+I_{p^i}(W;V_i|T,T_i).
\end{IEEEeqnarray}

Consider event $\mathcal{B}_4$:
\begin{align}
& \Pr \big[ \mathcal{B}_4 | \mathcal{H}=\h_1\big]  
\nonumber\\
& \leq  \sum_{\substack{m,\ell}}\;\;\sum_{m',\ell'} \Pr\Big[ (S^n(2;m,\ell),X^n)\in \mathcal{T}_{\mu/2}^n(p^2_{SX}),\;\; (T^n,W^n(2;m))\in \mathcal{T}_{\mu/2}^n(p^2_{TW}),\;\; W^n(2;m) \textnormal{ is sent},\nonumber \\
& \hspace{3cm}
\;\;(S^n(1;m',\ell'),Y_1^n)\in \mathcal{T}_{\mu}^n(p^1_{SY_1}),   \;\; (T^n,T_1^n,W^n(1;m'),V_1^n)\in \mathcal{T}_{\mu}^n(p^1_{TT_1WV_1}) \nonumber\\
&\hspace{6.3cm}H_{\text{tp}(S^n(1;m',\ell'),Y_1^n)}(S|Y_1) = \min_{\tilde{\ell}} H_{\text{tp}(S^n(1;m',\tilde{\ell}),Y_1^n)}(S|Y_1) \;\;\big|\mathcal{H}=\h_1\Big]\nonumber\\
& \stackrel{(a)}{\leq } \sum_{\substack{m,\ell}}\;\;\sum_{m',\ell'} \Pr\Big[ (S^n(2;m,\ell),X^n)\in \mathcal{T}_{\mu/2}^n(p^2_{SX}),\;\;(S^n(1;m',\ell'),Y_1^n)\in \mathcal{T}_{\mu}^n(p^1_{SY_1}),\nonumber\\[-2ex]
&\hspace{6cm} H_{\text{tp}(S^n(1;m',\ell'),Y_1^n)}(S|Y_1) = \min_{\tilde{\ell}} H_{\text{tp}(S^n(1;m',\tilde{\ell}),Y_1^n)}(S|Y_1) \big|\mathcal{H}=\h_1\Big]\nonumber\\
& \hspace{2.1cm} \cdot \Pr\Big[  (T^n,T_1^n,W^n(1;m'),V_1^n)\in \mathcal{T}_{\mu}^n(p^1_{TT_1WV_1}),\;   (T^n,W^n(2;m))\in \mathcal{T}_{\mu/2}^n(p^2_{TW}) \big|  \nonumber \\
 & \hspace{12cm}\; W^n(2;m) \textnormal{ is sent}, \; \mathcal{H}=\h_1\Big]\nonumber\\
& \stackrel{(b)}{\leq } 2^{n (R_1+R'_1+R_2+R'_2)}\cdot \!\!\!\!\!\!\! \max_{\substack{\pi_{SS'XY_1}:\\ |\pi_{SX}-p^2_{SX}|<\mu/2\\ |\pi_{S'Y_1}-p^1_{SY_1}|<\mu\\ H_{\pi}(S'|Y_1)\leq H_{\pi}(S|Y_1)}}\!2^{-n\left(D\left(\pi_{SS'XY_1}\| p^2_{S} p^1_{S'}q^{1}_{XY_1}\right)-\mu\right)}  \nonumber \\[1.2ex]
 & \hspace{1cm} \cdot  \max_{\substack{\pi_{TT_1W'WV_1}:\\ |\pi_{TW}-{p}^2_{TW}|<\mu/2\\|\pi_{TT_1W'V_1}-{p}^1_{TT_1WV_1}|<\mu}}\!\!\!\!\!\!\ 2^{-n\left(D\left(\pi_{TT_1W'W V_1}\| p_{T}p^1_{T_1|T}p^1_{W'|TT_1}p^2_{W|T} \Gamma_{V_1|W}\right)-\mu\right)},\label{b6firsts}
\end{align}
where  $(a)$ holds because the channel code is drawn independently of the source code and $(b)$ holds by Sanov's theorem.

Define
\begin{align}
\tilde{\theta}_{\mu,\text{cross},1}&:=\min_{\substack{\pi_{SS'XY_1}:\\ |\pi_{SX}-p^2_{SX}|<\mu/2\\ |\pi_{S'Y_1}-p^1_{SY_1}|<\mu\\ H_{\pi}(S'|Y_1)\leq H_{\pi}(S|Y_1)}}\!  D\left(\pi_{SS'XY_1}\| p^2_{S} p^1_{S'}q^{1}_{XY_1}\right) \nonumber \\
 & + \min_{\substack{\pi_{TT_1W'WV_1}:\\ |\pi_{TW}-{p}^2_{TW}|<\mu/2\\|\pi_{TT_1W'V_1}-{p}^1_{TT_1WV_1}|<\mu}} D\left(\pi_{TT_1W'W V_1}\| p_{T}p^1_{T_1|T}p^1_{W'|TT_1}p^2_{W|T} \Gamma_{V_1|W}\right)-R_1-R_2-R'_1-R'_2-2\mu, \label{b6s}
\end{align}
and notice that 
\begin{IEEEeqnarray}{rCl}
\tilde{\theta}_{\mu,\text{cross},1} & \stackrel{(\eqref{rate-BC1}\&\eqref{rate-BC2})}{=}&  \min_{\substack{\pi_{SS'XY_1}:\\ |\pi_{SX}-p^2_{SX}|<\mu/2\\ |\pi_{S'Y_1}-p^1_{SY_1}|<\mu\\ H_{\pi}(S'|Y_1)\leq H_{\pi}(S|Y_1)}}\!  D\left(\pi_{SS'XY_1}\| p^2_{S} p^1_{S'}q^{1}_{XY_1}\right) \nonumber \\[2ex]
 && + \min_{\substack{\pi_{TT_1W'WV_1}:\\ |\pi_{TW}-{p}^2_{TW}|<\mu\\|\pi_{TT_1W'V_1}-{p}^1_{TT_1WV_1}|<\mu}} D\left(\pi_{TT_1W'W V_1}\| p_{T}p^1_{T_1|T}p^1_{W'|TT_1}p^2_{W|T} \Gamma_{V_1|W}\right)-I_{p^1}(S;X)-I_{p^2}(S;X)-4\mu\nonumber\\[6.5ex]
 &  \stackrel{(c)}{=}&  \min_{\substack{\pi_{SS'XY_1}:\\ |\pi_{SX}-q^1_{SX}|<\mu/2\\ |\pi_{S'Y_1}-p^1_{SY_1}|<\mu\\ H_{\pi}(S'|Y_1)\leq H_{\pi}(S|Y_1)}}\!  D\left(\pi_{SS'XY_1}\| q^1_{S} p^1_{S'}q^{1}_{XY_1}\right) \nonumber \\[2ex]
 && + \min_{\substack{\pi_{TT_1W'WV_1}:\\ |\pi_{TW}-{q}^1_{TW}|<\mu\\|\pi_{TT_1W'V_1}-{p}^1_{TT_1WV_1}|<\mu}} D\left(\pi_{TT_1W'W V_1}\| p_{T}p^1_{T_1|T}p^1_{W'|TT_1}q^1_{W|T} \Gamma_{V_1|W}\right)-I_{p^1}(S;X)-I_{q^1}(S;X)-4\mu\nonumber\\[6.5ex]
&\stackrel{\text{(CR)}}{=} & \min_{\substack{\pi_{SS'XY_1}:\\ |\pi_{SX}-q^1_{SX}|<\mu/2\\ |\pi_{S'Y_1}-p^1_{SY_1}|<\mu\\ H_{\pi}(S'|Y_1)\leq H_{\pi}(S|Y_1)}}\!   \Big[ D\left(\pi_{SXY_1}\|q^1_{S|X}q^{1}_{XY_1}\right)+\mathbb{E}_{\pi_{SXY_1}}\left[D(\pi_{S'|SXY_1}\|p^1_{S'}) \right] \Big]-I_{p^1}(S;X)\nonumber\\[2ex]
&&
+ \min_{\substack{\pi_{TT_1W'WV_1}\colon \\ |\pi_{TW}-q^1_{TW}|<\mu\\|\pi_{TT_1W'V_1}-{p}^1_{TT_1WV_1}|<\mu}} \Big[ D(\pi_{TT_1W'W} \|p^1_{TT_1}p^1_{W'|TT_1}q^1_{W|T}) \nonumber \\
 && \hspace{4.5cm} +\mathbb{E}_{TT_1W'W}\left[ D(\pi_{V_1|TT_1W'W}\|\pi_{V_1|TT_1}) + D(\pi_{V_1|TT_1}\|\Gamma_{V_1|W}) \right]   \Big]
-4\mu\nonumber\\[3.5ex]
&\stackrel{\text{(DP)}}{\geq} &
 \min_{\substack{\pi_{SS'XY_1}:\\ |\pi_{SX}-q^1_{SX}|<\mu/2\\ |\pi_{S'Y_1}-p^1_{SY_1}|<\mu\\ H_{\pi}(S'|Y_1)\leq H_{\pi}(S|Y_1)}}\!   \Big[ D\left(\pi_{SXY_1}\|q^1_{S|X}q^{1}_{XY_1}\right)+\mathbb{E}_{\pi_{Y_1}}\left[D(\pi_{S'|Y_1}\|p^1_{S'}) \right] \Big]-I_{p^1}(S;X) \nonumber\\[2ex]
&&+ \min_{\substack{\pi_{TT_1W'WV_1}\colon\\ |\pi_{TW}-{q}^1_{TW}|<\mu\\ |\pi_{TT_1W'V_1}-{p}^1_{TT_1WV_1}|<\mu}} \Big[ \mathbb{E}_{\pi_{TT_1W'}}\left[D(\pi_{V_1|TT_1W'}\| \pi_{V_1|TT_1})+D(\pi_{V_1|TT_1} \| \Gamma_{V_1|W}) \right]-4\mu\nonumber\\[6ex]
&\stackrel{(d)}{=}&\min_{\substack{\pi_{SXY_1}:\\ |\pi_{Y_1}-p^1_{Y_1}|<\mu\\ H_{p^1}(S|Y_1)\leq H_{\pi}(S|Y_1)}}\!    \mathbb{E}_{q^1_{XS}}\left[ D\left(\pi_{Y_1|XS}\|q^{1}_{Y_1|X}\right)\right] +I_{p^1}(S;Y_1) - I_{p^1}(S;X) \nonumber \\[2ex]
& &  + I_{p^1}(V_1;W|T,T_1)+\min_{\substack{\pi_{TT_1WV_1}:\\ |\pi_{TW}-q^1_{TW}|<\mu\\|\pi_{TT_1V_1}-{p}^1_{TT_1V_1}|<\mu}} \mathbb{E}_{\pi_{TT_1W}}\left[D(p^1_{V_1|TT_1} \| \Gamma_{V_1|W}) \right] -\delta_3(\mu)\nonumber\\[3ex]
&=:& \theta_{\mu,\text{cross},1}-\delta_3(\mu)\label{t6finals}
\end{IEEEeqnarray}
for a function $\delta_3(\mu)$ that goes to zero as $\mu\to 0$. Here $(c)$ holds because the condition $p_X^1 \neq p_X^2$ implies that $\h_1 = \hnot_2$ and thus $p^2 =q^1$, and $(d)$ holds by the constraints in the minimizations.

 Combining \eqref{b6firsts}, \eqref{b6s} and \eqref{t6finals}, establishes:
\begin{align}
 \Pr \big[ \mathcal{B}_4 \big| \mathcal{H}=\h_1\big]   \leq 2^{-n\left(\theta_{\mu,\text{cross},1}-\delta_3(\mu)\right)}.\label{dec-ffs6}
\end{align}

The proof of the theorem is concluded by familiar arguments.

\bibliographystyle{IEEEtran}
\bibliography{references}

\end{document}